\documentclass[aps,10pt,prd,twocolumn,nofootinbib,superscriptaddress,preprintnumbers,a4paper,showpacs]{revtex4-1}
\pdfoutput=1
\usepackage[T1]{fontenc}

\usepackage{subfig}
\usepackage{graphicx}
\usepackage{amsmath,amssymb,amsthm,amsfonts}
\usepackage{multirow,array,bm,bbm,esint}
\usepackage{epsf}
\usepackage{float}
\usepackage{slashed}
\newcommand{\psib}{{\overline{\psi}}}
\newcommand{\Qb}{{\overline{Q}}}

\newcommand{\cN}{{\cal N}}
\newcommand{\cO}{{\cal O}}

\def\hf{\frac{1}{2}}
\def\qtr{\frac{1}{4}}

\def\nn{\nonumber}
\def\bec{\begin{center}}
\def\eec{\end{center}}
\def\beq{\begin{equation}}
\def\eeq{\end{equation}}
\def\bea{\begin{eqnarray}}
\def\eea{\end{eqnarray}}

\begin{document}

\title{Complex Langevin Simulations of Zero-dimensional Supersymmetric Quantum Field Theories}

\author{Anosh Joseph}
\email{anoshjoseph@iisermohali.ac.in}

\author{Arpith Kumar}
\email{arpithk@iisermohali.ac.in}

\affiliation{Department of Physical Sciences, Indian Institute of Science Education and Research (IISER) Mohali, Knowledge City, Sector 81, SAS Nagar, Punjab 140306, India}

\date{\today}

\begin{abstract}

We investigate the possibility of spontaneous supersymmetry breaking in a class of zero-dimensional ${\cal N} = 2$ supersymmetric quantum field theories, with complex actions, using complex Langevin dynamics and stochastic quantization. Our simulations successfully capture the presence or absence of supersymmetry breaking in these models. The expectation value of the auxiliary field under twisted boundary conditions was used as an order parameter to capture spontaneous supersymmetry breaking in these models.

\end{abstract}


\maketitle

\tableofcontents

\section{Introduction}
\label{sec:intro}

We can investigate numerous nonperturbative features of quantum field theories using lattice regularized form of the field theory path integral. Monte Carlo methods can be used to reliably extract the physics of such systems. The fundamental idea behind path integral Monte Carlo is to generate field configurations with a probability weight given by the exponential of the negative of the action (in Euclidean spacetime) and then compute the path integral by statistically averaging these importance sampled ensemble of field configurations. However, when the action is complex, for example, when studying QCD at finite density or with a theta term, Chern-Simons gauge theories or chiral gauge theories, it is not straightforward to apply path integral Monte Carlo. In these cases we encounter a {\it complex action problem} or {\it sign problem}. The basic aim of complex Langevin method \cite{Klauder:1983nn, Klauder:1983zm, Klauder:1983sp, Parisi:1984cs} is to overcome this problem by extending the idea of stochastic quantization for ordinary field theoretic systems with real actions to the cases with complex actions. This also leads to complexification of the real dynamical field variables that appear in the original path integral. We can define a stochastic process for the complexified field variables by Langevin equation with a complex action. Then the expectation values in the original path integral are calculated from an average of corresponding quantities over this stochastic process\footnote{Another recently proposed method, which is also based on complexification of the original real field variables, is the Lefschetz thimble method \cite{Cristoforetti:2012su, Fujii:2013sra, DiRenzo:2015foa, Tanizaki:2015rda, Fujii:2015vha, Alexandru:2015xva}.}. See Ref. \cite{Damgaard:1987rr} for a pedagogical review on this method and Ref. \cite{Berger:2019odf} for a recent review in the context of the sign problem in quantum many-body physics.

Complex Langevin dynamics has been used successfully in various models in the recent past \cite{Berges:2005yt, Berges:2006xc, Berges:2007nr, Bloch:2017sex, Aarts:2008rr, Pehlevan:2007eq, Aarts:2008wh, Aarts:2009hn, Aarts:2010gr, Aarts:2011zn}. There have also been studies of supersymmetric matrix models based on complex Langevin dynamics \cite{Ito:2016efb, Ito:2016hlj, Anagnostopoulos:2017gos}. In Ref. \cite{Basu:2018dtm} the authors used complex Langevin simulations to observe Gross-Witten-Wadia \cite{Gross:1980he, Wadia:2012fr, Wadia:1980cp} transitions in large-$N$ matrix models. In this paper, we make use of complex Langevin dynamics to study certain classes of zero-dimensional $\cN = 2$ supersymmetric quantum field theories with complex actions. 

The central theme of stochastic quantization is that expectation values of observables are obtained as equilibrium values of a stochastic process. In Langevin dynamics, this is implemented by evolving the system in a fictitious time direction, $\tau$, subject to a stochastic noise. We could think of applying Langevin dynamics when the actions under consideration are complex. In such cases, the field variables become complexified during Langevin evolution since the gradient of the action, the {\it drift term}, is complex. 

The complex Langevin equation in Euler discretized form reads
\beq
\phi (\tau + \Delta \tau) = \phi (\tau) - \Delta \tau \left( \frac{\delta S[\phi]}{\delta \phi (\tau)} \right) + \sqrt{\Delta \tau} ~\eta (\tau),
\eeq
where $\Delta \tau$ is the Langevin time step, and $\eta (\tau)$ is a Gaussian noise satisfying
\beq
\langle \eta (\tau) \rangle = 0, ~~\langle \eta (\tau) \eta (\tau ') \rangle = 2 \delta_{\tau \tau  '}.
\eeq
In our simulations, we use real Gaussian stochastic noise to tame excursions in the imaginary directions of the field configurations \cite{Aarts:2009uq, Aarts:2011ax, Nagata:2015uga}.

For an arbitrary operator $\cO$, we can define a noise averaged expectation value
\beq
\left \langle \cO[\phi(\tau)] \right \rangle_\eta = \int d\phi P[\phi(\tau)] \cO[\phi],
\eeq
where the probability distribution $P[\phi(\tau)]$ satisfies the Fokker-Planck equation
\beq
\frac{\partial P[\phi(\tau)]}{\partial \tau} = \frac{\delta}{\delta \phi(\tau)} \left( \frac{\delta}{\delta \phi(\tau)} + \frac{\delta S[\phi]}{\delta \phi(\tau)} \right) P[\phi(\tau)].
\eeq

When the action is real, it can be shown that in the limit $\tau \to \infty$, the stationary solution of the Fokker-Planck equation
\beq
P[\phi] \sim \exp \left( - S[\phi] \right)
\eeq
will be reached guaranteeing convergence of the Langevin dynamics to the correct equilibrium distribution. When the action is complex we will end up in a not so easy situation. The drift term will be complex and thus if we consider Langevin dynamics based on the above equation we will end up with complexified fields: $\phi = {\rm Re} \phi + i {\rm Im} \phi$.  We can still consider Langevin dynamics with complex probabilities \cite{Parisi:1984cs, Klauder:1985kq, Klauder:1985ks, Gausterer:1986gk} but proofs towards convergence to the complex weight, $\exp(-S)$, will be non-trivial. 

The paper is organized as follows. In Sec. \ref{sec:bosonic} we apply complex Langevin dynamics to a class of zero-dimensional bosonic field theories with complex actions, to compute expectation values of correlators and then compare them with analytical results. We discuss supersymmetry breaking in a zero-dimensional model with $\cN = 2$ supersymmetry and with a general form of the superpotential in Sec. \ref{sec:susy-breaking-mm}. In Sec. \ref{sec:various-sps}, using complex Langevin dynamics, we explore supersymmetry breaking in these models with real and complex actions for different forms of superpotentials. In Sec. \ref{sec:concl-f-dirs} we conclude and provide possible future directions. In Appendix. \ref{app:FP-correctness} we study a correctness criterion of our simulations using the Fokker-Planck operator. In Appendix. \ref{app:drift-decay} we study reliability of our simulations by examining the probability distributions of the magnitude of the drift terms. In Appendix. \ref{app:data-tables} we provide the set of simulation data tables.

\section{Bosonic models with complex actions}
\label{sec:bosonic}

Let us consider actions of zero-dimensional quantum field theories derived from a general potential of the form
\beq
\label{eq:0d-bosonic}
W(\phi) = - \frac{g}{(2 + \delta)} (i \phi)^{(2 + \delta)},
\eeq
with $\phi$ being a real scalar field, $g$ a coupling parameter and $\delta$ a real number. 

A class of (Euclidean) scalar quantum field theories, that are not symmetric under parity reflection, has been investigated in the literature using the above form of the potential \cite{Bender:1997ps}. We can, for example, write down a two-dimensional Euclidean Lagrangian of the form
\beq
{\cal L} = \hf (\partial_\mu \phi)^2 + \hf m^2 \phi^2 + W(\phi) ~~~~(\delta > -2),
\eeq
for a scalar field with mass $m$.

Such theories are very interesting from the point of view that they exhibit non-Hermitian Hamiltonians. Even more interesting is that there is numerous evidence that these theories possess energy spectra that are real and bounded below.

One can think of making the above Lagrangian supersymmetric by adding the right amount of fermions. The supersymmetric two-dimensional Lagrangian takes the form
\beq
{\cal L} = \hf (\partial_\mu \phi)^2 + \hf i \psib \slashed{\partial} \psi + \hf \psib W''(\phi) \psi + \hf \left[ W'(\phi) \right]^2,
\eeq   
where $\psi,\psib$ are Majorana fermions.

This supersymmetric Lagrangian also breaks parity symmetry. It would be interesting to ask whether the breaking of parity symmetry induces a breaking of supersymmetry. This question was answered in Ref. \cite{Bender:1997ps}. There, through a perturbative expansion in $\delta$, the authors found that supersymmetry remains unbroken in this model. We could think of performing nonperturbative investigations on SUSY breaking in this model using complex Langevin method. We leave this investigation for future work \cite{ToAppear:2019}. (Clearly, a nonperturbative investigation based on path integral Monte Carlo fails since the action of this model can be complex, in general.) 

Let us consider the 0-dimensional version of the bosonic Lagrangian with $m = 0$. The Euclidean action is the same as the one given in Eq. \eqref{eq:0d-bosonic}
\beq
S = -\frac{g}{N} (i\phi)^N,
\eeq
where $N = 2 + \delta$.

The partition function of this model is
\bea
Z &=& \frac{1}{2 \pi} \int_{-\infty}^{\infty} d\phi~ e^{-S} \\
&=& \frac{1}{2 \pi} \int_{-\infty}^{\infty} d\phi~ \exp{\left[ \frac{g}{N} (i\phi)^N \right]}.
\eea

We can look at the $k$-point correlation functions, $G_k$ of this model. We have
\bea
G_k = \left \langle \phi^k \right \rangle &=& \frac{1}{Z} \frac{1}{2 \pi} \int_{-\infty}^{\infty} d\phi ~\phi^k ~ \exp{ \left[ \frac{g}{N} (i\phi)^N \right] } \nn \\
&=& \frac{\int_{-\infty}^{\infty}d\phi~ \phi^k ~\exp{ \left[ \frac{g}{N} (i\phi)^N \right] } }{ \int_{-\infty}^{\infty} d\phi ~\exp{\left[ \frac{g}{N} (i\phi)^N \right] } }.
\eea

\begin{figure*}[htp]

\subfloat[Case $N=3$]{\includegraphics[width=3.2in]{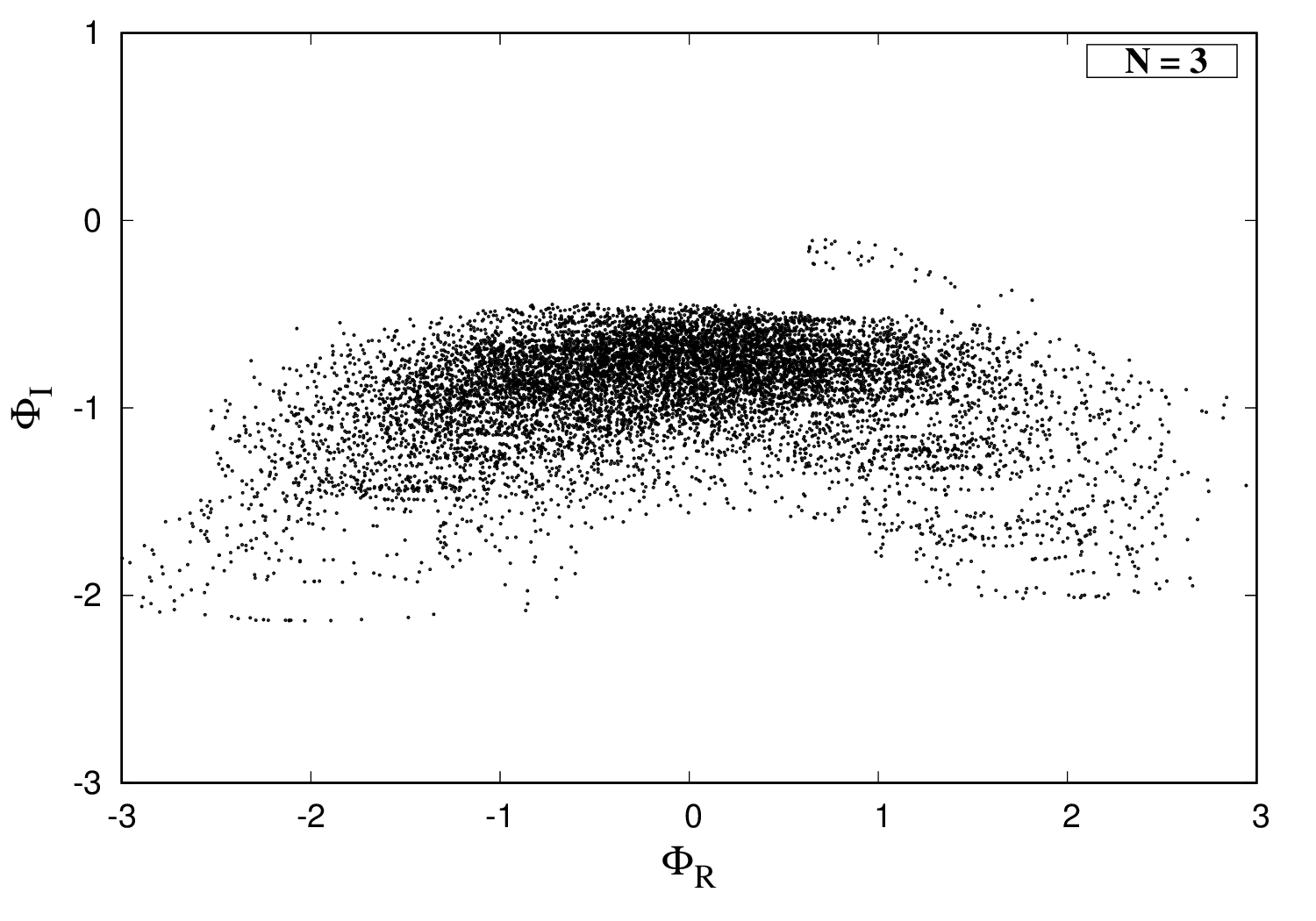}}
\subfloat[Case $N=4$]{\includegraphics[width=3.2in]{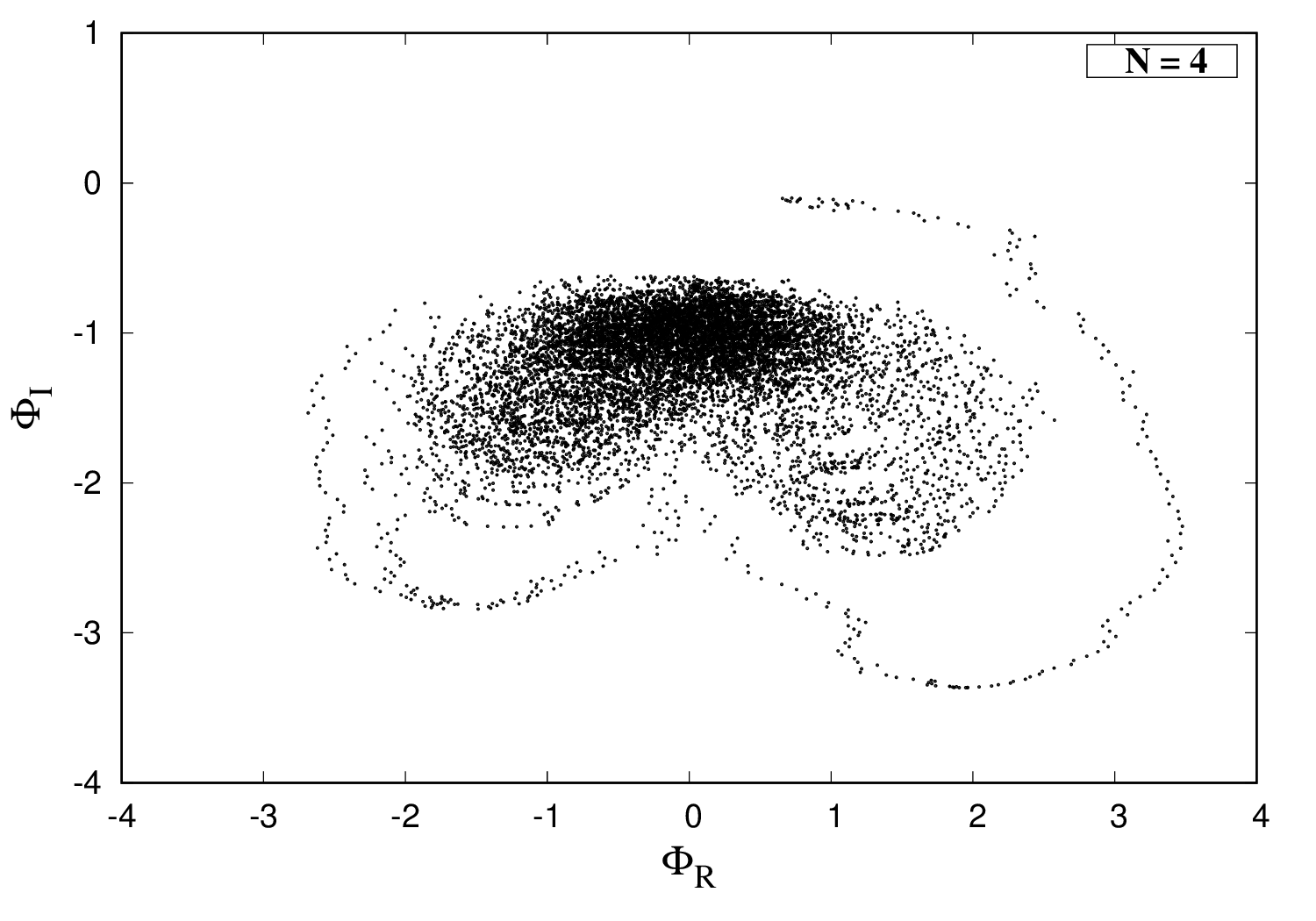}}

\caption{Scatter plot of complexified field configurations on the $\phi_R - \phi_I$ plane for the 0-dimensional $ -\frac{g}{N} \left(i\phi\right)^N$ theory with $g = 0.5$. Black dots represent the trajectories of the fields during complex Langevin evolution. (Left) Case $N=3$. The field configuration starts at point $(0.5, -0.1)$ and with the aid of a stochastic noise, it drifts towards the equilibrium configuration, forming a cloud averaging around $0.0 - i 0.9185$. (Right) Case $N=4$. The field starts at point $(0.5, -0.1)$ and with the aid of a stochastic noise, it drifts towards the equilibrium configuration, forming a cloud averaging around $0.0 -i 1.163$.}
\label{fig:n3-n4-clouds}

\end{figure*}

The one-point correlation function, $G_1$ can be evaluated as \cite{Bender:1999ek}
\bea
G_1 = -i \left(\frac{4N}{g}\right)^{1/N} \frac{ \Gamma\left(\frac{1}{N} + \hf\right) \cos \left(\frac{\pi}{N}\right) } {\sqrt{\pi}},
\eea
and the two-point correlation function, $G_2$ as
\bea
G_2 = \left(\frac{N}{g}\right)^{2/N} \frac{\Gamma\left(\frac{3}{N}\right) \left[\sin^2\left(\frac{\pi}{N}\right)- 3 \cos^2\left(\frac{\pi}{N}\right)\right]}{\Gamma\left(\frac{1}{N}\right)}.
\eea

Similarly we can compute higher moments of $\phi$. In Table \ref{tab:bosonic} we compare our results from complex Langevin simulations for $G_1$ and $G_2$ with their corresponding analytical results.

\begin{table*}[t]
\centering
\begin{tabular}{| c | c | c | c | c |} 
\hline\hline
$ ~~N~~ $  &	$~~~~~~~~~G_1^{\rm exact}~~~~~~~$  & $~~~~~~~~~~~~~~~~ G_1^{\rm cL}~~~~~~~~~~~~~~ $ 						& $ ~~~~~~~~~~~ G_2^{\rm exact}~~~~~~~ $ 		& $~~~~~~~~~~~~~~ G_2^{\rm cL}~~~~~~~~~~~~~~$ \\ [1.5ex] 
\hline
\hline
$3$	  		& $0.0 - i 0.9185$	&	$-0.0003(12) - i 0.9225(4) $  &	$ - $				 &				$ - $	\\ [0.5ex]
\hline
$4$   		& $0.0 - i 1.1630$	&	$-0.0005(8) - i 1.1678(4)$	  & $-0.9560 + i 0.0$	 &	$-0.9602 (6) -i 0.0009(24)$	\\ [0.5ex]
\hline
\end{tabular}
\caption{\label{tab:bosonic}The simulated values of the correlation functions $G_1$ and $G_2$ obtained from complex Langevin dynamics for 0-dimensional $-\frac{g}{N} (i\phi)^N$ theory for $N = 3, 4$. The simulations were performed with coupling parameter $g = 0.5$, adaptive Langevin step size $\Delta \tau \leq 0.02$, thermalization steps $N_{\rm therm} = 10^4$, generation steps $N_{\rm gen} = 10^6$ and measurements taken every $100$ steps. We have used an average of $10^2$ such simulation chains with random initial configurations. The table compares these numerically simulated values with the exact results.}
\end{table*}

\begin{figure}[H]
\centering
\includegraphics[width=3.2in]{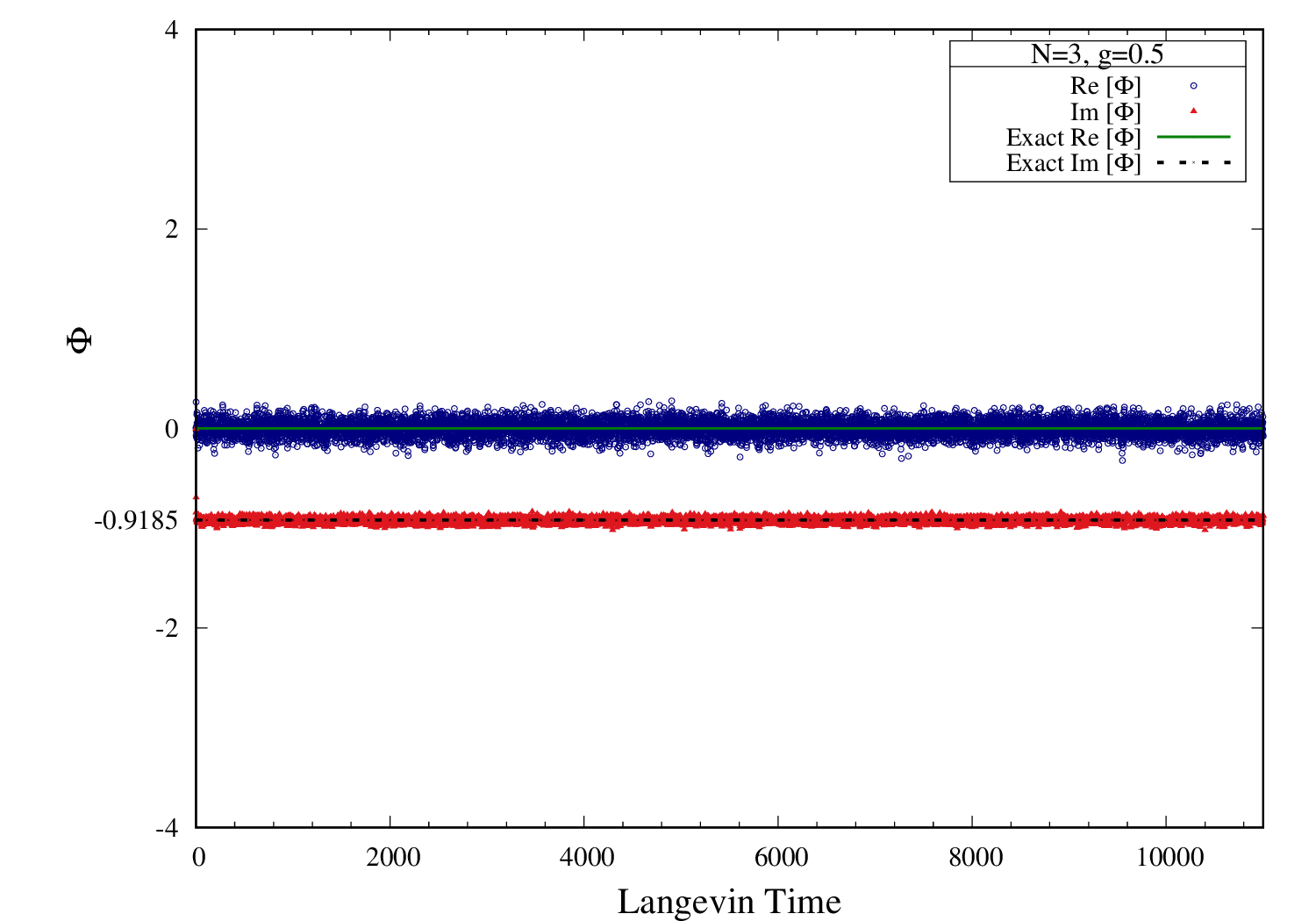}
\caption{Langevin time history of the field variable (one-point correlation function $G_1$) for the $i \frac{g}{3} \phi^3$ theory at coupling parameter $g = 0.5$. Simulations were performed with adaptive Langevin step size $\Delta \tau \leq 0.02$, generation steps $N_{\rm gen} = 10^6$ and measurements taken every $100$ steps. Simulated field configurations are an average of $10^2$ such simulation chains with random initialization. Solid and dashed lines represent the exact values.}
\label{fig:n3-history}
\end{figure}

\begin{figure*}[htp]

\subfloat[One-point correlation function]{\includegraphics[width=3.2in]{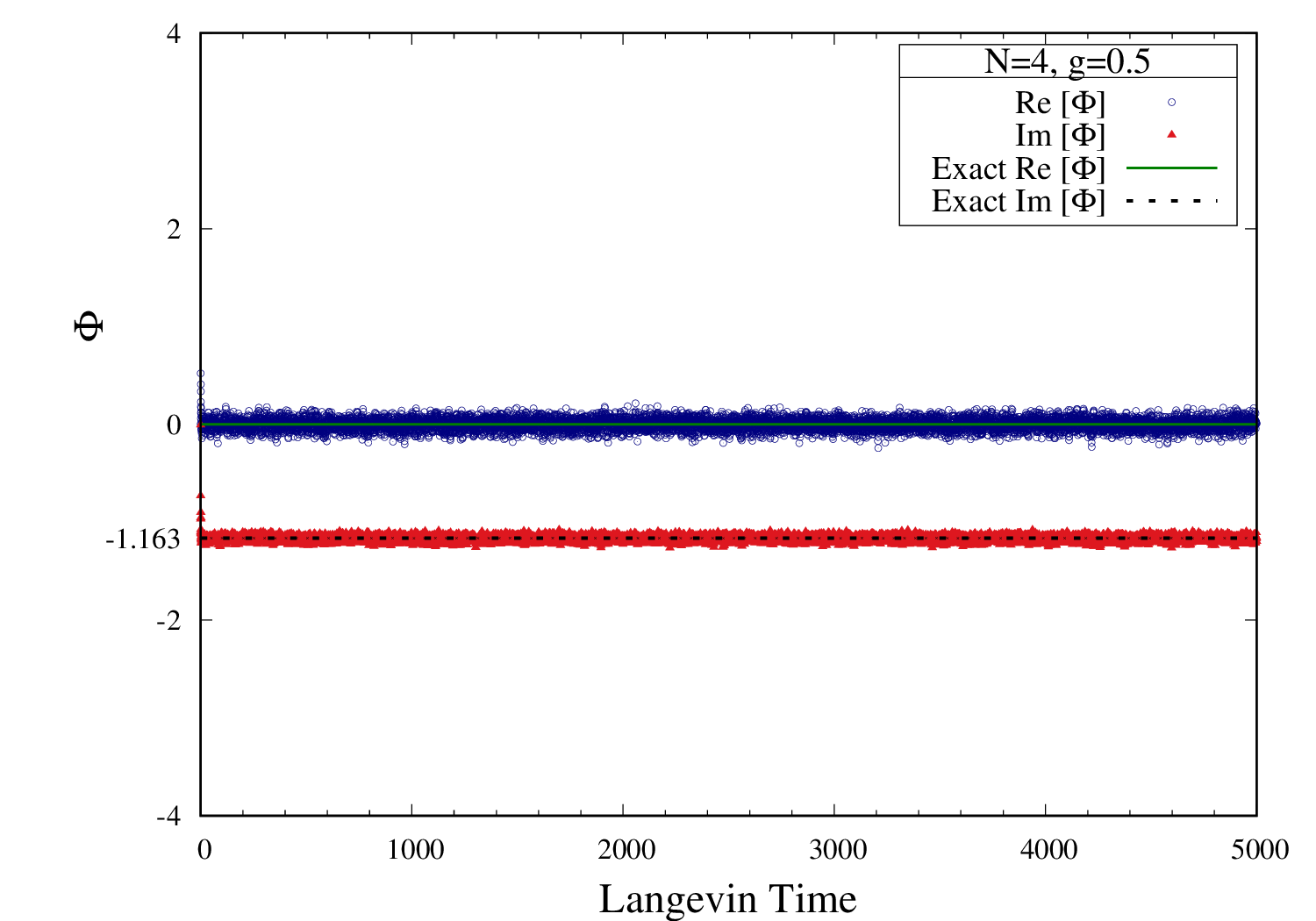}}
\subfloat[Two-point correlation function]{\includegraphics[width=3.2in]{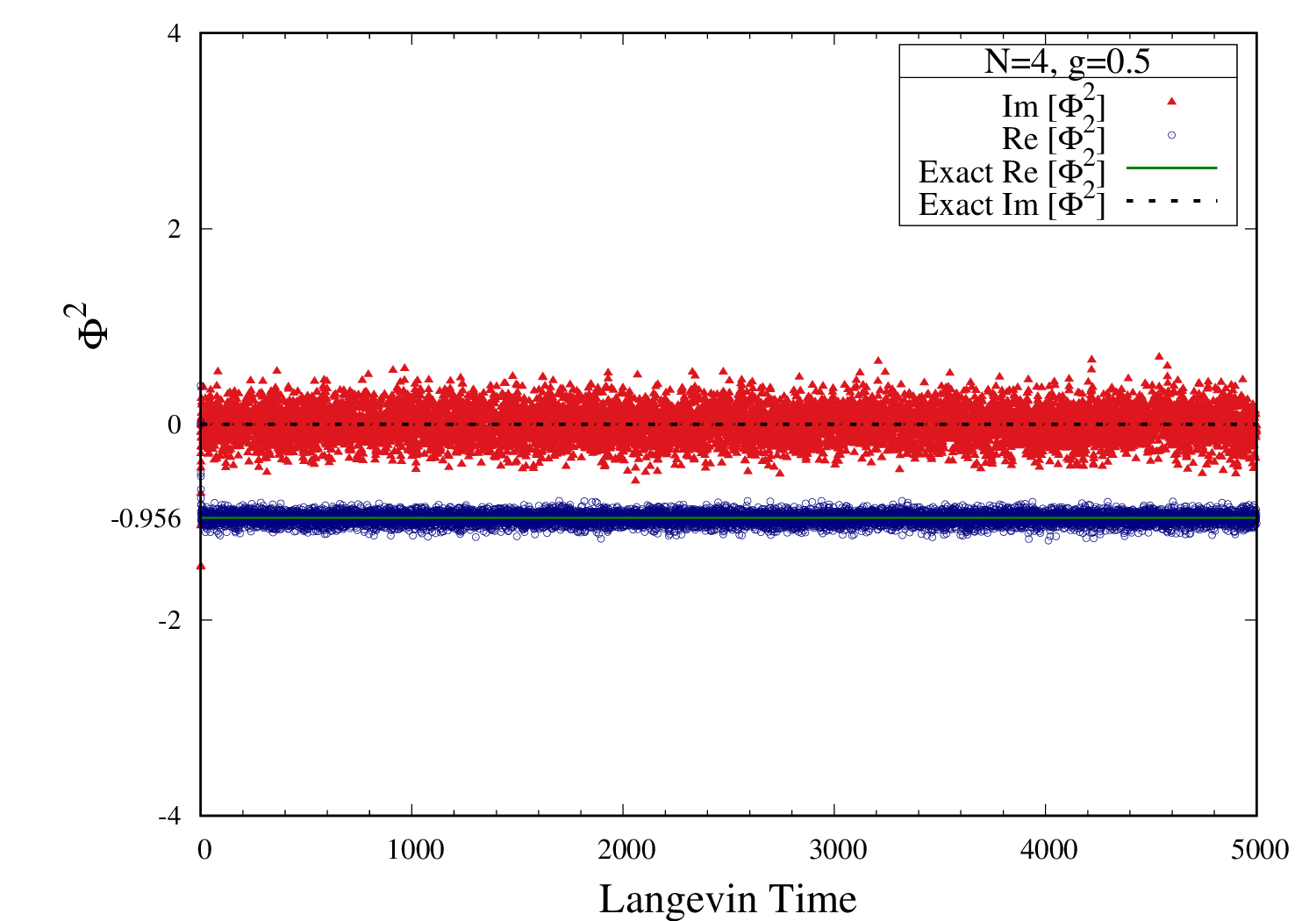}}

\caption{Langevin time history of one-point (Left) and two-point (Right) correlation functions for the $- \frac{g}{4} \phi^4$ theory at fixed coupling constant $g = 0.5$. Simulations were performed with adaptive Langevin step size $\Delta \tau \leq 0.02$, generation steps $N_{\rm gen} = 10^6$ and measurements taken every $100$ steps. Simulated field configurations are an average of $10^2$ such simulation chains with random field initialization. Solid and dashed lines represent the exact values.}
\label{fig:n4-history}

\end{figure*}

In Fig. \ref{fig:n3-n4-clouds} we show the complexified $\phi$ field configurations on the complex $\phi_R - \phi_I$ plane as it evolves in Langevin time. The Langevin time history of $G_1$ for the case $N=3$ is shown in Fig. \ref{fig:n3-history}. In Fig. \ref{fig:n4-history} we show the Langevin time history of $G_1$ and $G_2$ for the case $N=4$.

\section{Supersymmetry breaking in zero-dimensional field theories}
\label{sec:susy-breaking-mm}

Let us consider a 0-dimensional supersymmetric model. For a general supersymmetric potential, $W(\phi)$, the action is given by
\beq
S = \hf B^2 + i B W' + \bar{\psi} W'' \psi,
\eeq
where $\phi$ is a bosonic field, $\psi$ and $\bar{\psi}$ are fermionic fields, and $B$ is an auxiliary field. The prime denotes derivative of the superpotential with respect to $\phi$. There is a symmetry in the above action that exchanges fermionic fields with bosonic fields and this symmetry is known as supersymmetry. We define two independent supersymmetry charges $Q$ and $\Qb$ corresponding to an $\cN = 2$ supersymmetry.  This action can be derived from dimensional reduction of a one-dimensional theory, that is, a supersymmetric quantum mechanics with two supercharges. 

We can see that the above action is invariant under the following supersymmetry transformations
\begin{subequations}
\label{eq:susy-transf-Q}
\begin{align}
        Q \phi  &= \psi,\\
        Q \psi &= 0, \\
        Q \bar{\psi} &= - iB, \\
	Q B &= 0,
\end{align}
\end{subequations}

and

\begin{subequations}
\label{eq:susy-transf-Qb}
\begin{align}
\Qb \phi &= - \bar{\psi}, \\
\Qb \bar{\psi} &= 0, \\
\Qb \psi &= - i B, \\
\Qb B &= 0. 
\end{align}
\end{subequations}

The supercharges $Q$ and $\Qb$ satisfy the algebra
\begin{subequations}
\begin{align}
\{ Q, Q \} &= 0, \\
\{ \Qb, \Qb \} &= 0, \\
\{ Q, \Qb \} &= 0.
\end{align}
\end{subequations}

We also note that the action can be expressed in $Q$- or $Q \Qb$- exact forms. That is,
\bea
S &=& Q \psib \left( \frac{i}{2} B - W' \right), \\ 
&=& Q \Qb \left( \hf \psib \psi + W \right).
\eea

The auxiliary field $B$ has been introduced for off-shell completion of the supersymmetry algebra. It is possible to integrate out this field using its equation of motion
\beq
B = -i W'.
\eeq

It is easy to show that the action is invariant under the two supersymmetry charges
\bea
Q S &=& 0, \\
\Qb S &=& 0.
\eea

The partition function of the model is
\bea
Z & =& \frac{1}{2 \pi} \int dB d\phi d\psi d\psib \ e^{-S} \nn \\
& =& \frac{1}{2 \pi} \int dB d\phi d\psi d\psib \nn \\
&& \times \exp\left[- \Big( \hf B^2 + i B W' + \psib W'' \psi \Big) \right].
\eea

Completing the square and integrating over the auxiliary field it becomes
\bea
Z & = & \frac{1}{\sqrt{2 \pi}} \int d\phi d\psi d\psib \ \exp \left[ - \left( \hf  {W'}^2 + \psib W'' \psi \right) \right].~~~~
\eea

Integrating over the fermions it takes the form
\bea
Z  &=& - \frac{1}{\sqrt{2 \pi}} \int d\phi \ W'' \ \exp{ \left[-\hf {W'}^2 \right] }.
\eea

When SUSY is broken, the supersymmetric partition function vanishes. In that case, the expectation values of observables normalized by the partition function could be ill-defined.

The expectation value of the auxiliary field $B$ is crucial in investigating SUSY breaking. It can be evaluated as
\bea
\langle B \rangle &=& \frac{1}{Z} \frac{1}{2 \pi}  \int dB d\phi d\psi d\psib \ B  \ e^{-S} \nn  \\
&=& \frac{1}{Z} \frac{i}{\sqrt{2 \pi}} \int d\phi \ W' \ {W}'' \ \exp {\left[-\hf {W'}^2 \right]} \nn \\
&=& - \frac{1}{Z} \frac{i}{\sqrt{2 \pi}} \int d\phi \ \frac{\partial}{\partial \phi} \left( \ \exp { \left[-\hf {W'}^2 \right]} \ \right).
\eea

Thus, in this model, the normalized expectation value of $B$ is indefinite (it is of the form $0/0$) when SUSY is broken.

In order to overcome this difficulty we can introduce an external field and then eventually take a limit where it goes to zero. We usually introduce some external field to detect spontaneous breaking of ordinary symmetry so that the ground state degeneracy is lifted to specify a single broken ground state. We take the thermodynamic limit of the theory, and after that, the external field is turned off. The value of the corresponding order parameter then would tell us if spontaneous symmetry breaking happens in the model or not. (Note that to detect the spontaneous magnetization in the Ising model, we use the external field as a magnetic field, and the corresponding order parameter then would be the expectation value of the spin operator.) We will also perform an analogues method to detect SUSY breaking in the system. Introduction of an external field can be achieved by changing the boundary conditions for the fermions to twisted boundary conditions.

\subsection{Theory on a one-site lattice}

Let us consider the above 0-dimensional theory as a dimensional reduction of a one-dimensional theory, which is a supersymmetric quantum mechanics. The action of the one-dimensional theory is an integral over a compactified time circle of circumference $\beta$ in Euclidean space. We have the action
\beq
S  =  \int_0^\beta d\tau \left[ \ \hf B^2 + iB \left(\dot{\phi} + W'\right) + \psib \left( \dot{\psi} + W'' \psi \right) \ \right]. 
\eeq

Here the dot denotes derivative with respect to Euclidean time $\tau \in [0, \beta]$. Note that the $\Qb$ supersymmetry will not be preserved in the quantum mechanics theory.

Let us discretize the theory on a one-dimensional lattice with $T$ sites, using finite differences for derivatives. We have the lattice action
\bea
S &=& \sum_{n=0}^{T-1} \ \Bigg[\ \hf B^2(n) +  i B(n) \Big( \phi (n+1) - \phi (n) + W' \Big) \nn \\
&& + \psib(n) \ \Big( \psi(n+1) - \psi(n)+ W'' \ \psi(n) \Big) \ \Bigg],
\eea
with $n$ denoting the lattice site. We have rescaled the fields and coupling parameters such that the lattice action is expressed in terms of dimensionless variables. The lattice action preserves one of the supercharges, $Q$. The $\Qb$ supersymmetry will not be a symmetry on the lattice when $T\geq 2$.

Let us consider the simplest case of one lattice point, that is, when $T = 1$. The action becomes
\bea
S &=& \Bigg[ \ \hf B^2(0) +  i B(0)  \Big( \phi (1) - \phi (0) + W'  \Big) \nn \\
&& + \psib(0)  \Big(  \psi(1) - \psi(0)+ W'' \ \psi(0) \Big) \ \Bigg],
\eea
where $\phi(1)$ and $\psi(1)$ are dependent on the boundary conditions. In the case of periodic boundary conditions, 
\begin{subequations}
\begin{align}
\phi(1) & =  \phi(0), \\
\psi(1) & = \psi(0), \\
\psib(1) & = \psib(0), \\
B(1) & =  B(0),
\end{align}
\end{subequations}
the action reduces to
\beq
S = \hf B^2 + i B W' + \psib W'' \psi.
\eeq

Thus the action for the 0-dimensional supersymmetric model with $\cN = 2$ supersymmetry is equivalent to the dimensional reduction of a one-dimensional theory (a supersymmetric quantum mechanics) with periodic boundary conditions. 

\subsection{Twisted boundary conditions}

Now, instead of periodic boundary conditions, let us introduce twisted boundary conditions for fermions (analogues to turning on an external field), with the motivation to regularize the indefinite form of the expectation values we encountered earlier\footnote{Twisted boundary conditions were considered in the context of supersymmetric models by Kuroki and Sugino in Refs. \cite{Kuroki:2009yg, Kuroki:2010au}.}. We have

\begin{subequations}
\begin{align}
\phi(1) & =  \phi(0), \\
\psi(1) & = e^{i\alpha} \psi(0), \\
\psib(1) & = e^{- i\alpha} \psib(0), \\
B(1) & =  B(0).
\end{align}
\end{subequations}

The action in this case has the form
\beq
S_\alpha =  \hf B^2 + i B W' + \psib \Big( e^{i \alpha} - 1 + W'' \Big) \psi.
\eeq

We see that supersymmetry is softly broken by the introduction of the twist $\alpha$
\beq
Q S_\alpha = -i \Qb S_\alpha = \psib \left( e^{i \alpha} - 1 \right) \psi.
\eeq 
In the limit $\alpha \to 0$ supersymmetry is recovered.

The partition function is
\bea
\label{eq:Z-twist}
Z_\alpha &=& \frac{1}{2 \pi} \int dB d\phi d\psi d\psib \ e^{ -S_\alpha } \nn \\
&=& - \frac{1}{\sqrt{2 \pi}} \int d\phi ~ \Big( e^{i \alpha} - 1 + W'' \Big) \nn \\
&& \times \exp \left[ - \hf W'^2 \right].
\eea

The expectation of auxiliary field $B$ is given by
\bea
\langle B \rangle_\alpha &=&  \frac{1}{Z_\alpha} \frac{1}{2 \pi} \int dB d\phi d\psi d\bar{\psi} \ B \ e^{ -S_\alpha }  \nn \\
&=&  \frac{1}{Z_\alpha} \frac{i}{\sqrt{2 \pi}} \int d\phi \ W' \Big( e^{i \alpha} - 1 + W'' \Big) \nn \\
&& \times \exp \left[ - \hf W'^2 \right].
\eea

It is important to note that the quantity $\langle B \rangle_\alpha$ is now well defined. Here, the external field $\alpha$ plays the role of a regularization parameter and it regularizes the indefinite form, $\langle B\rangle = 0/0$, of the expectation value under periodic boundary conditions and leads to the non-trivial result. Vanishing expectation value of auxiliary field, $\langle B \rangle_\alpha$ in the limit $\alpha \to 0$ indicates that SUSY is not broken, while a non-zero value indicates SUSY breaking. 

We can write down the effective action of the model with twisted boundary conditions as
\beq
S_\alpha^{~\text{eff}} = \hf  W'^2 -  \text{ln} \left[ e^{i \alpha} - 1 + W'' \right].
\eeq
The drift term needed for the application of complex Langevin method in Sec. \ref{sec:various-sps} has the form
\bea
\label{eq:clm-drift}
\frac{\partial S_\alpha^{~\text{eff}}}{\partial \phi} &=& \frac{\partial}{\partial \phi}  \left( \hf  W'^2 - \ln \left[ e^{i \alpha} - 1 + W'' \right] \right) \nn \\
&=& W' W'' - \frac{W'''}{\Big ( e^{i \alpha} - 1 + W''\Big)}.
\eea

\section{Models with various superpotentials}
\label{sec:various-sps}

In this section, we investigate spontaneous supersymmetry breaking in various zero-dimensional models using complex Langevin method. Wherever possible, we also compare our numerical results with corresponding analytical results.

\subsection{Double-well potential}

Let us begin with a case where the action is real. We consider the case when the derivative of the superpotential is a double-well potential 
\beq
W' = g \ (\phi^2 + \mu^2),
\eeq
where $g$ and $\mu$ are two parameters in the theory.
\newline

When $\mu^2 > 0$, the classical minimum is given by the field configuration $\phi = 0$ with energy
\beq
E_0 = \hf g^2 \mu^4 > 0,
\eeq
implying spontaneous SUSY breaking.

The ground state energy can be computed as the expectation value of the bosonic action at the classical minimum
\bea
E_0 \Big|_{\phi =0} &=& \langle S_B \rangle \nn \\
&=& \hf B^2 +iB W' \nn \\
&=& - \hf (W')^2 + (W')^2 = \hf (W')^2\Big|_{\phi=0} \nn \\
&=& \hf g^2 \mu^4.
\eea

We can also see from SUSY transformations
\bea
Q \psib &=& -g \mu^2, \\
\Qb \psi &=& -g \mu^2,
\eea
that SUSY is broken in the model.

The twisted partition function is
\begin{widetext}
\bea
Z_\alpha &=&- \frac{1}{ \sqrt{2 \pi} } \int_{-\infty}^\infty d\phi ~ \Big( e^{i \alpha} -1 + W'' \Big) \exp \left[ - \hf (W')^{2} \right] \nn \\
&=&- \frac{1}{\sqrt{2 \pi}} \int_{-\infty}^\infty d\phi ~ \Big( e^{i \alpha} - 1 + 2g   \phi \Big) \exp \left[ - \hf g^2 (\phi^2 + \mu^2)^2 \right] \nn \\
&=& -\frac{\mu}{2 \sqrt{\pi} } \left( e^{i \alpha} - 1 \right) ~ e^{- \qtr g^2 \mu^4 } ~\text{ Bessel~K} \left( \qtr, \frac{g^2 \mu^4}{4} \right) ~~~  \forall ~~ \text{Re} \left( g^2 \right) > 0 ~{\rm and} ~\text{Re} \left( g^2 \mu^2 \right) > 0.
\eea
\end{widetext}

When $\alpha \rightarrow 0$ we have
\beq
Z_\alpha \Big|_{\alpha = 0} = 0.
\eeq

Hence, SUSY is broken for $W' = g \ (\phi^2 + \mu^2)$.

Let us consider the observable
\begin{widetext}
\bea
\langle B \rangle_\alpha &=& -\frac{1}{Z_\alpha} \frac{1}{\sqrt{2 \pi}} \int_{-\infty}^\infty d\phi ~ \left(-iW'\right)~ \Big( e^{i \alpha} -1 + W'' \Big) \exp \left[ - \hf W'^{2} \right] \nn \\
&=& -ig \ \frac{ \int_{-\infty}^\infty d\phi ~(\phi^2 + \mu^2) ~\exp \left[ - \hf g^2 (\phi^2 + \mu^2)^{2} \right] }{ \int_{-\infty}^\infty d\phi ~\exp \left[ - \hf g^2 (\phi^2 + \mu^2)^{2} \right] }.
\eea
\end{widetext}

The above expression, once evaluated, becomes
\begin{widetext}
\bea
\langle B \rangle_\alpha &=& - \frac{i}{2} g \mu^2 \frac{ \Big(\text{Bessel K} \left( \qtr, \frac{g^2 \mu^4}{4} \right) + \text{Bessel K} \left(\frac{3}{4}, \frac{g^2 \mu^4}{4} \right) \Big)}{\text{Bessel K} \left(\qtr, \frac{g^2 \mu^4}{4} \right)} ~~~ \forall ~~ \text{Re}\left( g^2 \right) > 0 ~{\rm and}~\text{Re} \left(g^2 \mu^2 \right) > 0. 
\eea
\end{widetext}

\begin{figure*}[htp]

\subfloat[Case $g=1$]{\includegraphics[width=3.2in]{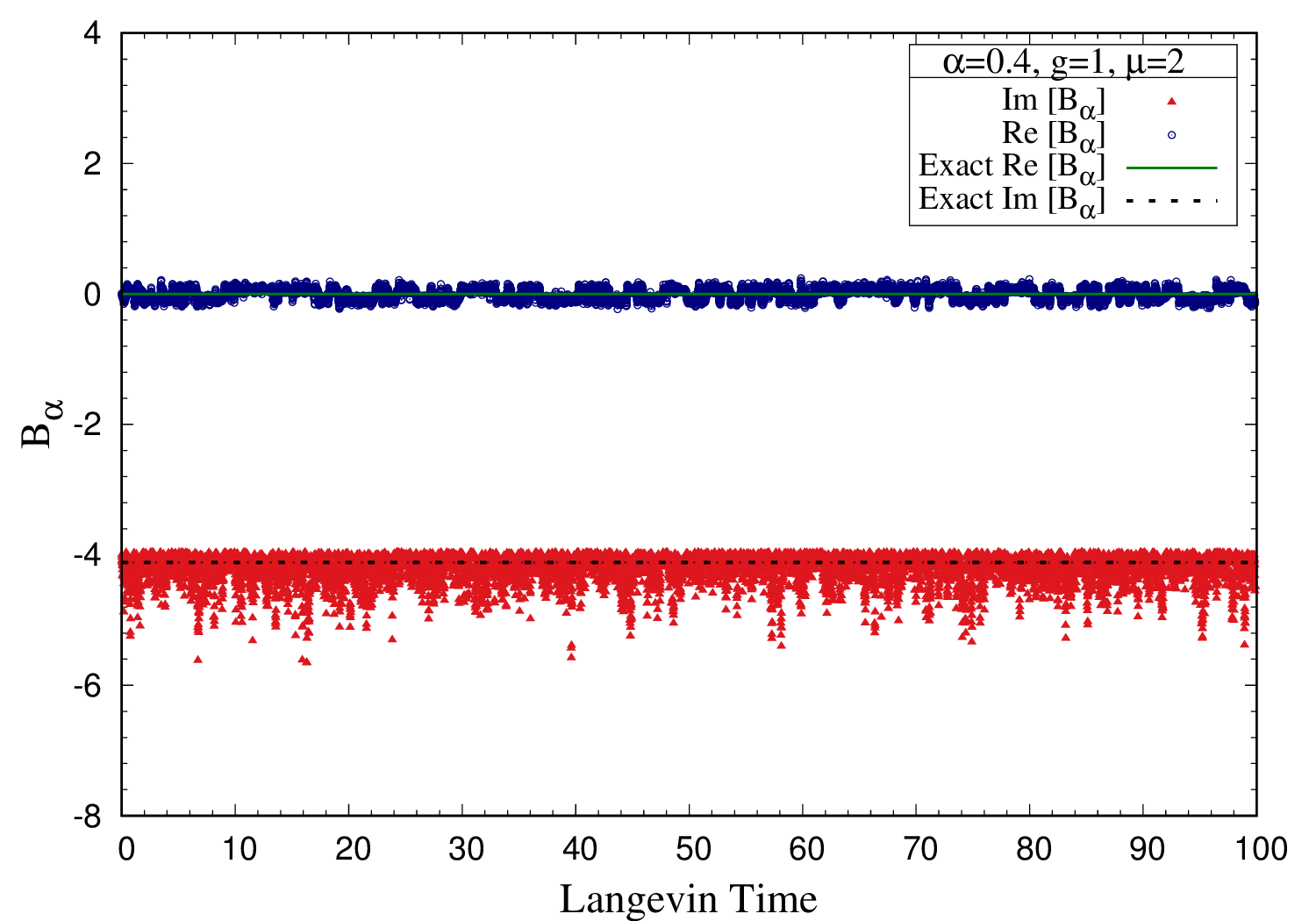}}
\subfloat[Case $g=3$]{\includegraphics[width=3.2in]{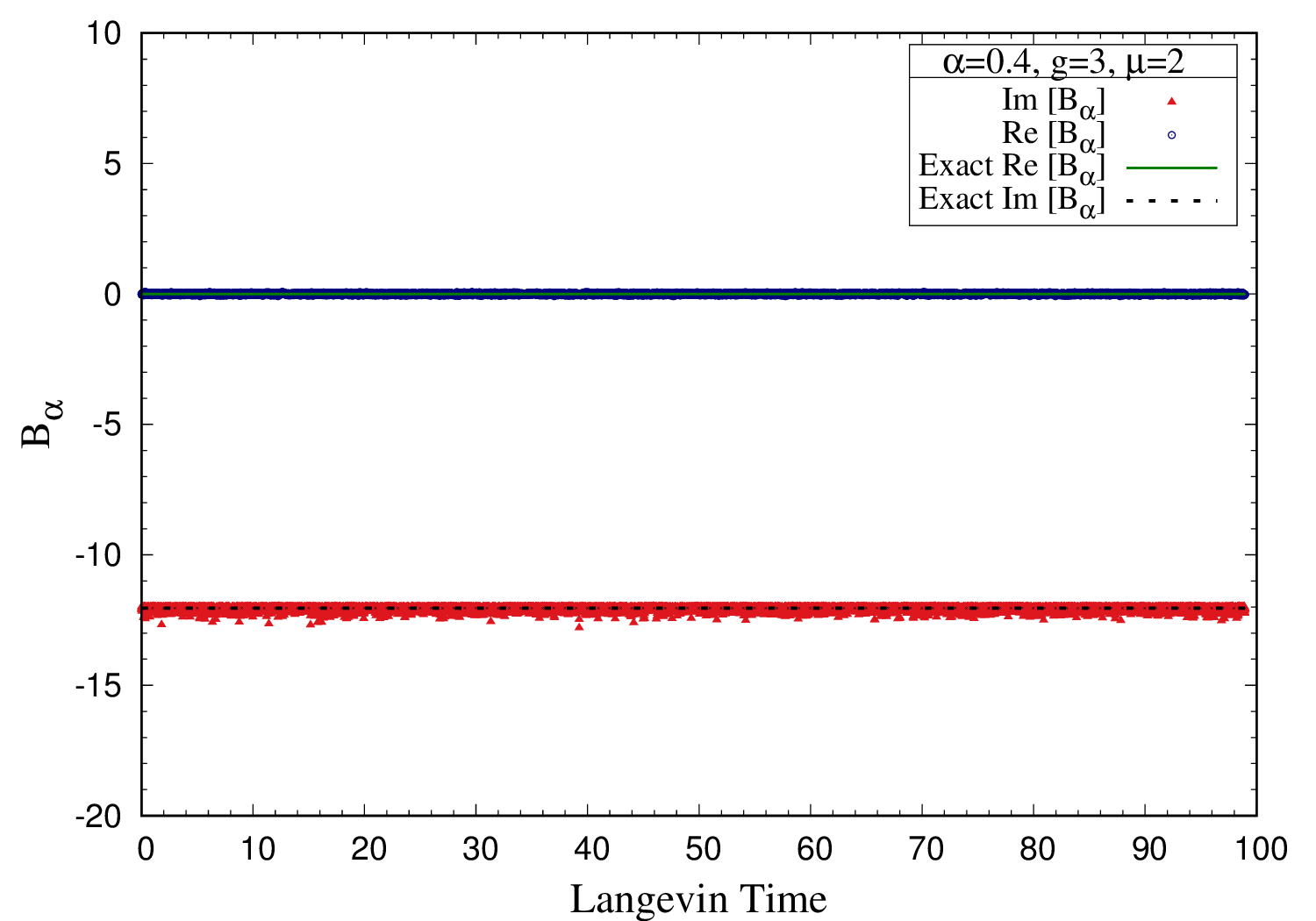}}

\caption{The observable $ B$ against Langevin time for regularization parameter $\alpha = 0.4$. Simulations were performed for superpotential $W' = g \ (\phi^2 + \mu^2)$ with $\mu = 2$. In these simulations, we have used adaptive Langevin step size $\Delta \tau \leq 10^{-4}$, generation steps $N_{\rm gen} = 10^6$ and measurements were taken every $100$ steps. (Left) Case $g=1$. The exact value is $\langle B \rangle = 0.0 - i 4.115$ corresponding to a system with broken SUSY.  (Right) Case $g=3$. The exact value is $\langle B \rangle = 0.0 - i 12.041$ again indicating that SUSY is broken in the model.}
\label{fig:dw-g1-g3-p0_mu2p0}

\end{figure*}

\begin{figure*}[htp]

\subfloat[Real part of $\langle B \rangle_\alpha$]{\includegraphics[width=3.2in]{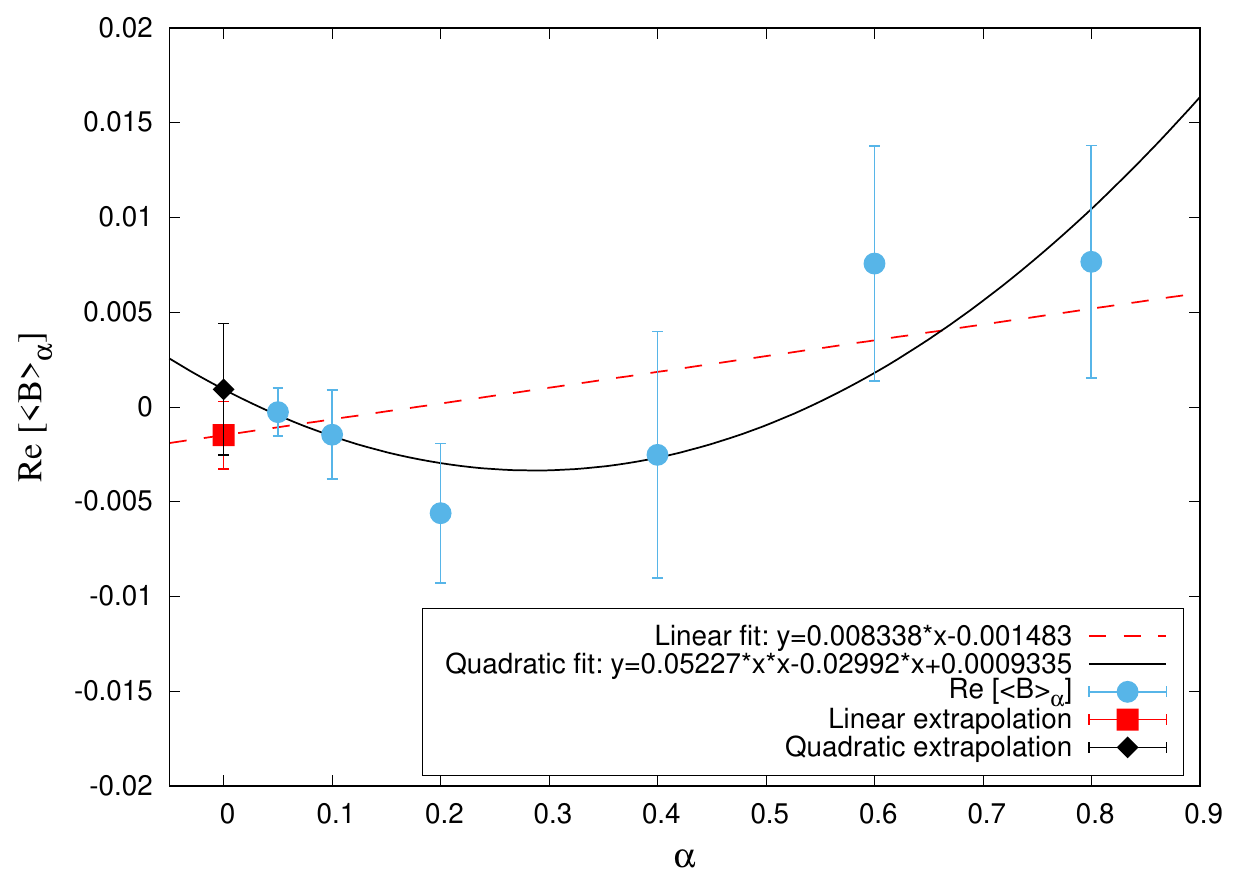}}
\subfloat[Imaginary part of $\langle B \rangle_\alpha$]{\includegraphics[width=3.2in]{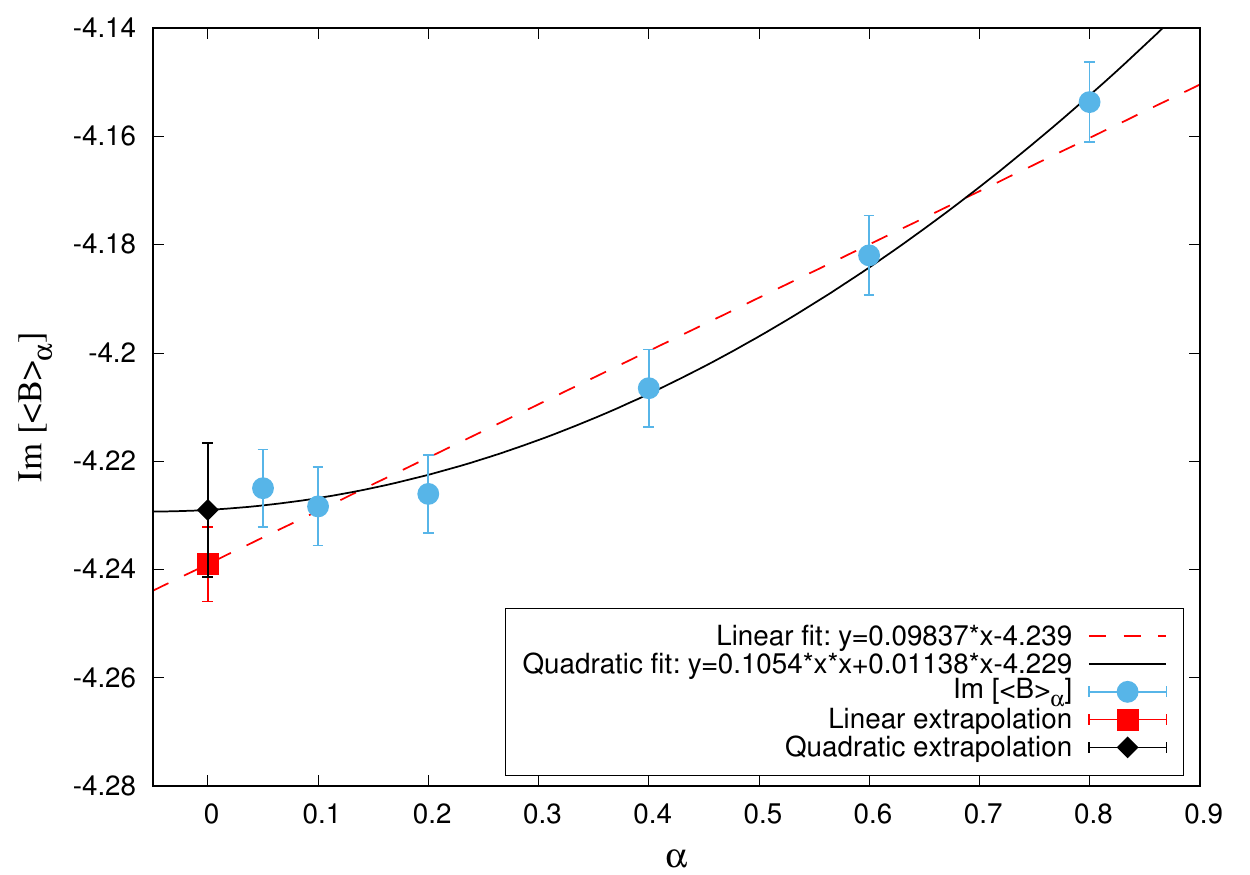}}

\caption{Plot of real (Left) and imaginary (Right) parts of $\langle B \rangle_\alpha$ against the regularization parameter, $\alpha$ for supersymmetric potential $W' = g \ (\phi^2 + \mu^2)$. Simulations were performed with $g = 1$ and $\mu = 2$. We have used adaptive Langevin step size $\Delta \tau \leq 10^{-4}$, thermalization steps $N_{\rm therm} = 10^{4}$, generation steps $N_{\rm gen} = 10^6$ and measurements were taken every $100$ steps. The dashed red lines are the linear fits to $\langle B \rangle_\alpha$ in $\alpha$, and filled red squares are the linear extrapolation values at $\alpha = 0$.  The solid black lines represent the quadratic fits to $\langle B \rangle_\alpha$ in $\alpha$, and filled black diamonds are the quadratic extrapolation values at $\alpha = 0$. The $\alpha \to 0$ limit values obtained from these plots are given in Table \ref{tab:sqw_B_mu2p0}.}
\label{fig:dw_fit_g1p0_mu2p0}

\end{figure*}

\begin{figure*}[htp]

\subfloat[Real part of $\langle B \rangle_\alpha$]{\includegraphics[width=3.2in]{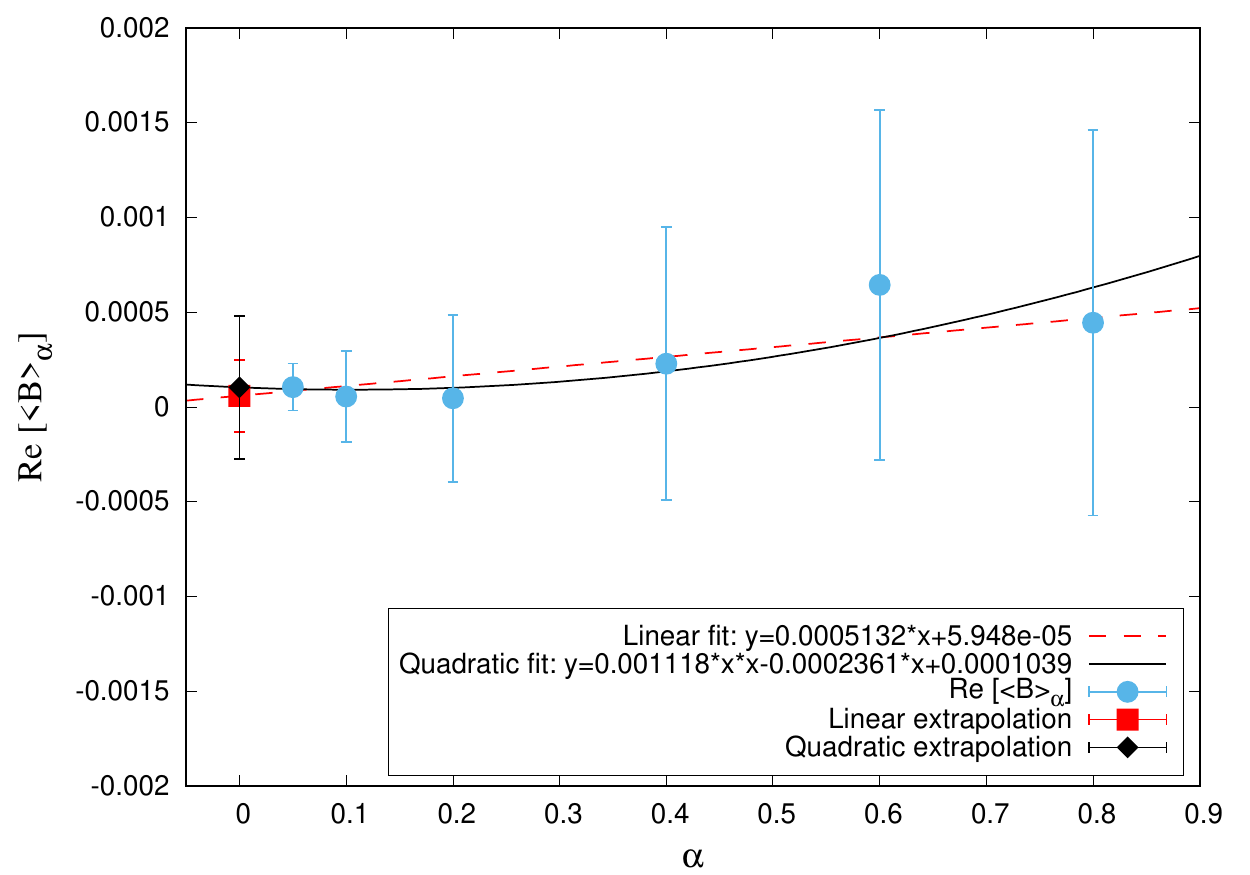}}
\subfloat[Imaginary part of $\langle B \rangle_\alpha$]{\includegraphics[width=3.2in]{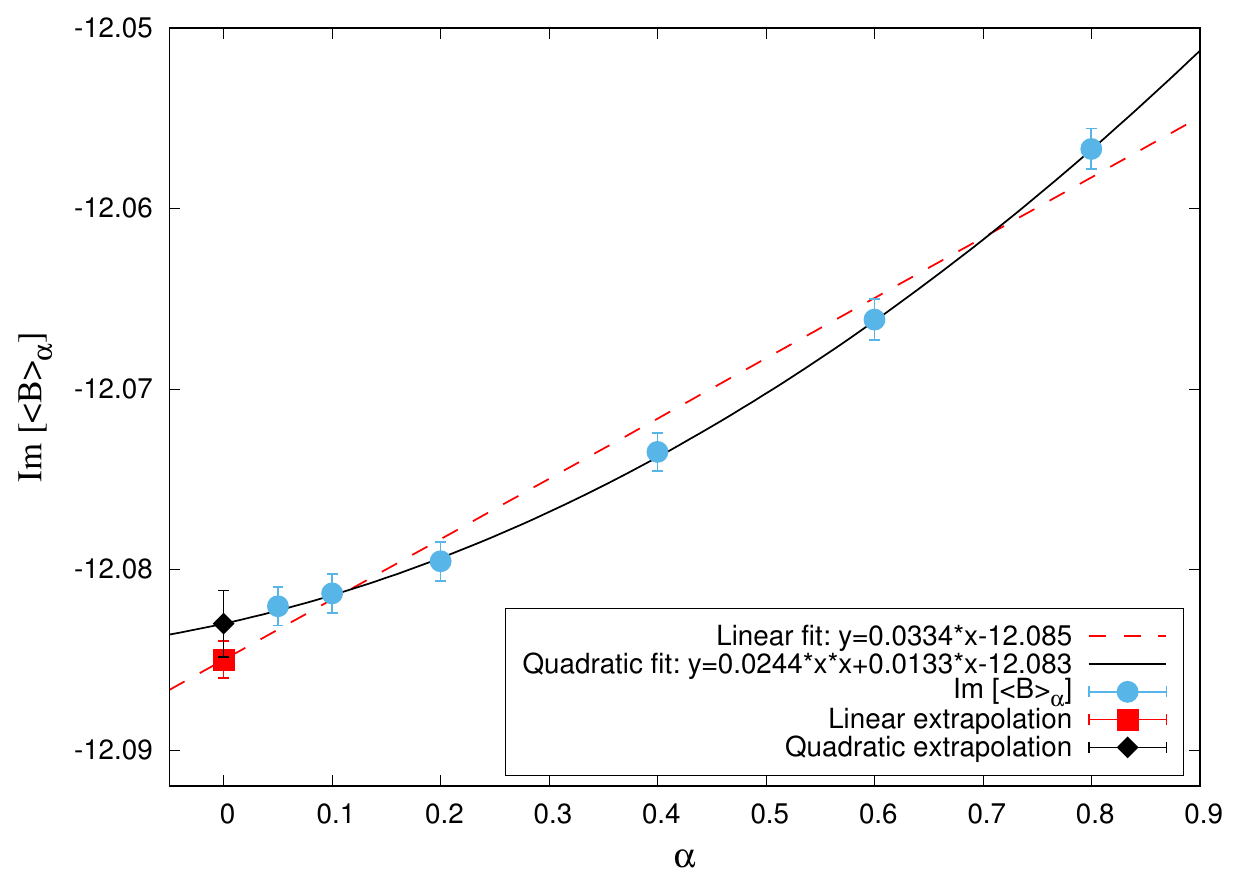}}

\caption{Plot of real (Left) and imaginary (Right) parts of $\langle B \rangle_\alpha$ against the regularization parameter, $\alpha$ for supersymmetric potential $W' = g \ (\phi^2 + \mu^2)$. Simulations were performed with $g = 3$ and $\mu = 2$. We have used adaptive Langevin step size $\Delta \tau \leq 10^{-4}$, thermalization steps $N_{\rm therm} = 10^{4}$, generation steps $N_{\rm gen} = 10^6$ and measurements were taken every $100$ steps. The dashed red lines are the linear fits to $\langle B\rangle_\alpha$ in $\alpha$, and filled red squares are the linear extrapolation values at $\alpha = 0$.  The solid black lines represent the quadratic fits to $\langle B \rangle_\alpha$ in $\alpha$, and filled black diamonds are the quadratic extrapolation values at $\alpha = 0$ .  The $\alpha \to 0$ limit values obtained from these plots are given in Table \ref{tab:sqw_B_mu2p0}.}
\label{fig:dw_fit_g3p0_mu2p0}

\end{figure*}

In Fig. \ref{fig:dw-g1-g3-p0_mu2p0} we show our results from Langevin simulations of this model. We show linear and quadratic extrapolations to $\alpha \to 0$ limit in Figs. \ref{fig:dw_fit_g1p0_mu2p0} and \ref{fig:dw_fit_g3p0_mu2p0}. The results are tabulated in Table \ref{tab:sqw_B_mu2p0}. The simulation results are in good agreement with the analytical predictions, and strongly suggest that SUSY is broken for this model. 

\begin{figure*}[htp]
	
	\subfloat[Case $g=1$]{\includegraphics[width=3.2in]{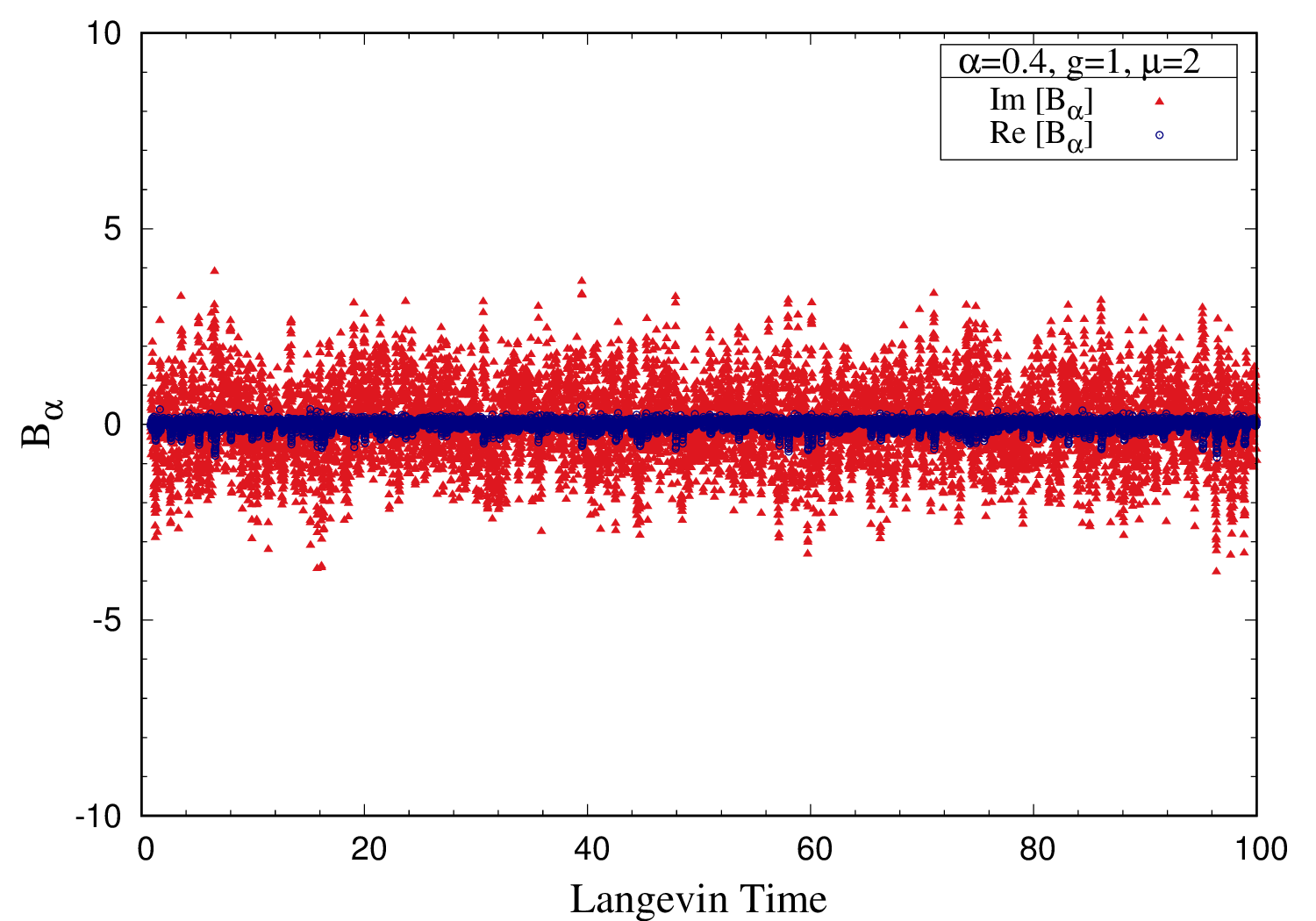}}
	\subfloat[Case $g=3$]{\includegraphics[width=3.2in]{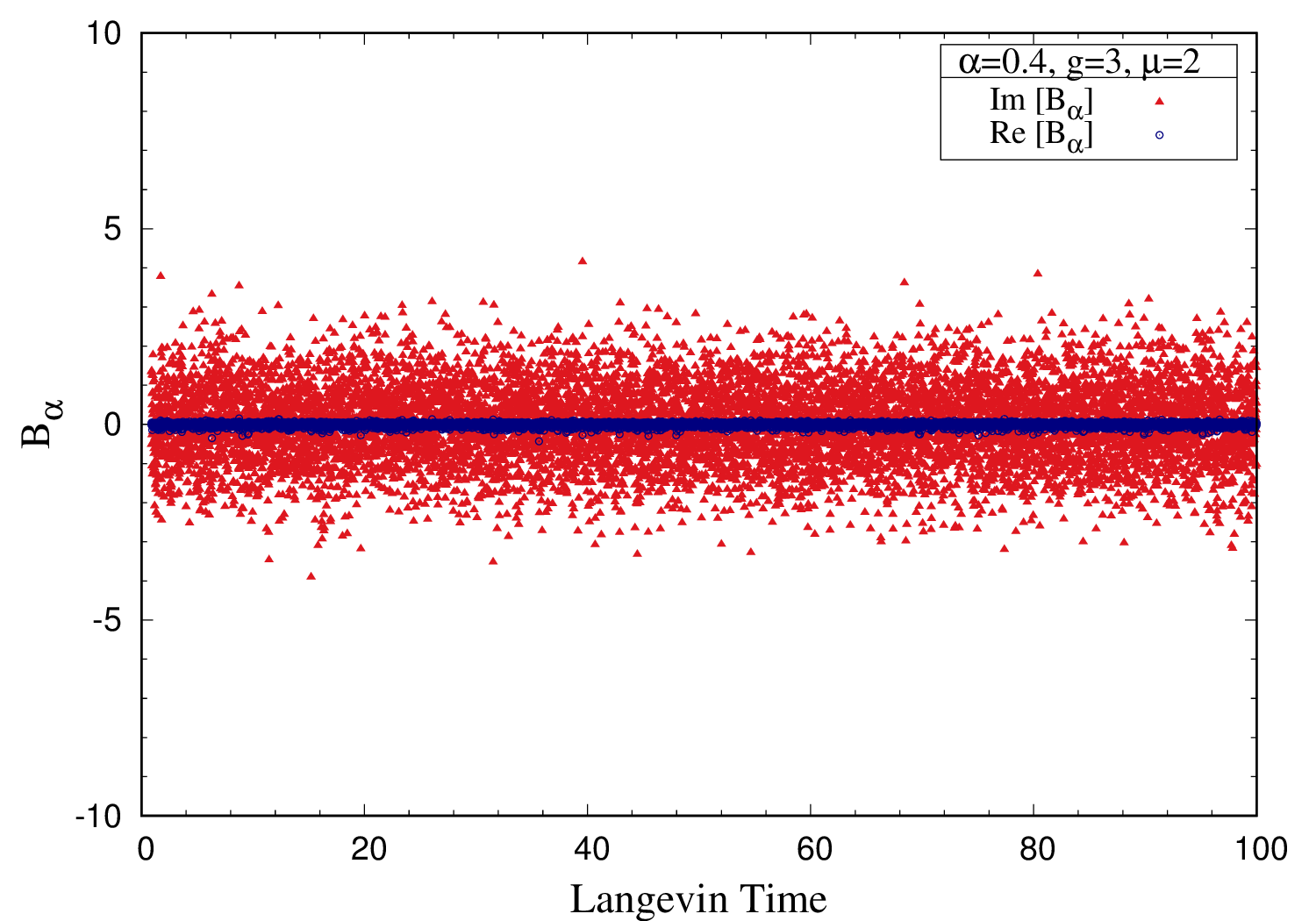}}
	
	\caption{The Langevin time history of $B$ for regularization parameter $\alpha = 0.4$. Simulations were performed for complex superpotential $W' = ig \ (\phi^2 + \mu^2)$ with $\mu = 2$. In these simulations, we have used adaptive Langevin step size $\Delta \tau \leq 10^{-4}$, thermalization steps $N_{\rm therm} = 10^{4}$, generation steps $N_{\rm gen} = 10^6$ and measurements were taken every $100$ steps. (Left) $g=1$. (Right) $g=3$. }
	\label{fig:isqw-g1-g3-p0_mu2p0}
	
\end{figure*}

\begin{figure*}[htp]
	
	\subfloat[Real part of $\langle B \rangle_\alpha$]{\includegraphics[width=3.2in]{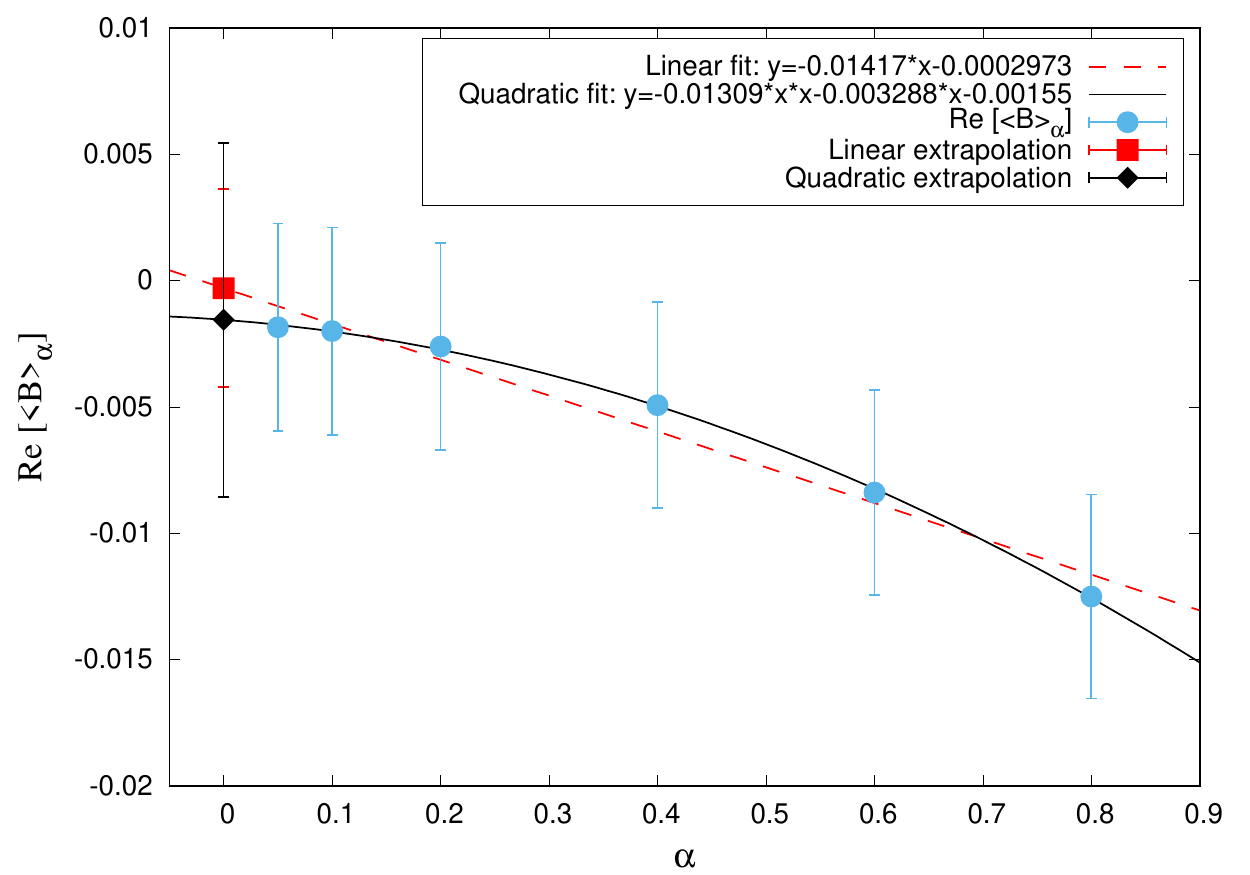}}
	\subfloat[Imaginary part of $\langle B \rangle_\alpha$]{\includegraphics[width=3.2in]{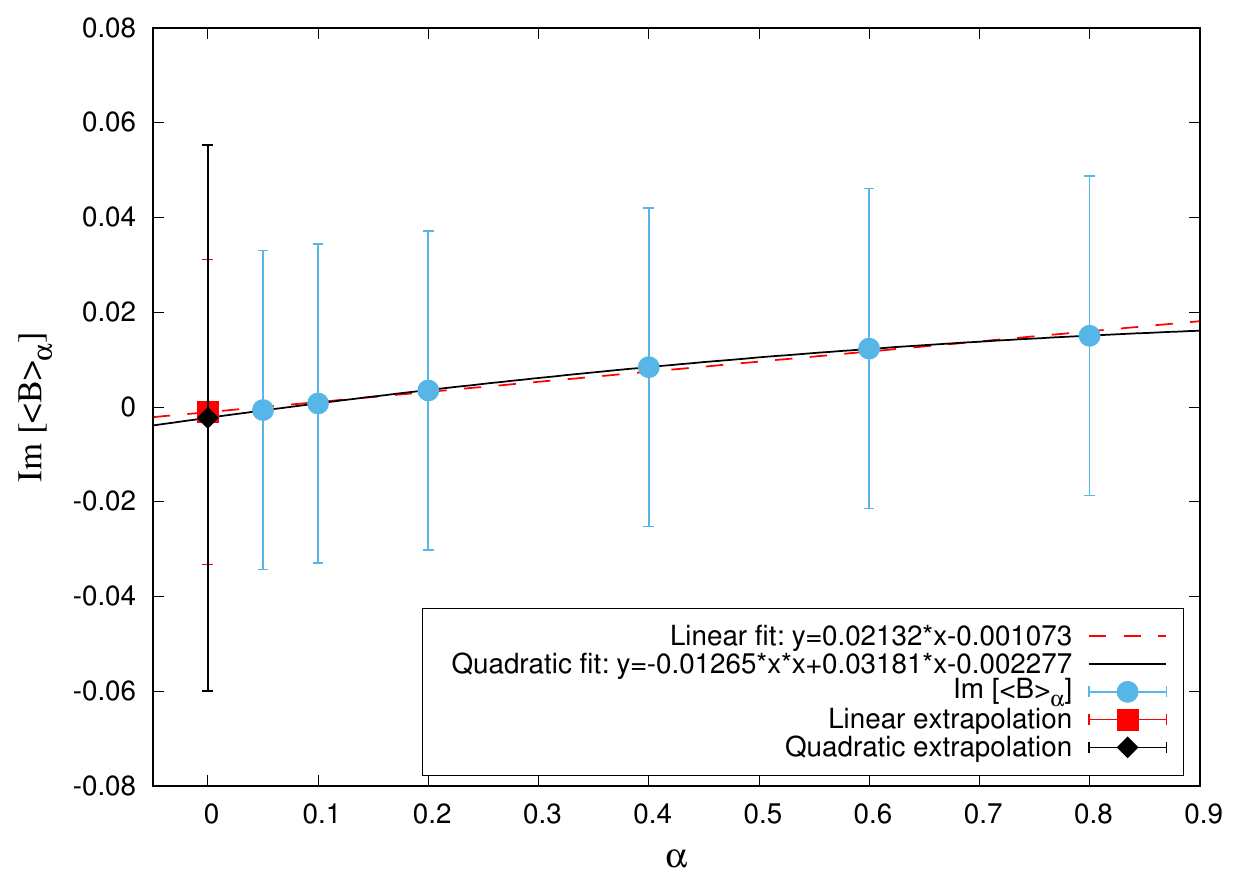}}
	
	\caption{Plot of real (Left) and imaginary (Right) parts of $\langle B \rangle_\alpha$ against the regularization parameter, $\alpha$ for supersymmetric potential $W' = ig \ (\phi^2 + \mu^2)$. The simulations were performed with parameters $g = 1$ and $\mu = 2$. We have used adaptive Langevin step size $\Delta \tau \leq 10^{-4}$, thermalization steps $N_{\rm therm} = 10^{4}$, generation steps $N_{\rm gen} = 10^6$ and measurements were taken every $100$ steps. The dashed red lines are the linear fits to $\langle B \rangle_\alpha$ in $\alpha$, and filled red squares are the linear extrapolation values at $\alpha = 0$.  The solid black lines represent the quadratic fits to $\langle B \rangle_\alpha$ in $\alpha$, and filled black diamonds are the quadratic extrapolation values at $\alpha = 0$. The $\alpha \to 0$ limit values obtained from these plots are given in Table \ref{tab:isqw_mu2p0}.}
	\label{fig:isqw_fit_g1p0_mu2p0}
	
\end{figure*}

\begin{figure*}[htp]
	
	\subfloat[Real part of $\langle B \rangle_\alpha$]{\includegraphics[width=3.2in]{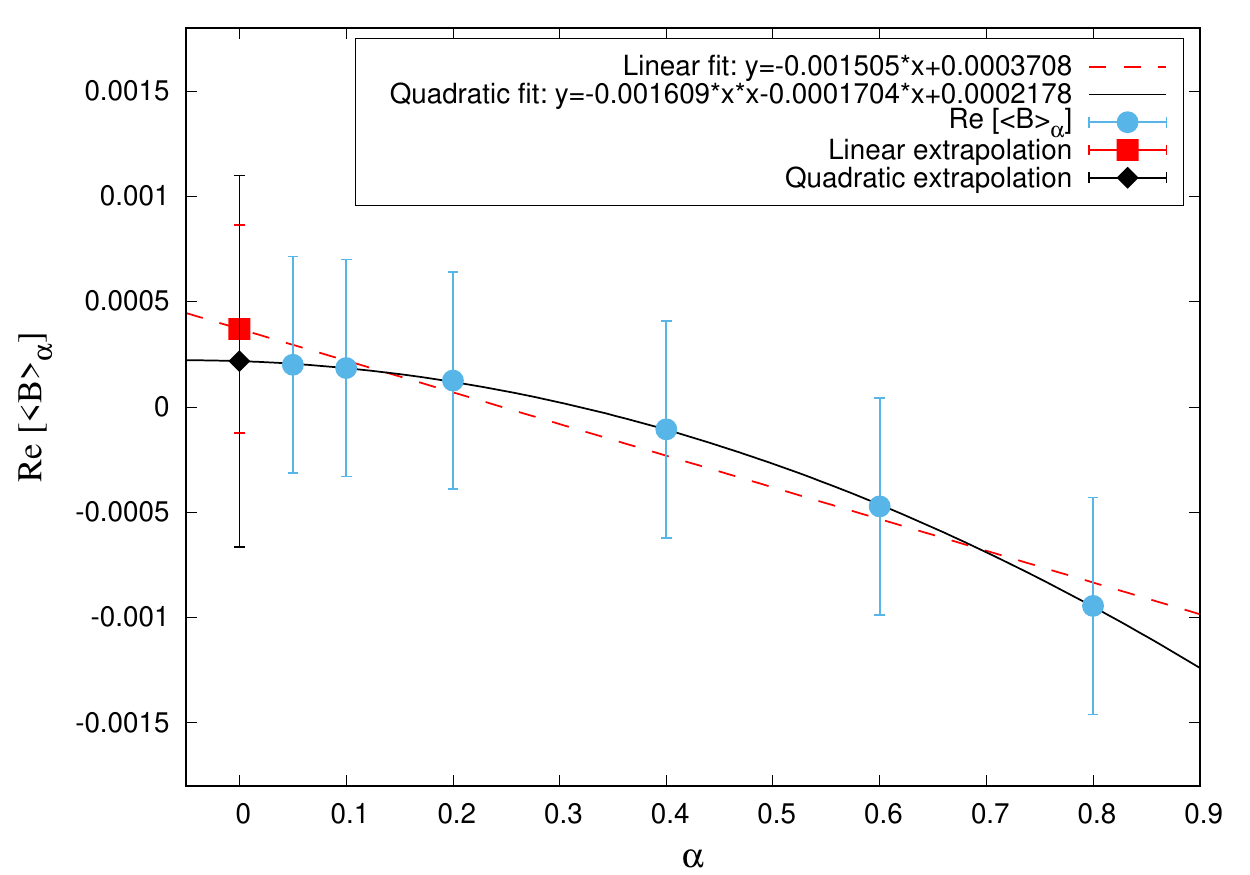}}
	\subfloat[Imaginary part of $\langle B \rangle_\alpha$]{\includegraphics[width=3.2in]{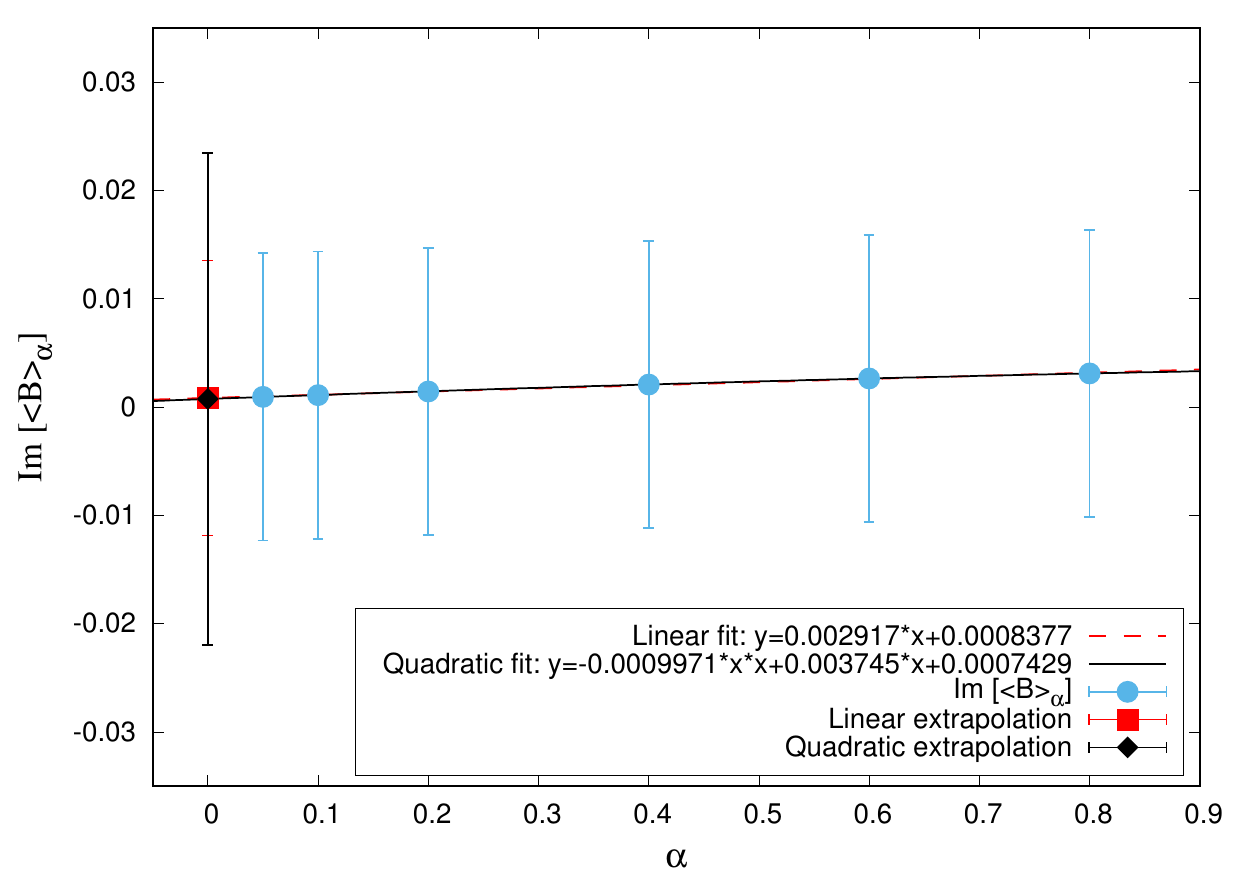}}
	
	\caption{Real (Left) and imaginary (Right) parts of $\langle B \rangle_\alpha$ against the regularization parameter $\alpha$ for supersymmetric potential $W' = ig \ (\phi^2 + \mu^2)$. The simulations were performed with $g = 3$ and $\mu = 2$. We have used adaptive Langevin step size $\Delta \tau \leq 10^{-4}$, thermalization steps $N_{\rm therm} = 10^{4}$, generation steps $N_{\rm gen} = 10^6$ and measurements were taken every $100$ steps. The dashed red lines are the linear fits to $\langle B \rangle_\alpha$ in $\alpha$, and filled red squares are the linear extrapolation values at $\alpha = 0$.  The solid black lines represent the quadratic fits to $\langle B \rangle_\alpha$ in $\alpha$, and filled black diamonds are the quadratic extrapolation values at $\alpha = 0$. The $\alpha \to 0$ limit values obtained from these plots are given in Table \ref{tab:isqw_mu2p0}.}
	\label{fig:isqw_fit_g3p0_mu2p0}
	
\end{figure*}

We also consider the case when the derivative of the superpotential is complex,
\beq
W' = ig \ (\phi^2 + \mu^2),
\eeq
where $g$ and $\mu$ are again two parameters in the theory. We show Langevin time history of the auxiliary $B$ field, and linear and quadratic extrapolations to $\alpha \to 0$ limit in Figs. \ref{fig:isqw-g1-g3-p0_mu2p0}, \ref{fig:isqw_fit_g1p0_mu2p0} and \ref{fig:isqw_fit_g3p0_mu2p0}, respectively. The results are tabulated in Table \ref{tab:isqw_mu2p0}. We have successfully simulated the complex double-well superpotential using complex Langevin and our results strongly suggest that SUSY is preserved for this model.

\begin{figure*}[htp]

\subfloat[$W' = g (\phi^2 + \mu^2)$]{\includegraphics[width=3.2in]{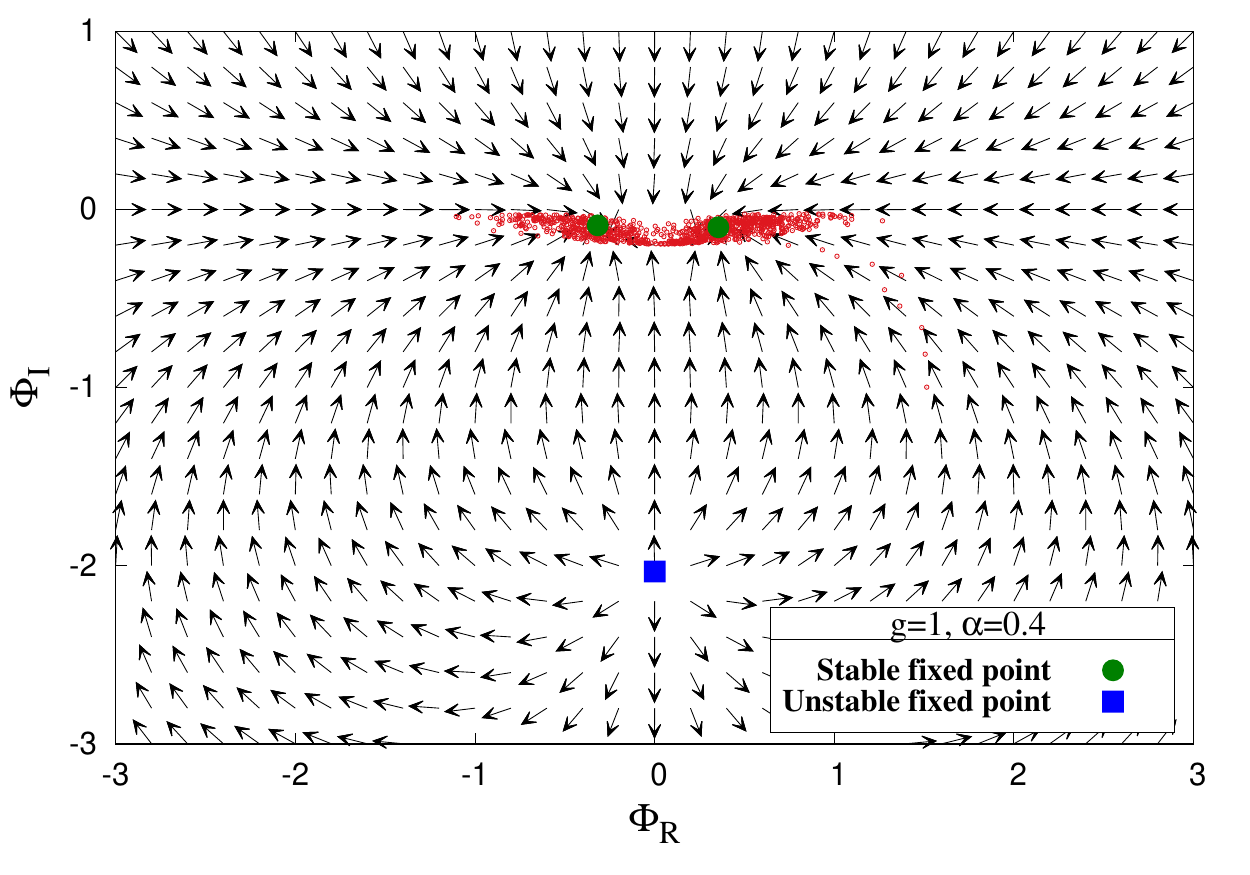}}
\subfloat[$W' = i g (\phi^2 + \mu^2)$]{\includegraphics[width=3.2in]{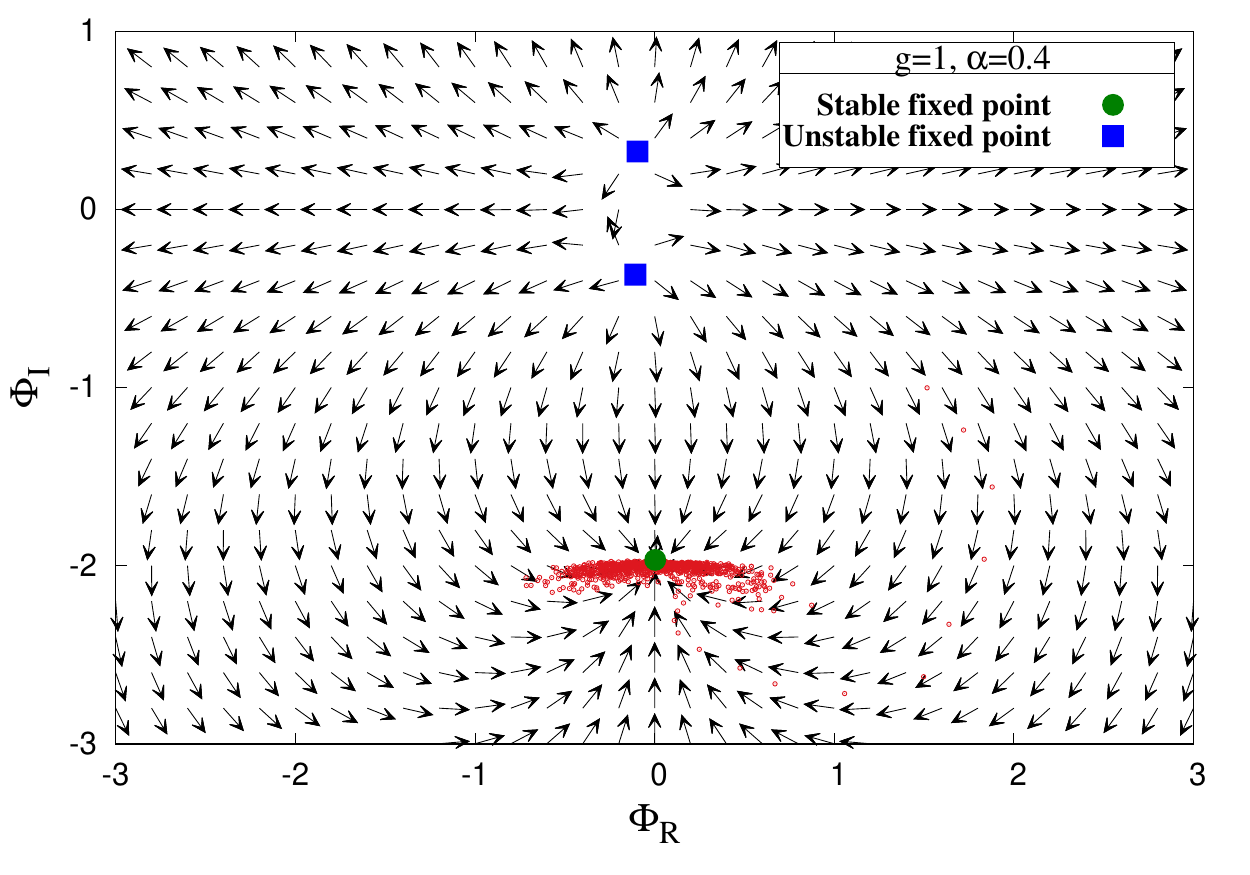}}

\caption{The scatter plot of field configurations (red dots) and classical flow diagram (arrows) on the $\phi_R - \phi_I$ plane. The red dots represent trajectories of the fields during Langevin evolution for superpotential (Left) $W' = g (\phi^2 + \mu^2)$ and (Right) $W' = i g (\phi^2 + \mu^2)$. In these simulations, we have used $g = 1.0, ~\mu = 2.0$ and $\alpha = 0.4$.  The first $10^5$ points are plotted with measurements taken every $10^2$ steps.  In both cases, the field start at point $(1.5, -1.0)$ and with the aid of a stochastic noise it drifts towards equilibrium configuration. Filled circles and squares, represent the stable and unstable fixed points, respectively.}
\label{fig:dw_flow}
\end{figure*}

The results mentioned above can be partly motivated by classical dynamics, that is, in the absence of stochastic noise. In Fig. \ref{fig:dw_flow}, we show the classical flow diagrams on the $\phi_R-\phi_I$ plane for the above discussed double-well models. The arrows indicate normalized drift term evaluated at the particular field point. In the same figure, we have also shown the scatter plot of complexified field configurations. These plots demonstrate how equilibrium configurations are attained during complex Langevin dynamics.

\subsection{General polynomial potential}
\label{subsec:gen-sps}

Let us extend our analyses to the case where the derivative of superpotential, $W'$, is a general polynomial of degree $k$,
\beq
W' = g_k \phi^k + g_{k-1} \phi^{k-1} + \cdots + g_0.
\eeq

The twisted partition function is written as
\begin{widetext}
\bea
\label{eq:gen-poly-Z}
Z_\alpha &=& -\frac{1}{ \sqrt{2 \pi} } \int_{-\infty}^\infty d\phi ~ \Big( e^{i \alpha} -1 + W'' \Big) \exp \left[ - \hf W'^{2} \right] \nn \\
&=& -\frac{(e^{i \alpha} -1)}{ \sqrt{2 \pi} } \int_{-\infty}^\infty d\phi ~  \exp \left[ - \hf W'^{2} \right]-\frac{1}{ \sqrt{2 \pi} } \int_{-\infty}^\infty d\phi ~  W'' \exp \left[ - \hf W'^{2} \right]. 
\eea
\end{widetext}

For the second term in the above equation, assuming the coefficients of the polynomial potential to be real, we have
\beq
\frac{1}{\sqrt{2 \pi}} \int_{-\infty}^\infty W'' e^{\left[- \hf W'^2\right]} = 
\begin{cases}
	{\rm sgn}(g_k)      & \ k: \text{ odd} \\
	0 & \ k: \text{ even}
\end{cases}
\eeq

Upon turning off the external field, the first term of Eq. \eqref{eq:gen-poly-Z} vanishes, hence 
\beq
\label{eq:Z-susy-poly-cond}
Z_\alpha \Big|_{\alpha \to 0} = 
\begin{cases}
	{-\rm sgn}(g_k)      & \ k: \text{ odd} \\
	0 & \ k: \text{ even}
\end{cases}
\eeq

Thus, for a general polynomial superpotential, $W'$ of the degree even (odd), the SUSY is broken (preserved).

The expectation value of the auxiliary $B$ field is
\begin{widetext}
\bea
\label{eq:B_poly}
\langle B \rangle_\alpha &=& - \frac{1}{Z_\alpha} \frac{1}{ \sqrt{2 \pi} } \int_{-\infty}^\infty d\phi ~ (-iW')\Big( e^{i \alpha} -1 + W'' \Big) \exp \left[ -\hf W'^{2} \right] \nn \\
&=&\frac{i}{Z_\alpha}\frac{(e^{i \alpha} -1)}{ \sqrt{2 \pi} } \int_{-\infty}^\infty d\phi ~W'  \exp \left[ - \hf W'^{2} \right] + \frac{i}{Z_\alpha}\frac{1}{ \sqrt{2 \pi} } \int_{-\infty}^\infty d\phi ~ W'~ W'' \exp \left[ - \hf W'^{2} \right].
\eea
\end{widetext}

The second term of Eq. \eqref{eq:B_poly} vanishes for a polynomial superpotential. ( Since we have twisted partition function in denominator, this term is not indefinite.) Hence, we have
\beq
\langle B \rangle_\alpha  = 
\begin{cases}
	\frac{i\frac{(e^{i \alpha} -1)}{ \sqrt{2 \pi} } \int_{-\infty}^\infty d\phi ~W'  e^{\left[ - \hf W'^{2} \right]} }{-\frac{(e^{i \alpha} -1)}{ \sqrt{2 \pi} } \int_{-\infty}^\infty d\phi ~ e^{\left[ - \hf W'^{2} \right]}~ - ~ {\rm sgn}(g_k) }  & k: \text{ odd}\\
	\\
	\frac{-i\int_{-\infty}^\infty d\phi ~W'  e^{\left[ - \hf W'^{2} \right]}}{\int_{-\infty}^\infty d\phi ~ e^{\left[ - \hf W'^{2} \right]}} & k: \text{ even}
\end{cases} 
\eeq

Now, turning external field off, $\alpha \to 0$,
\beq
\label{eq:B-poly-susy-cond}
\langle B \rangle_\alpha \Big|_{\alpha \to 0} =
\begin{cases}
	~~~~0    & k: \text{ odd} \\
	\frac{-i\int_{-\infty}^\infty d\phi ~W'  e^{\left[ - \hf W'^{2} \right]}}{\int_{-\infty}^\infty d\phi ~e^{\left[ - \hf W'^{2} \right]}} \neq 0 & k: \text{ even}
\end{cases}
\eeq

The above expression confirms that SUSY is preserved (broken) for odd (even) degree of derivative of a real general polynomial superpotential.

Let us consider polynomial superpotential with real coefficients. In this case the above argument for SUSY breaking is valid. Later, we will also discuss a specific case of complex polynomial potential. For simplicity we assume  that $g_k = g_{k-1} = \cdots = g_0 = 1$, then for $k = 3, 4$ we have
\beq
W'[k=3] = \phi^3 + \phi^2 + \phi +1,
\eeq
and
\beq
W'[k=4] = \phi^4 + \phi^3 + \phi^2 + \phi +1.
\eeq

\begin{figure*}[htp]
	
	\subfloat[$k = 3$]{\includegraphics[width=3.2in]{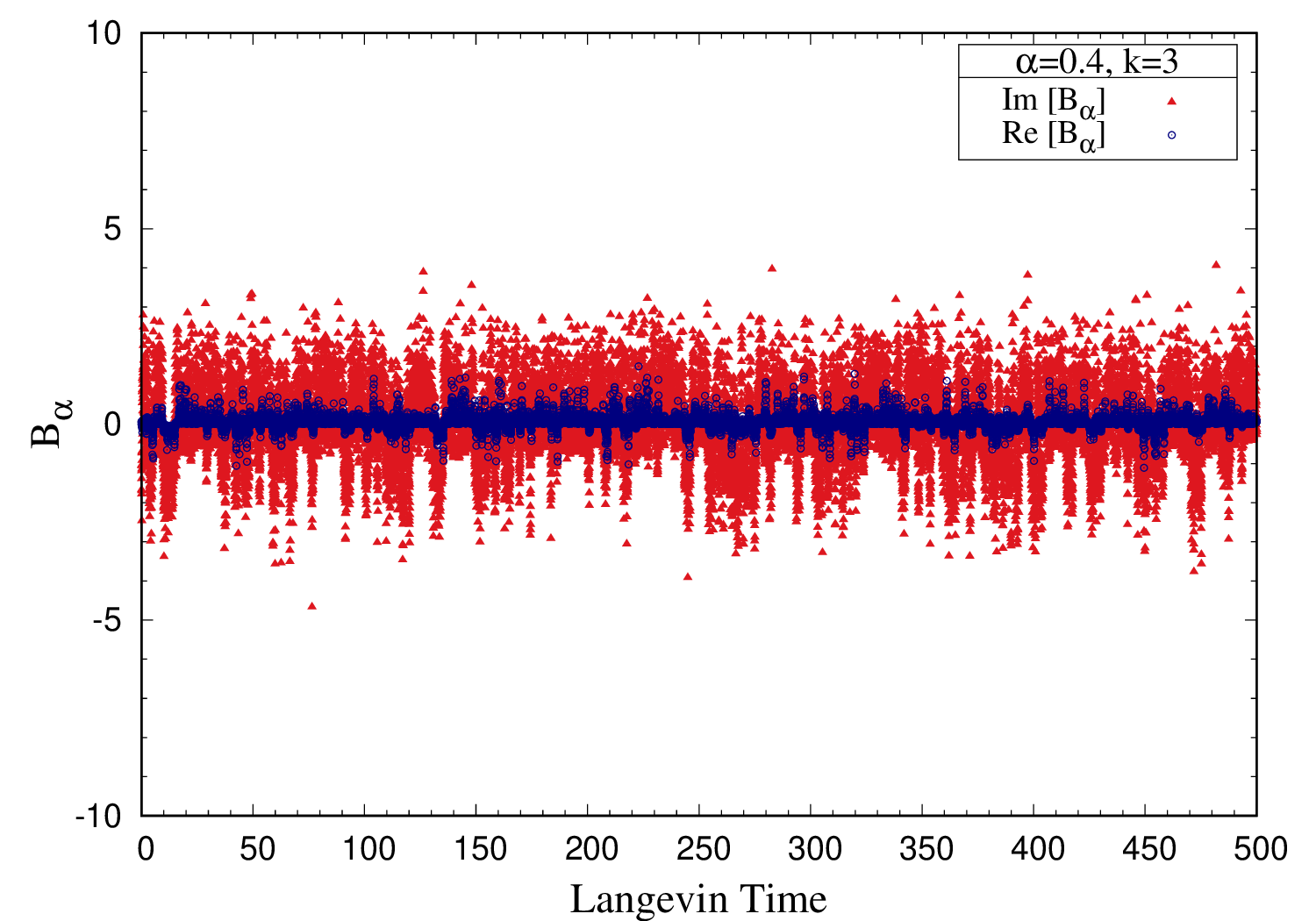}}
	\subfloat[$k = 4$]{\includegraphics[width=3.2in]{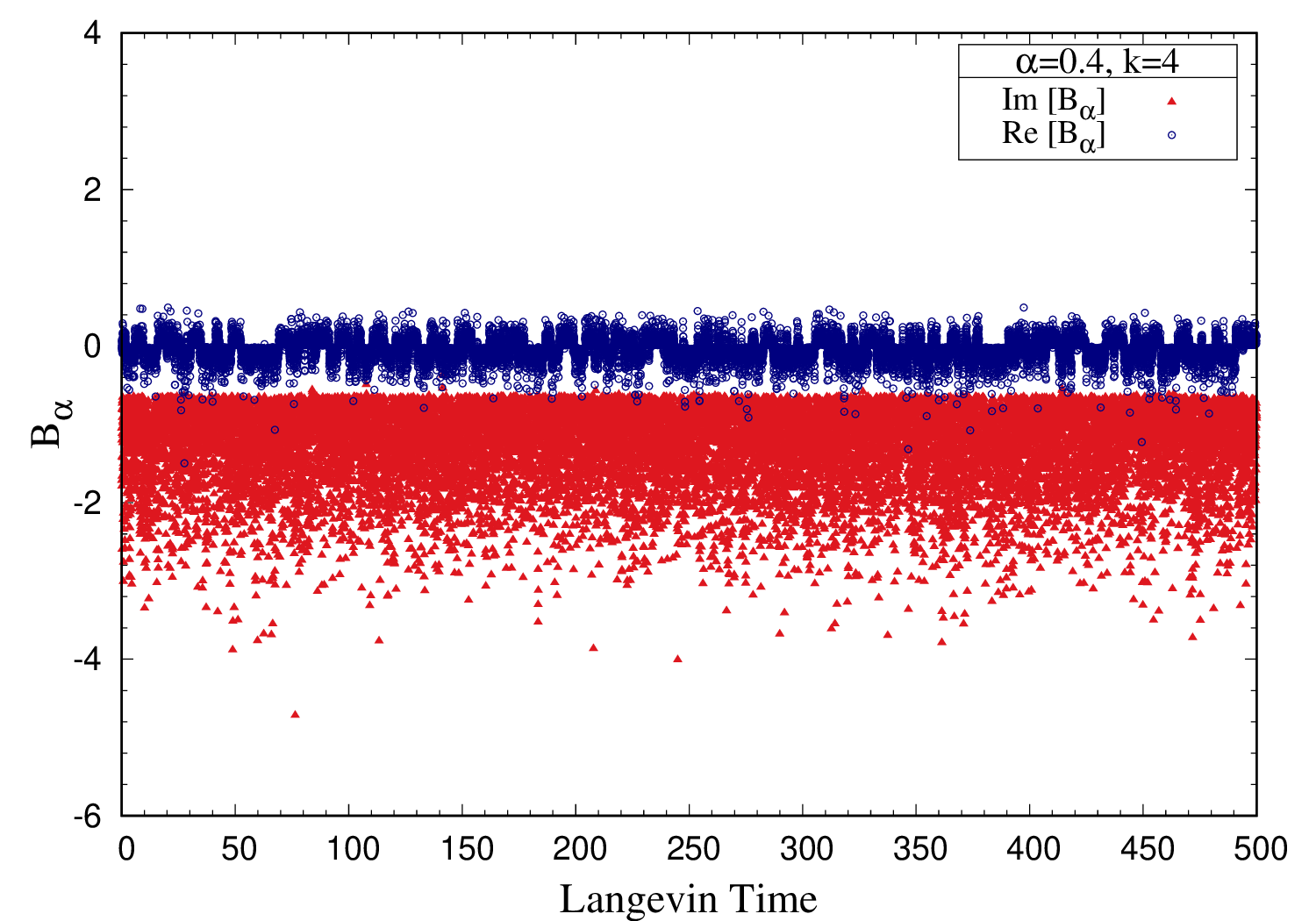}}
	
	\caption{Langevin time history of the field $B$ for $\alpha = 0.4$. Simulations were performed for superpotential $W'(\phi) = g_k \phi^k + g_{k-1} \phi^{k-1} + \cdots + g_0$ with $g_k = g_{k-1} = \cdots = g_0 = 1$. In these simulations, we have used adaptive Langevin step size $\Delta \tau \leq 5 \times 10^{-5}$, generation steps $N_{\rm gen} = 10^7$ and measurements were taken every $500$ steps. (Left) $k = 3$ case. (Right) $k = 4$ case.}
	\label{fig:B-poly-history}
	
\end{figure*}

\begin{figure*}[htp]
	
	\subfloat[ Real part of $ \langle B \rangle_\alpha$ for $k=3$]{\includegraphics[width=3in]{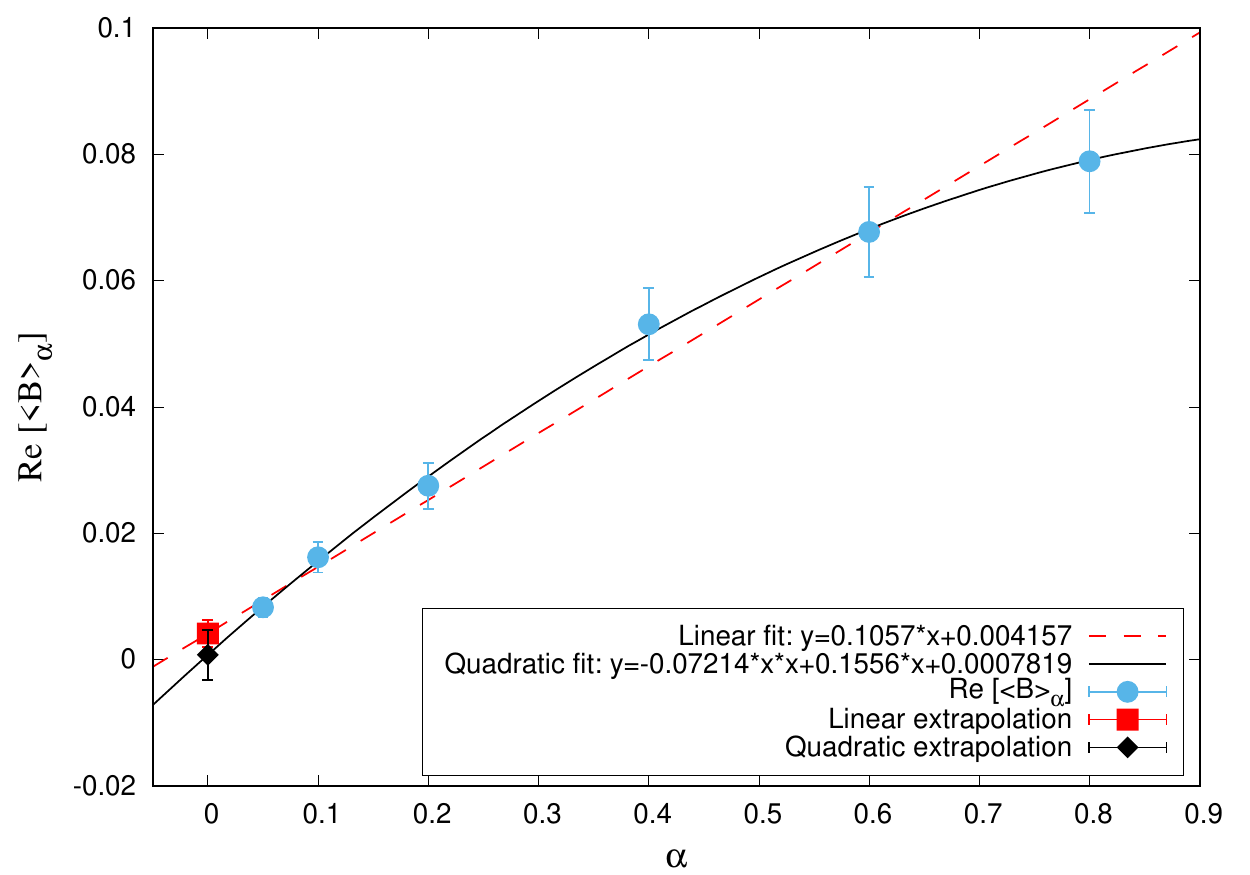}}
	\subfloat[ Imaginary part of $ \langle B\rangle_\alpha$ for $k=3$]{\includegraphics[width=3in]{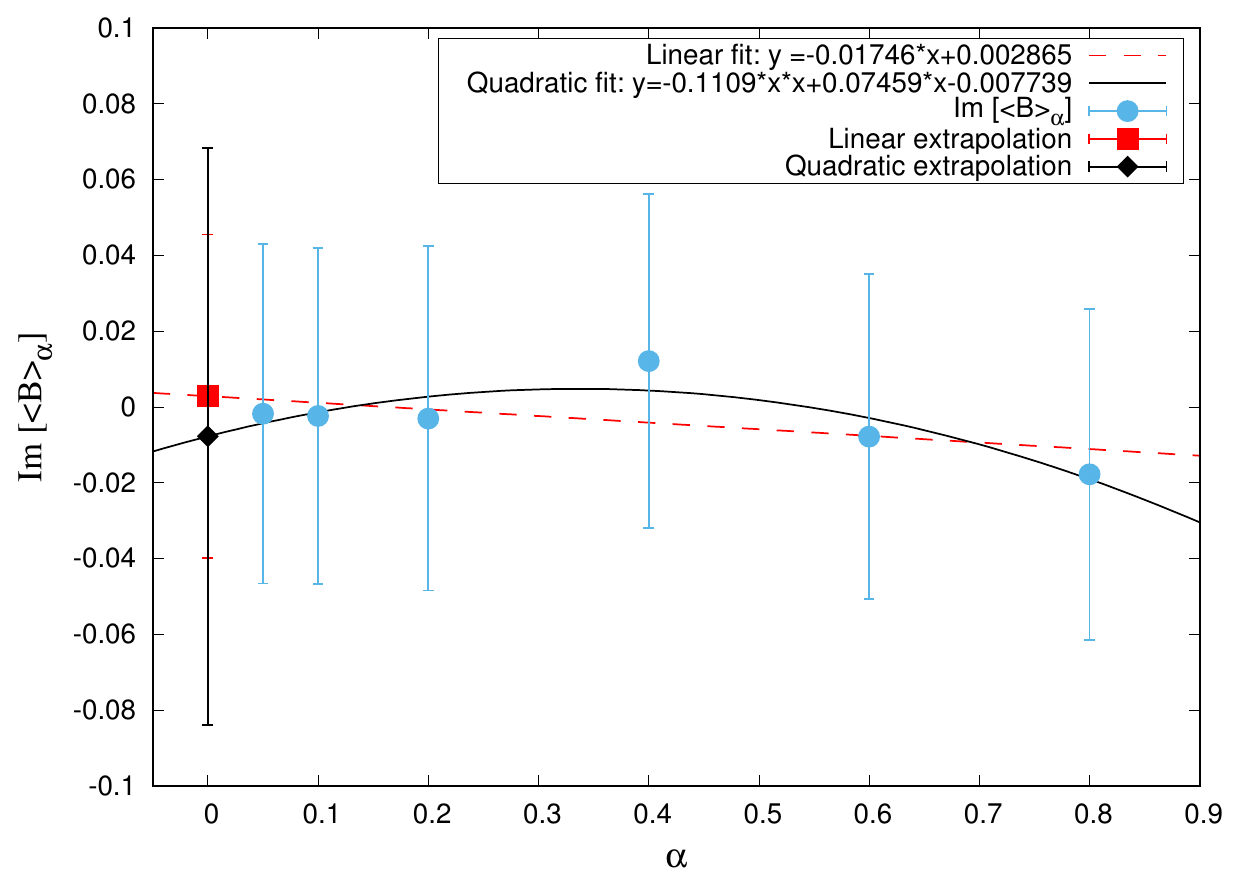}}
	
	\subfloat[ Real part of $ \langle B \rangle_\alpha$ for $k=4$]{\includegraphics[width=3.0in]{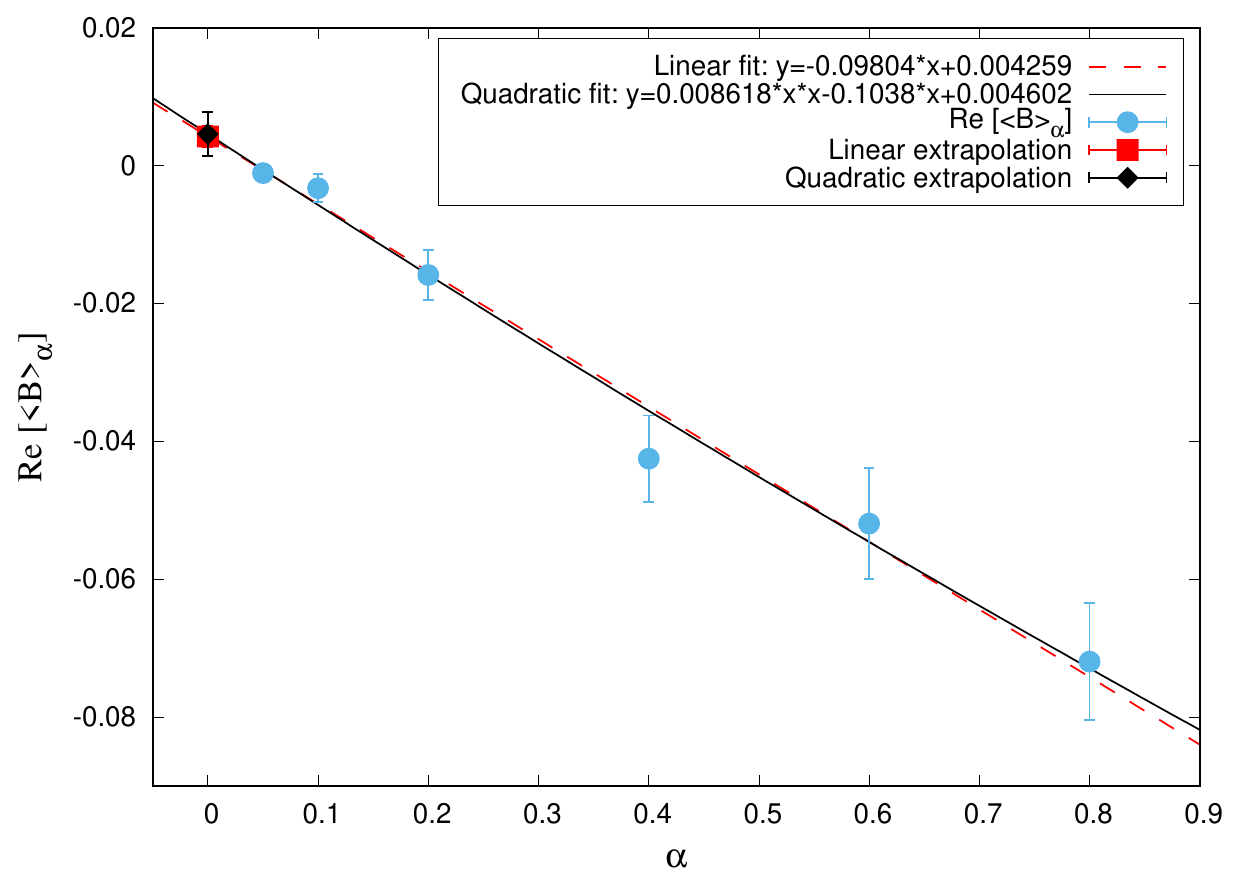}}
	\subfloat[ Imaginary part of $ \langle B\rangle_\alpha$ for $k=4$]{\includegraphics[width=3.0in]{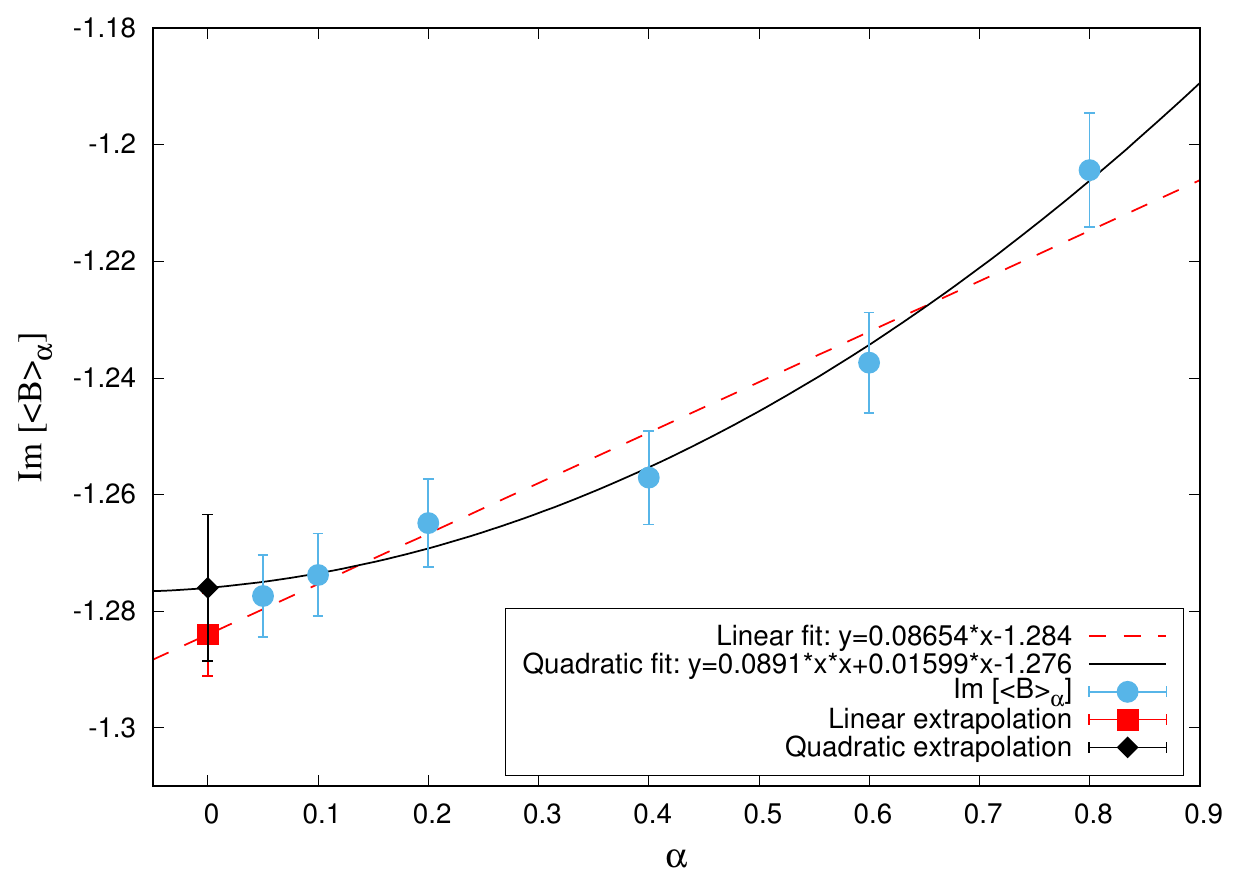}}
	
	\caption{The expectation value, $\langle B \rangle_\alpha$ against the regularization parameter, $\alpha$ for superpotential $W'(\phi) = g_k \phi^k + g_{k-1} \phi^{k-1} + \cdots + g_0$ with $g_k =g_{k-1}=\cdots=g_0= 1$. (Top-Left) Real part and (Top-Right) imaginary part of $\langle B \rangle_\alpha$ for $k = 3$. (Bottom-Left) Real part and (Bottom-Right) imaginary part of $\langle B \rangle_\alpha$ for $k = 4$. The simulations were performed with adaptive Langevin step size $\Delta \tau \leq 5\times10^{-5}$, thermalization steps $N_{\rm therm} = 5 \times 10^4$, generation steps $N_{\rm gen} = 10^7$ and measurements taken every 500 steps. The dashed red lines are the linear fits to $\langle B \rangle_\alpha$ in $\alpha$, and red dots are the linear extrapolation value at $\alpha = 0$.  The solid black lines represent the quadratic fits to $\langle B \rangle_\alpha$ in $\alpha$, and black dots are the quadratic extrapolation value at $\alpha = 0$. The $\alpha \to 0$ limit values obtained from these plots are given in Table \ref{tab:real-gen-poly}.}
	\label{fig:B_fit_poly}
\end{figure*}

We have learned from Eq. \eqref{eq:Z-susy-poly-cond} and \eqref{eq:B-poly-susy-cond} that SUSY is broken (preserved) for $k = 4 ~ (k = 3)$. In Fig. \ref{fig:B-poly-history} we show Langevin time history of $\langle B \rangle_\alpha$ for the above two polynomial models. We show linear and quadratic extrapolations to $\alpha \to 0$ limit in Fig. \ref{fig:B_fit_poly}. The results are tabulated in Table \ref{tab:real-gen-poly}. The simulation results are in good agreement with the corresponding analytical predictions.

\begin{figure*}[htp]
	
	\subfloat[$g=1$]{\includegraphics[width=3.2in]{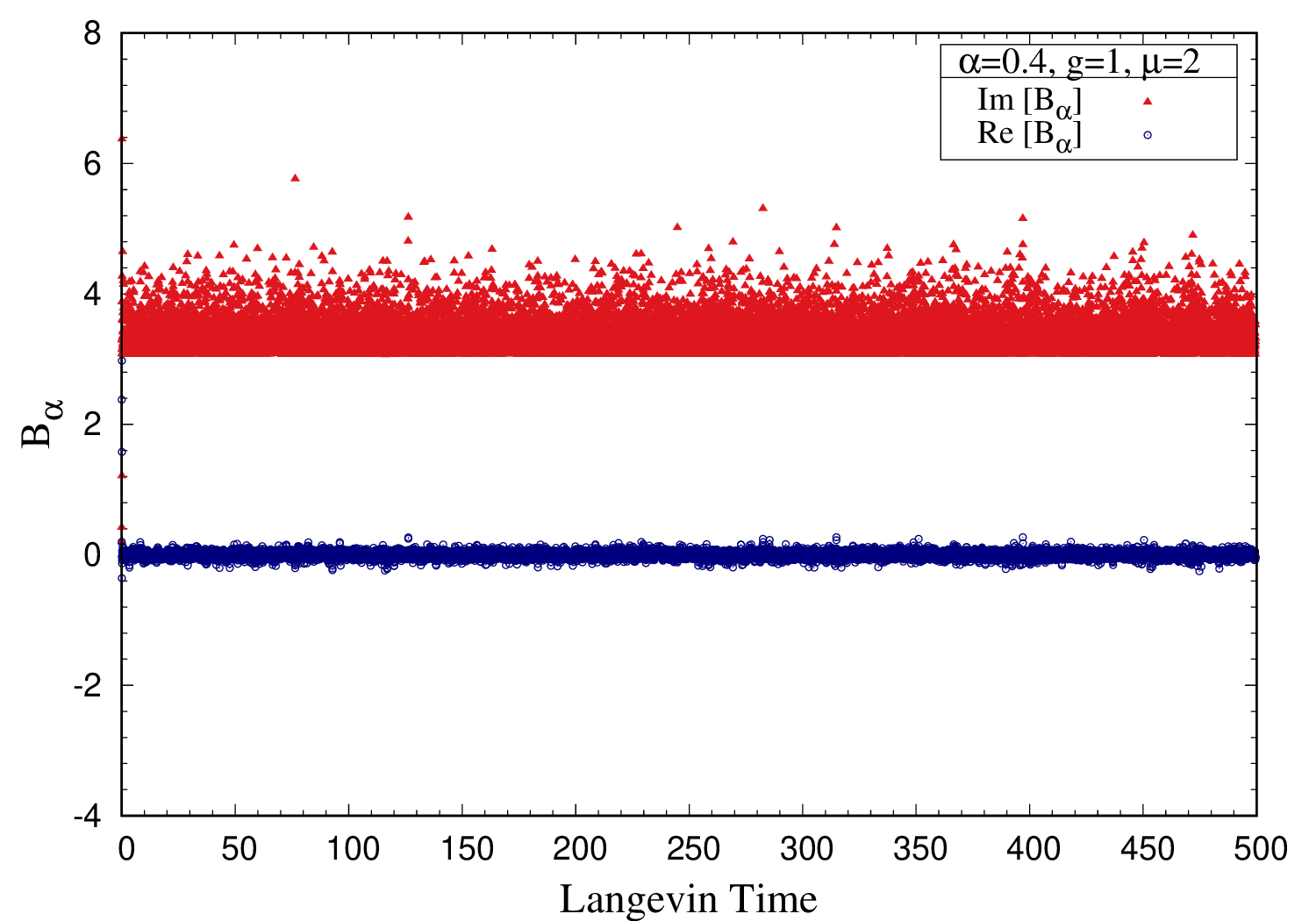}}
	\subfloat[$g = 3$]{\includegraphics[width=3.2in]{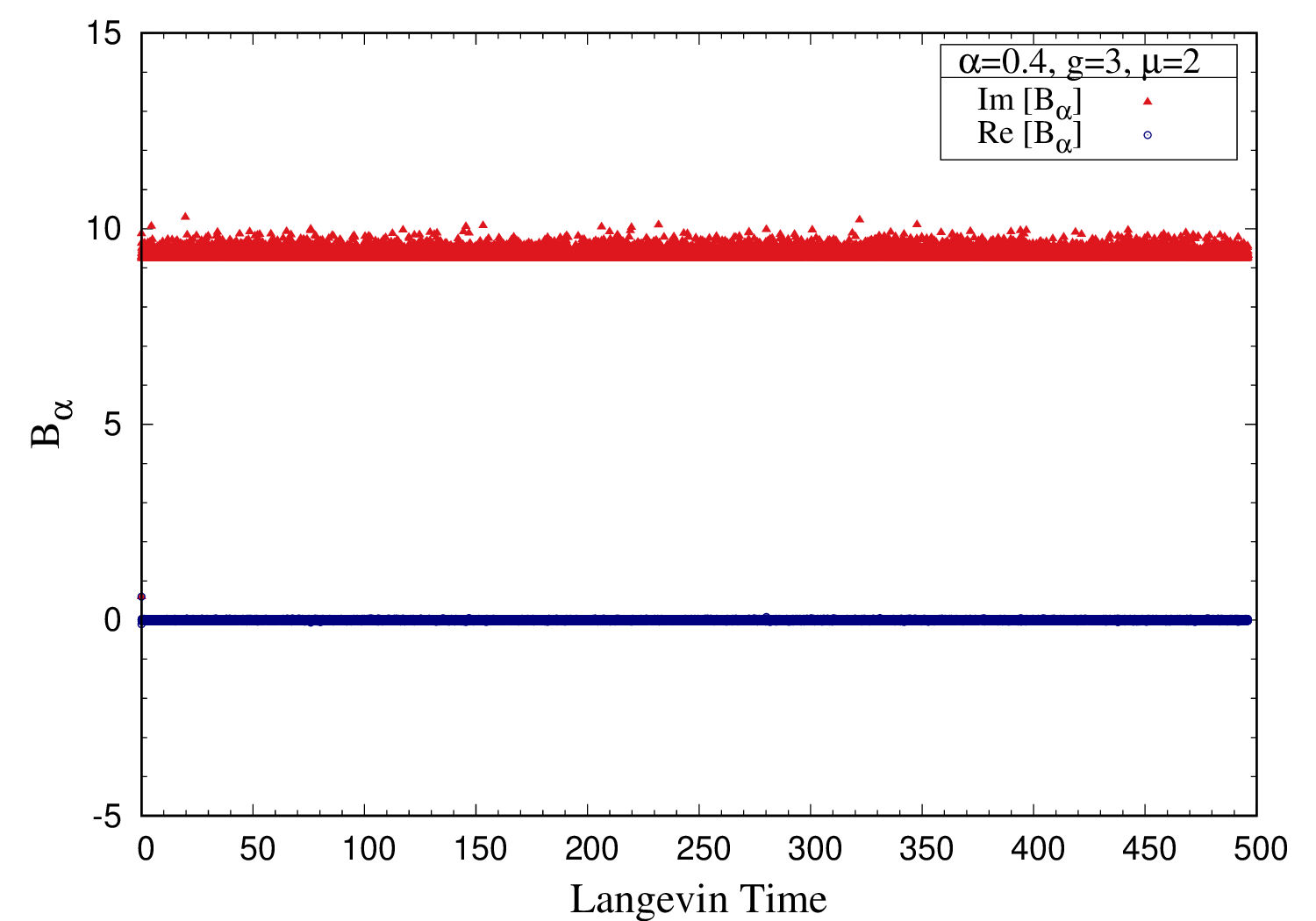}}

	\caption{The Langevin time history of $B$ for $\alpha = 0.4$. The simulations were performed for superpotential $W'(\phi) =  ig\phi (\phi^2 +\mu^2) $ with $\mu=2$. In these simulations, we have used adaptive Langevin step size $\Delta \tau \leq 5\times10^{-5}$, generation steps $N_{\rm gen} = 10^7$ and measurements were taken every $500$ steps. (Left) $g = 1$ case. (Right) $g = 3$ case. }
	\label{fig:sqw_iphi-g1-g3-p0_mu2p0}
	
\end{figure*}

\begin{figure*}[htp]
	
	\subfloat[Real part of $\langle B \rangle_\alpha$]{\includegraphics[width=3in]{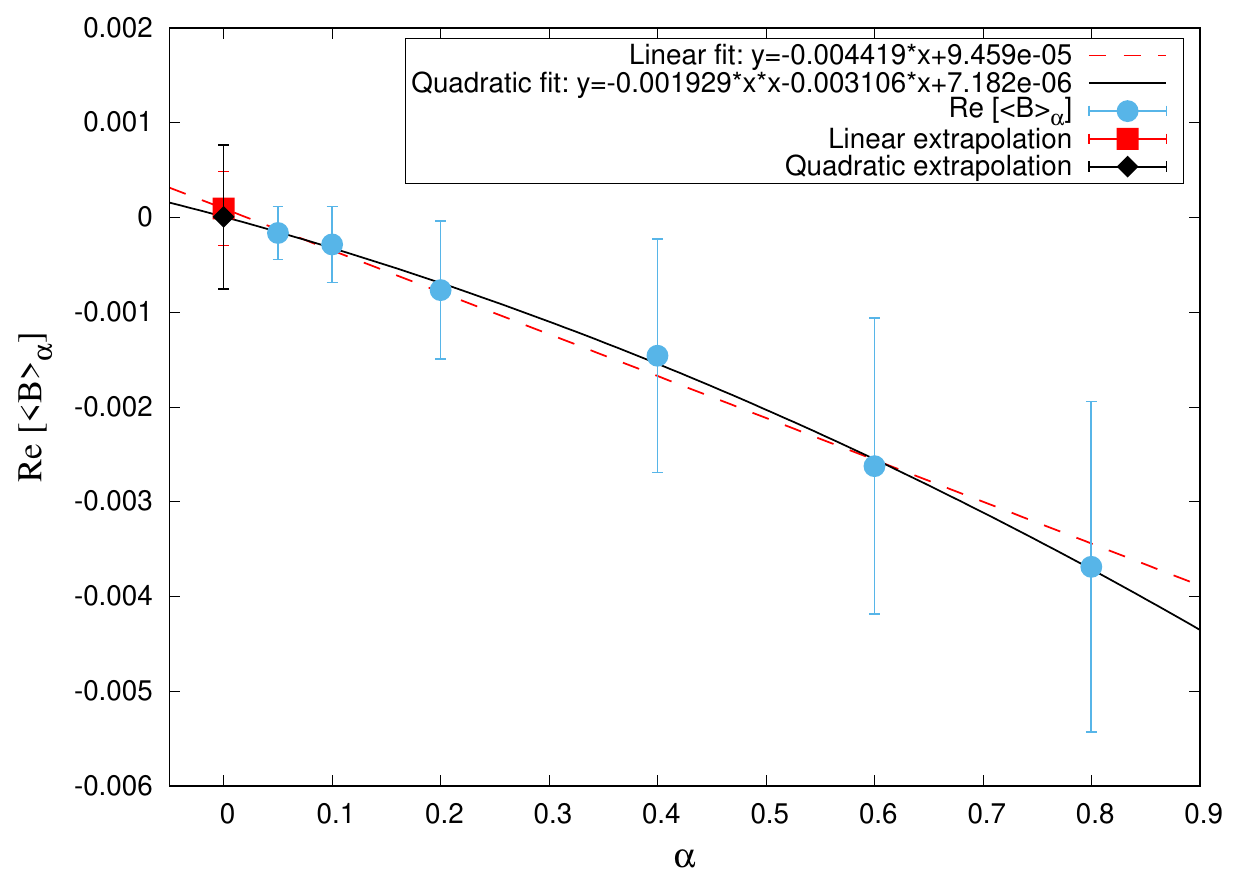}}
	\subfloat[Imaginary part of $\langle B \rangle_\alpha$]{\includegraphics[width=3in]{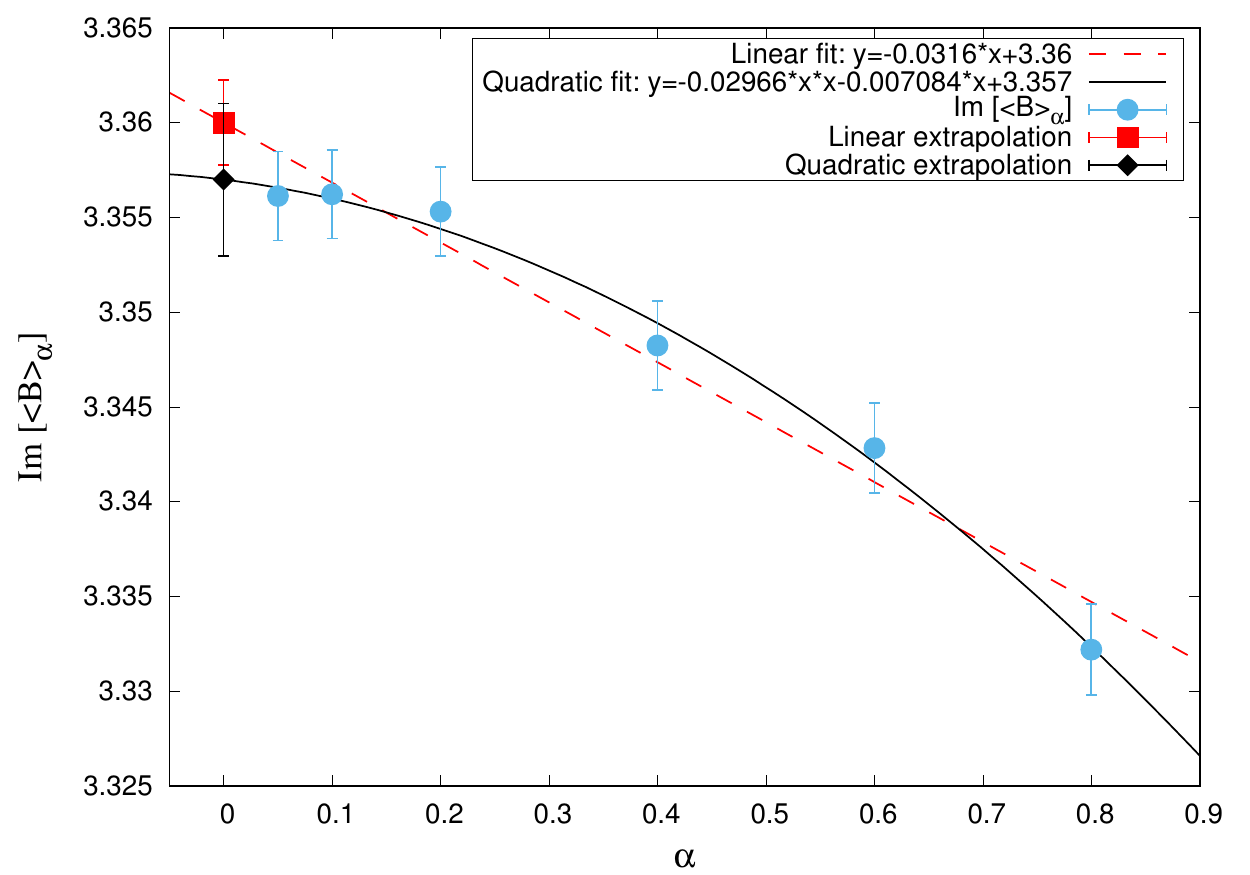}}
	
	\caption{Real (Left) and imaginary (Right) parts of $\langle B \rangle_\alpha$ against the regularization parameter, $\alpha$ for supersymmetric potential $W' = ig\phi \ (\phi^2 + \mu^2)$. Simulations were performed with $g = 1$ and $\mu = 2$. We have used adaptive Langevin step size $\Delta \tau \leq 5\times10^{-5}$, thermalization steps $N_{\rm therm} =  5 \times 10^{4}$, generation steps $N_{\rm gen} = 10^7$ and measurements were taken every $500$ steps. The dashed red lines are the linear fits to $\langle B \rangle_\alpha$ in $\alpha$, and red dots are the linear extrapolation value at $\alpha = 0$.  The solid black lines represent the quadratic fits to $\langle B \rangle_\alpha$ in $\alpha$, and black dots are the quadratic extrapolation value at $\alpha = 0$. The $\alpha \to 0$ limit values obtained from these plots are given in Table \ref{tab:sqw_iphi_mu2p0}.}
	\label{fig:sqw_iphi_fit_g1p0_mu2p0}
	
\end{figure*}

\begin{figure*}[htp]
	
	\subfloat[Real part of $\langle B \rangle_\alpha$]{\includegraphics[width=3.0in]{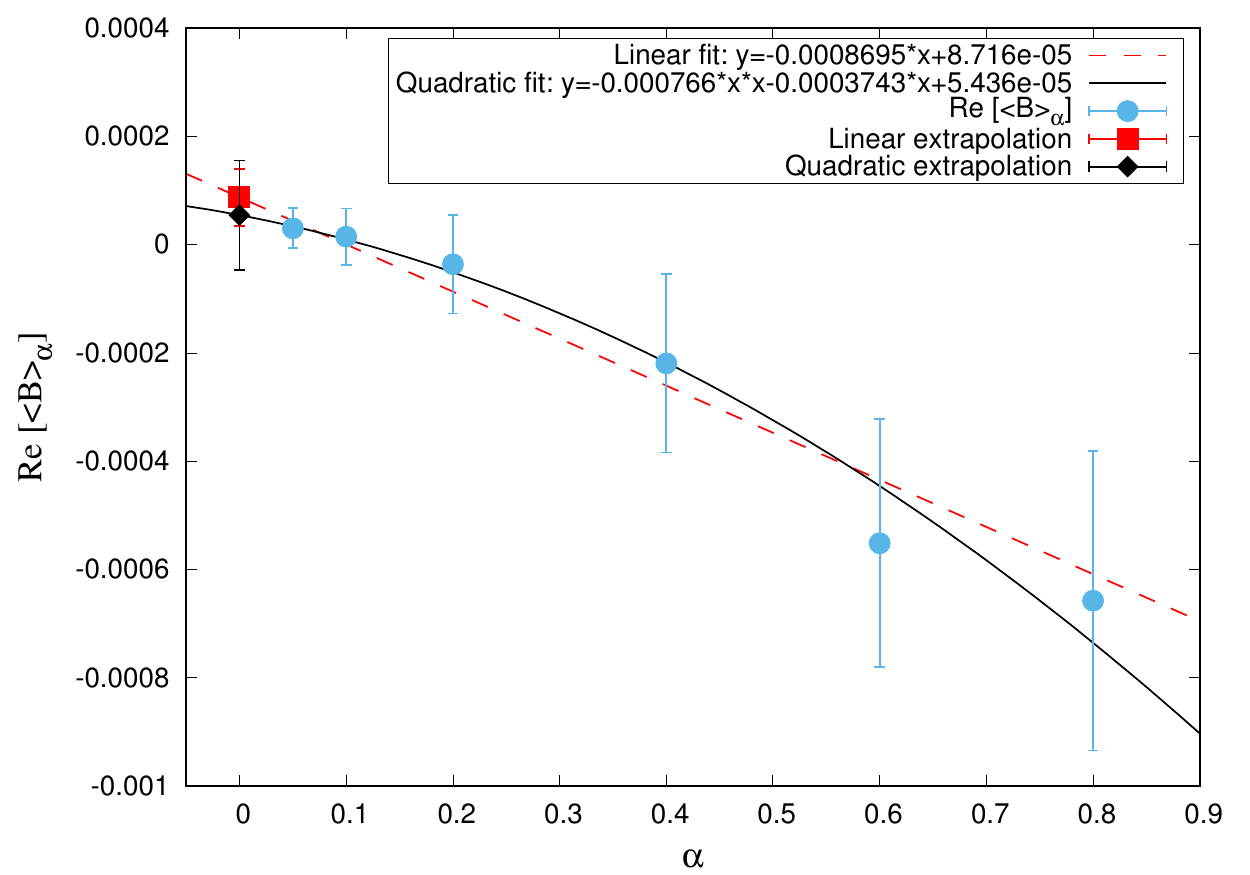}}
	\subfloat[Imaginary part of $\langle B \rangle_\alpha$]{\includegraphics[width=3.0in]{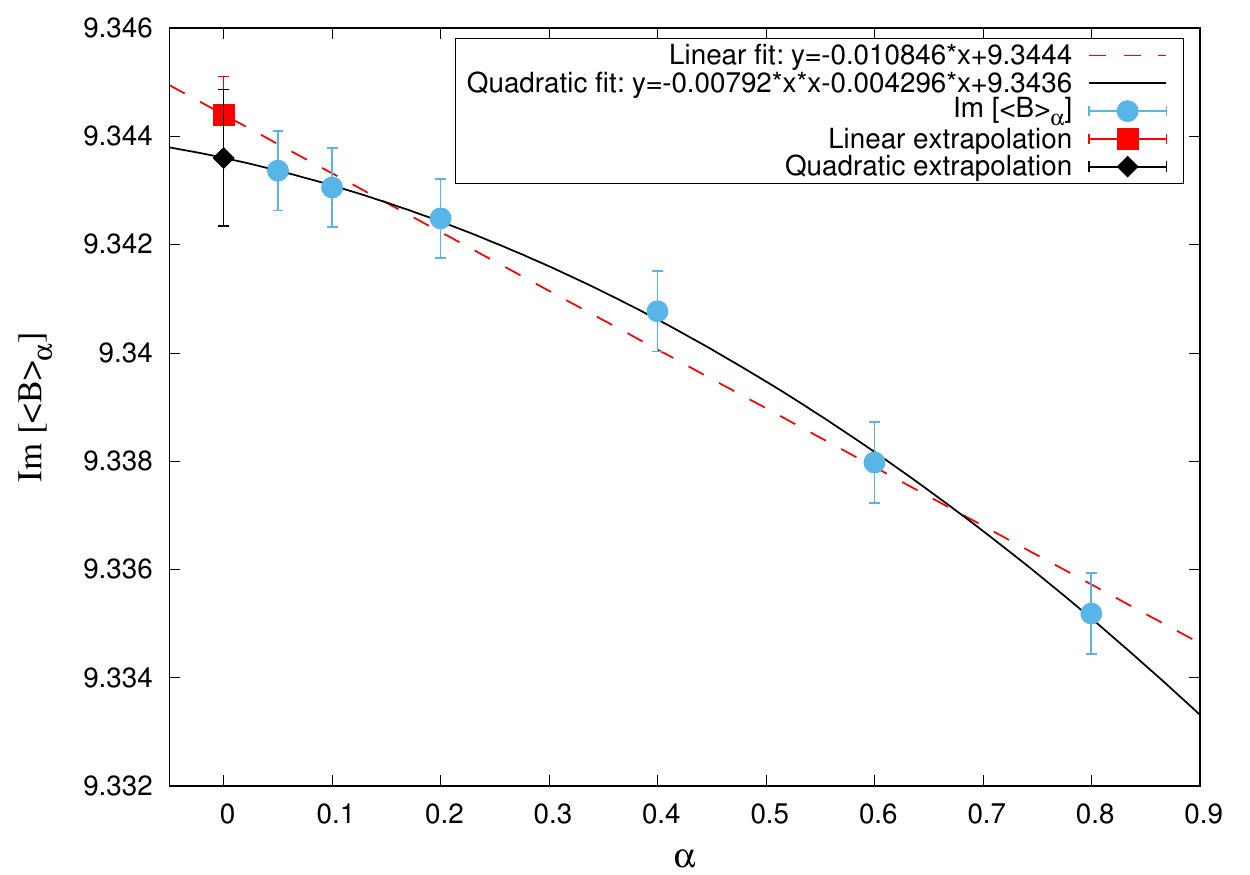}}
	
	\caption{Real (Left) and imaginary (Right) parts of $\langle B \rangle_\alpha$ against the regularization parameter, $\alpha$ for supersymmetric potential $W' = ig\phi \ (\phi^2 + \mu^2)$. Simulations were performed with $g = 3$ and $\mu = 2$. We have used adaptive Langevin step size $\Delta \tau \leq 5\times10^{-5}$, thermalization steps $N_{\rm therm} =  5 \times 10^{4}$, generation steps $N_{\rm gen} = 10^7$ and measurements were taken every $500$ steps. The dashed red lines are the linear fits to $\langle B \rangle_\alpha$ in $\alpha$, and red dots are the linear extrapolation value at $\alpha = 0$.  The solid black lines represent the quadratic fits to $\langle B \rangle_\alpha$ in $\alpha$, and black dots are the quadratic extrapolation value at $\alpha = 0$. The $\alpha \to 0$ limit values obtained from these plots are given in Table  \ref{tab:sqw_iphi_mu2p0}.}
	\label{fig:sqw_iphi_fit_g3p0_mu2p0}
	
\end{figure*}

Now, let us consider the case with complex polynomial superpotential. We modify the real double-well potential discussed in the previous section as follows,
\bea
W' &=& ig\phi (\phi^2 +\mu^2).
\eea
 In this complex potential case, the argument given in Eq. \eqref{eq:Z-susy-poly-cond} and \eqref{eq:B-poly-susy-cond} are not valid. We investigate SUSY breaking using complex Langevin dynamics. In Fig. \ref{fig:sqw_iphi-g1-g3-p0_mu2p0}, we show Langevin time history of the auxiliary $B$ field for regularization parameter, $\alpha = 0.4$. We show linear and quadratic extrapolations to $\alpha \to 0$ limit in Figs. \ref{fig:sqw_iphi_fit_g1p0_mu2p0} and \ref{fig:sqw_iphi_fit_g3p0_mu2p0}. The results are tabulated in Table \ref{tab:sqw_iphi_mu2p0}. Our simulation results imply that expectation value of auxiliary field, $\langle B \rangle_\alpha$ does not vanish in the limit, $\alpha \to 0$. Hence SUSY is broken in this model.

\subsection{$\mathcal{PT}$-symmetric models inspired $\delta$-potentials}
\label{subsec7}

Let us consider the superpotential
\beq
W(\phi) = - \frac{g}{(2 + \delta)} (i \phi)^{(2 + \delta)},
\eeq
which is the same as the one we considered earlier for the case of the bosonic models.

The twisted partition function takes the form
\bea
Z_\alpha &=& -\frac{1}{\sqrt{2 \pi}} \int_{-\infty}^\infty d\phi \ \Big( e^{i \alpha} - 1 + W'' \Big) \nn \\
&& \times e^{ \left[ - \hf W'^2 \right] } \nn \\
&=& -\frac{1}{\sqrt{2 \pi}} \int_{-\infty}^\infty d\phi \ \Big( e^{i \alpha} - 1 + g (1 + \delta) (i \phi)^{\delta} \Big) \nn \\
&& \times e^{ \left[ \hf g^2 (i \phi)^{2(1+\delta)} \right] }.
\eea

The expectation value of the auxiliary field is
\bea
\langle B \rangle_\alpha &=&- \frac{1}{Z_\alpha} \frac{1}{\sqrt{2 \pi}} \int_{-\infty}^\infty d\phi \ (-iW') \Big( e^{i \alpha} - 1 + W'' \Big) \nn \\&& \times e^{ \left[ - \hf W'^2 \right] } \nn \\
&=& \frac{1}{Z_\alpha} \frac{1}{\sqrt{2 \pi}} \int_{-\infty}^\infty d\phi \  g  (i \phi)^{1 + \delta} \nn \\
&& \times \Big( e^{i \alpha} - 1 + g (1 + \delta) (i\phi)^\delta \Big) e^{ \left[ \hf g^2 (i\phi)^{2(1+\delta)} \right] }.~~
\eea

Let us consider various integer cases of $\delta$ and check whether SUSY is broken or preserved in these cases. 

For the case, $\delta =0$ one can easily perform analytical evaluations. We have the twisted partition function
\bea
Z_\alpha [\delta = 0] &=& -\frac{1}{\sqrt{2 \pi}} \int_{-\infty}^\infty d\phi \ \Big( e^{i \alpha} - 1 + g \Big) \ e^{ \left[ - \hf g^2 \phi^2 \right] } \nn \\
&=& -\frac{1}{\sqrt{2 \pi}} \Big( e^{i \alpha} - 1 + g \Big) \sqrt{\frac{2 \pi}{g^2}}.
\eea

Turning the external field off, $\alpha \to 0$, we get a non-zero value for the partition function
\beq
Z_{\alpha=0} [\delta = 0] = -\frac{1}{\sqrt{2 \pi}} g \sqrt{\frac{2 \pi}{g^2}} = -1,
\eeq
implying that SUSY is preserved in the system.

Also we have
\bea
\langle B \rangle_\alpha   [\delta = 0] &=&  \frac{1}{Z_\alpha} \frac{1}{\sqrt{2 \pi}} \int_{-\infty}^\infty d\phi \ (ig \phi) \nn \\
&& \times \Big( e^{i \alpha} - 1 + g \Big) ~ e^{ \left[ -\hf g^2 \phi^{2} \right] } \nn \\
&=&-\frac{ ig \int_{-\infty}^\infty d\phi \ \phi \Big( e^{i \alpha} - 1 + g \Big) e^{ \left[ - \hf g^2 \phi^2 \right] }}{ \Big( e^{i \alpha} - 1 + g \Big) \sqrt{\frac{2 \pi}{g^2}}} \nn \\
&=& -\frac{i g \int_{-\infty}^\infty d\phi \ \phi \ \exp \left[ - \hf g^2 \phi^2 \right]}{ \sqrt{\frac{2 \pi}{g^2}}} \nn \\
&=& 0.
\eea

Since $\langle B \rangle_\alpha   [\delta = 0] = 0$, we infer that SUSY is preserved in the theory when $\delta = 0$.

For the case $\delta =2$, we have
the twisted partition function
\bea
Z_\alpha [\delta = 2] &=&- \frac{1}{\sqrt{2 \pi}} \int_{-\infty}^\infty d\phi \ \Big( e^{i \alpha} - 1 - 3g  \phi^{2} \Big) \nn \\
&& \times e^{\left[ -\hf g^2 \phi^{6} \right]} \nn \\
&=& -\frac{\Big( e^{i \alpha} - 1\Big)}{\sqrt{2 \pi}} \int_{-\infty}^\infty d\phi \ e^{\left[ -\hf g^2  \phi^{6} \right]} \nn \\
&& +  \frac{3g}{\sqrt{2 \pi}} \int_{-\infty}^\infty d\phi \ \phi^{2} \ e^{\left[ -\hf g^2  \phi^{6} \right]}.
\eea

Turning the external field off, $\alpha \to 0$, we get a non-zero partition function
\bea
Z_{\alpha = 0} [ \delta = 2 ] &=&  \frac{3g}{\sqrt{2 \pi}} \int_{-\infty}^\infty d\phi \ \phi^{2} \ e^{\left[ -\hf g^2  \phi^{6} \right]} \nn \\
&=& 1,
\eea
indicating that SUSY is preserved in the system.

The expectation value of the $B$ field is
\bea
\langle B \rangle_\alpha [\delta = 2] &=&   -\frac{1}{Z_\alpha} \frac{ig}{\sqrt{2 \pi}} \int_{-\infty}^\infty d\phi \ \phi^{3} \nn \\
&& \times \left( e^{i \alpha} - 1 - 3g \phi^2 \right) e^{\left[ -\hf g^2 \phi^{6} \right]} \nn \\ 
&=&0,
\eea
confirming that SUSY is preserved for the case $\delta=2$. One can perform similar calculations for the case $\delta = 4$ and show that SUSY is preserved in the theory.

We simulate the $\delta$-potential using complex Langevin dynamics for $\delta = 1, 2, 3$ and $4$. The drift term coming from the $\delta$-potential is
\bea
\frac{\partial S_\alpha^{~\text{ eff}}}{\partial \phi} &=& \frac{\partial}{\partial \phi}  \left( \hf  W'^2 - \ln \left[ e^{i \alpha} - 1 + W'' \right] \right) \nn \\
&=& W' W'' - \frac{W'''}{\Big ( e^{i \alpha} - 1 + W''  \Big)} \nn \\
&=& - i g^2 (1+ \delta) (i \phi)^{2\delta +1} \nn \\
&& - \frac{i g \delta (1+\delta)  (i \phi)^{\delta - 1}}{\Big(   e^{i \alpha} - 1 + g (1+\delta) (i \phi)^{\delta}  \Big)}.
\eea

\begin{figure*}[htp]

\subfloat[$\delta = 1$]{\includegraphics[width=3.2in]{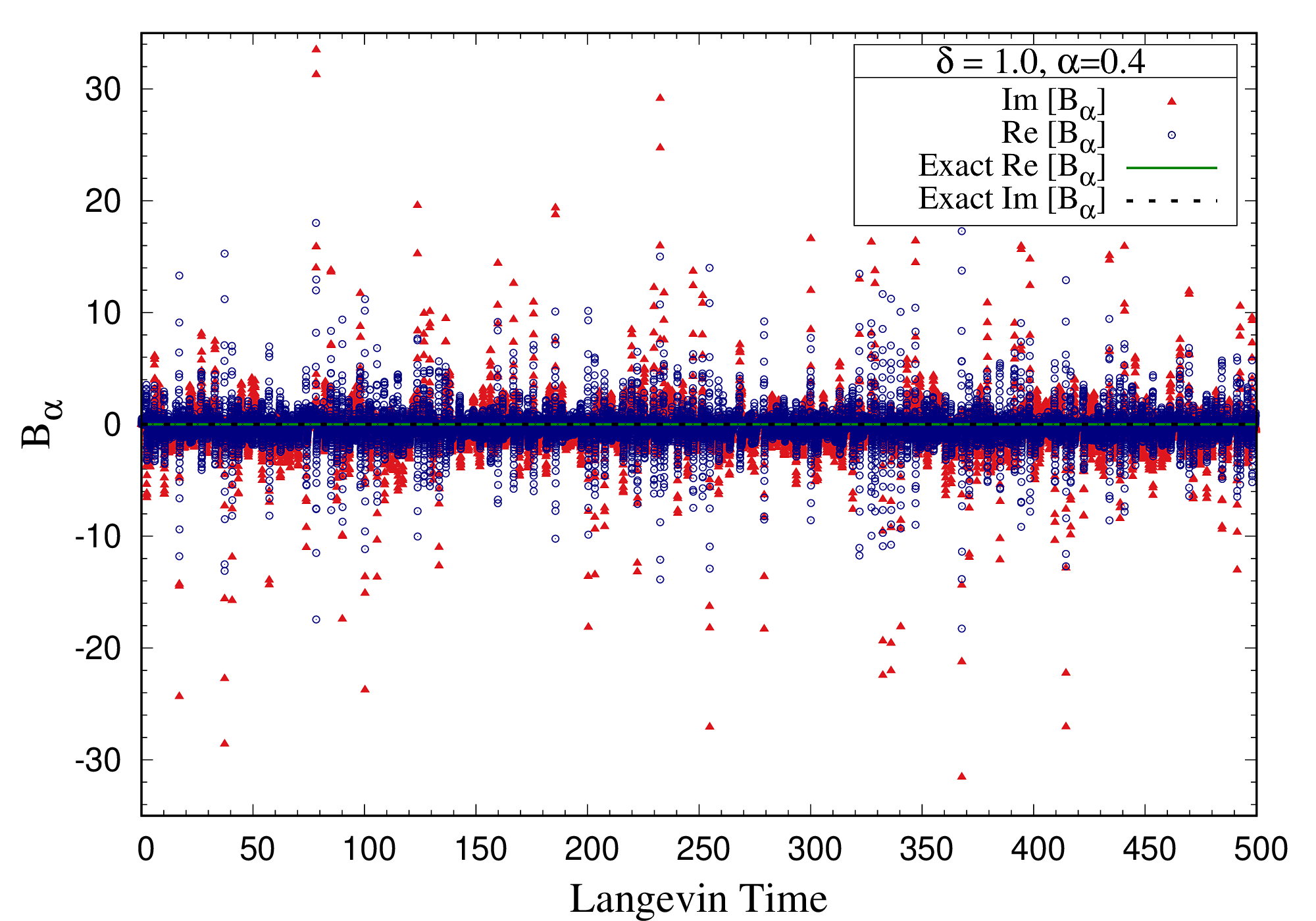}}
\subfloat[$\delta = 2$]{\includegraphics[width=3.2in]{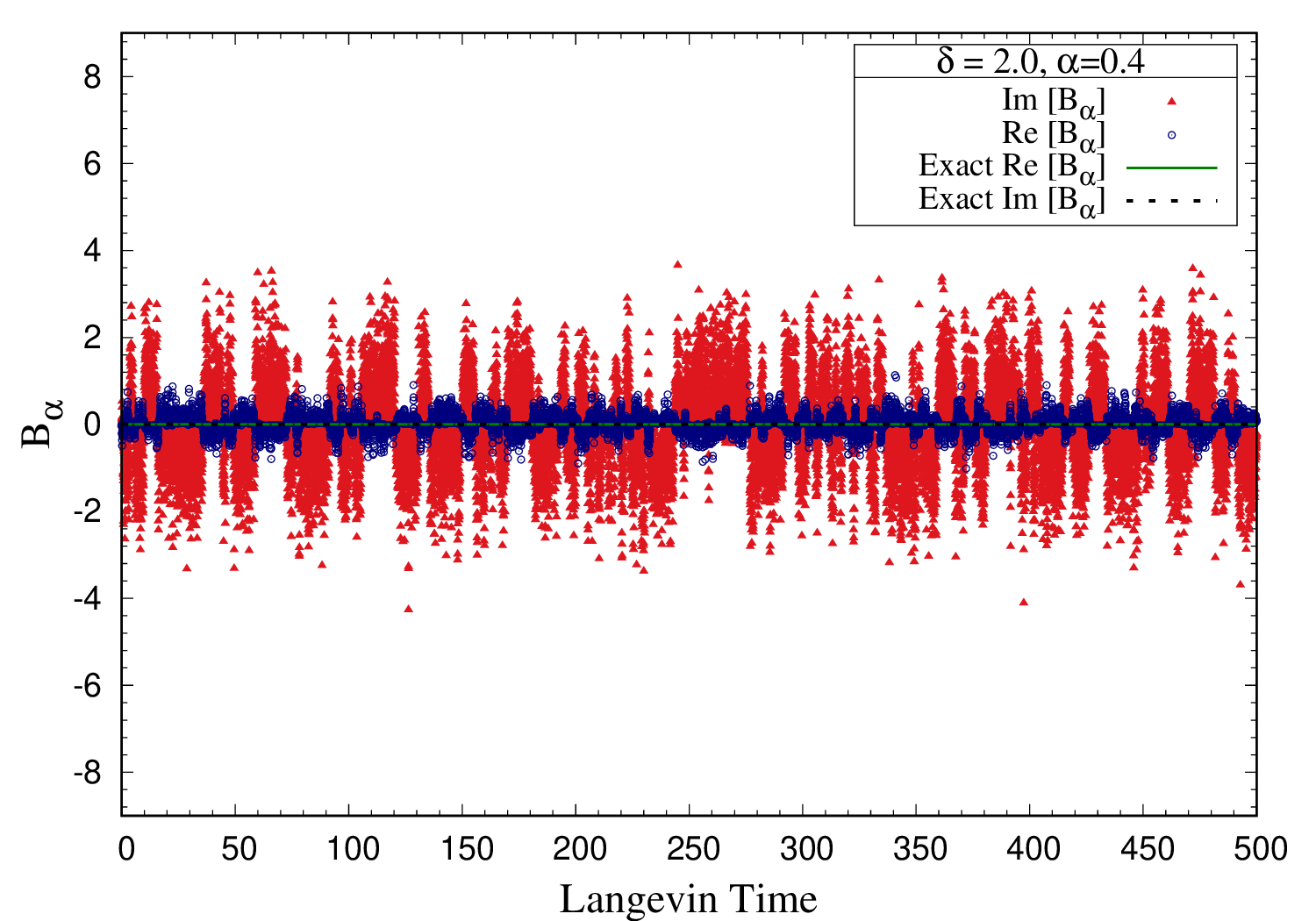}}

\subfloat[$\delta = 3$]{\includegraphics[width=3.2in]{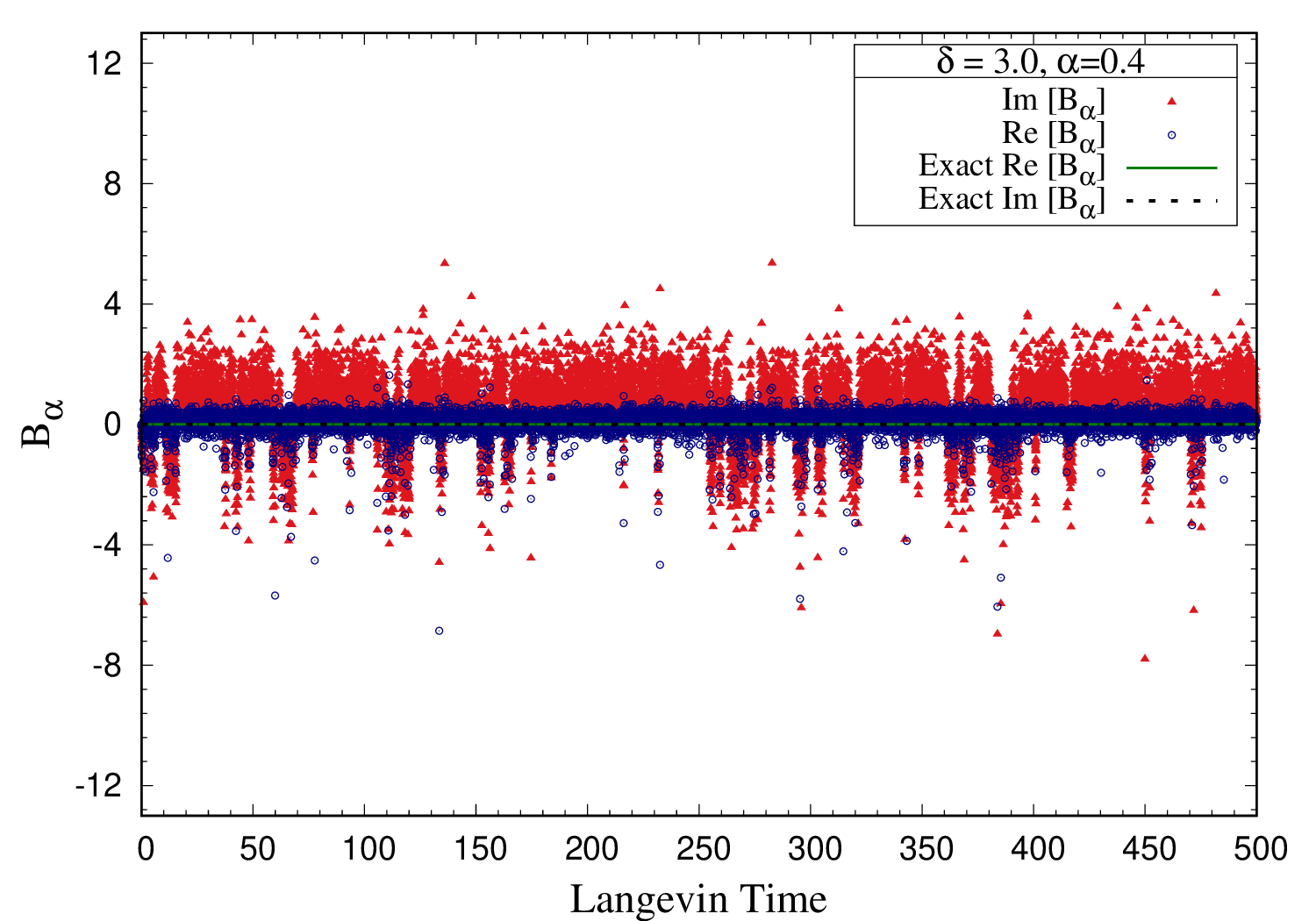}}
\subfloat[$\delta = 4$]{\includegraphics[width=3.2in]{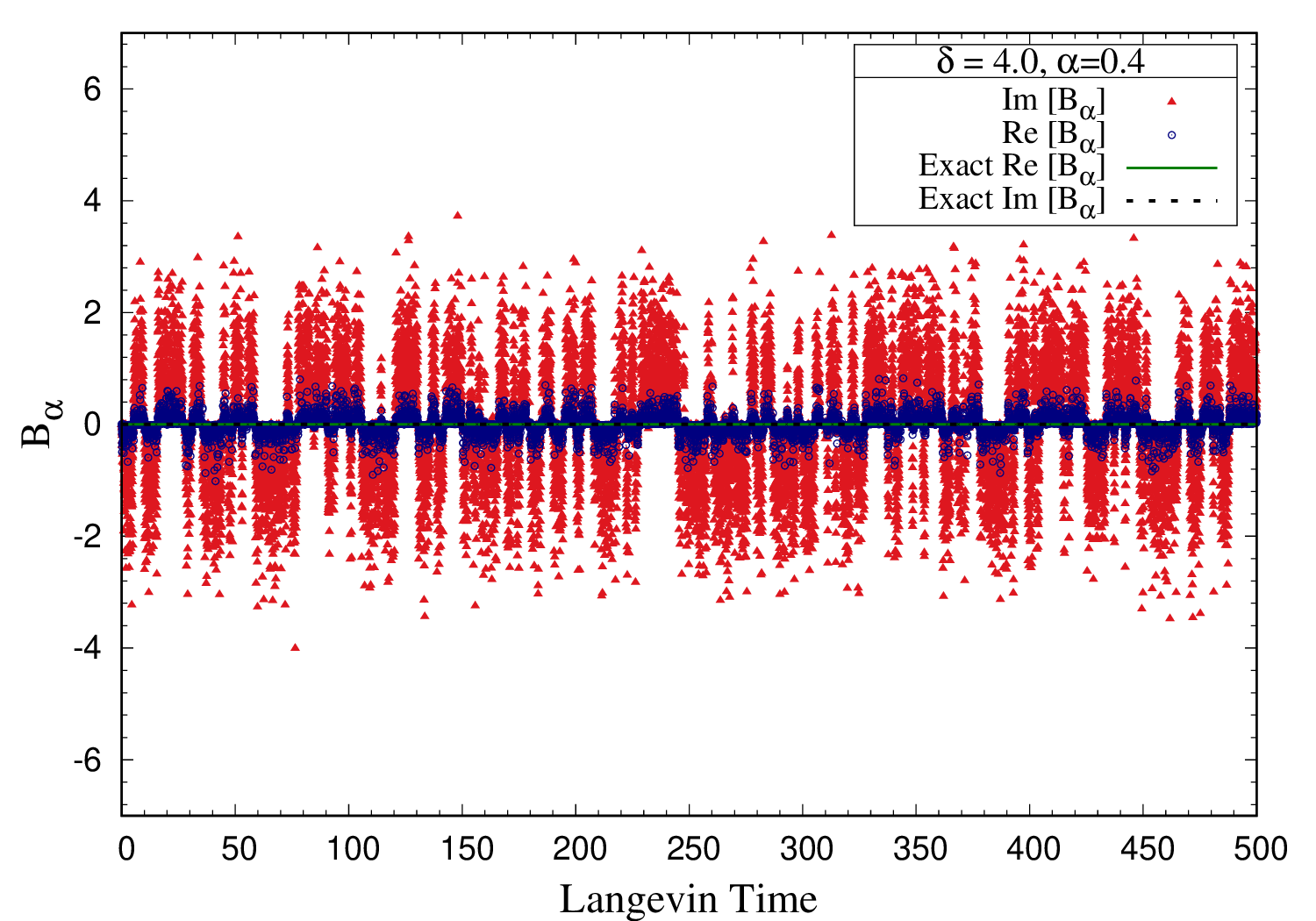}}

\caption{The Langevin time history of field $B$ for $\alpha = 0.4$. The simulations were performed for superpotential $W'(\phi) = -ig (i \phi)^{(1+\delta)}$ with $g=0.5$. In these simulations, we have used adaptive Langevin step size $\Delta \tau \leq 5\times10^{-5}$, generation steps $N_{\rm gen} = 10^7$ and measurements were taken every 500 steps. The plots show $\delta = 1$ case (Top-Left), $\delta = 2$ case (Top-Right), $\delta = 3$ case (Bottom-Left) and $\delta = 4$ case (Bottom-Right).}
\label{fig:B-delta-history}

\end{figure*}

\begin{figure*}[htp]

\subfloat[Real part of $ \langle B\rangle_\alpha$ for $\delta = 1$]{\includegraphics[width=3.0in]{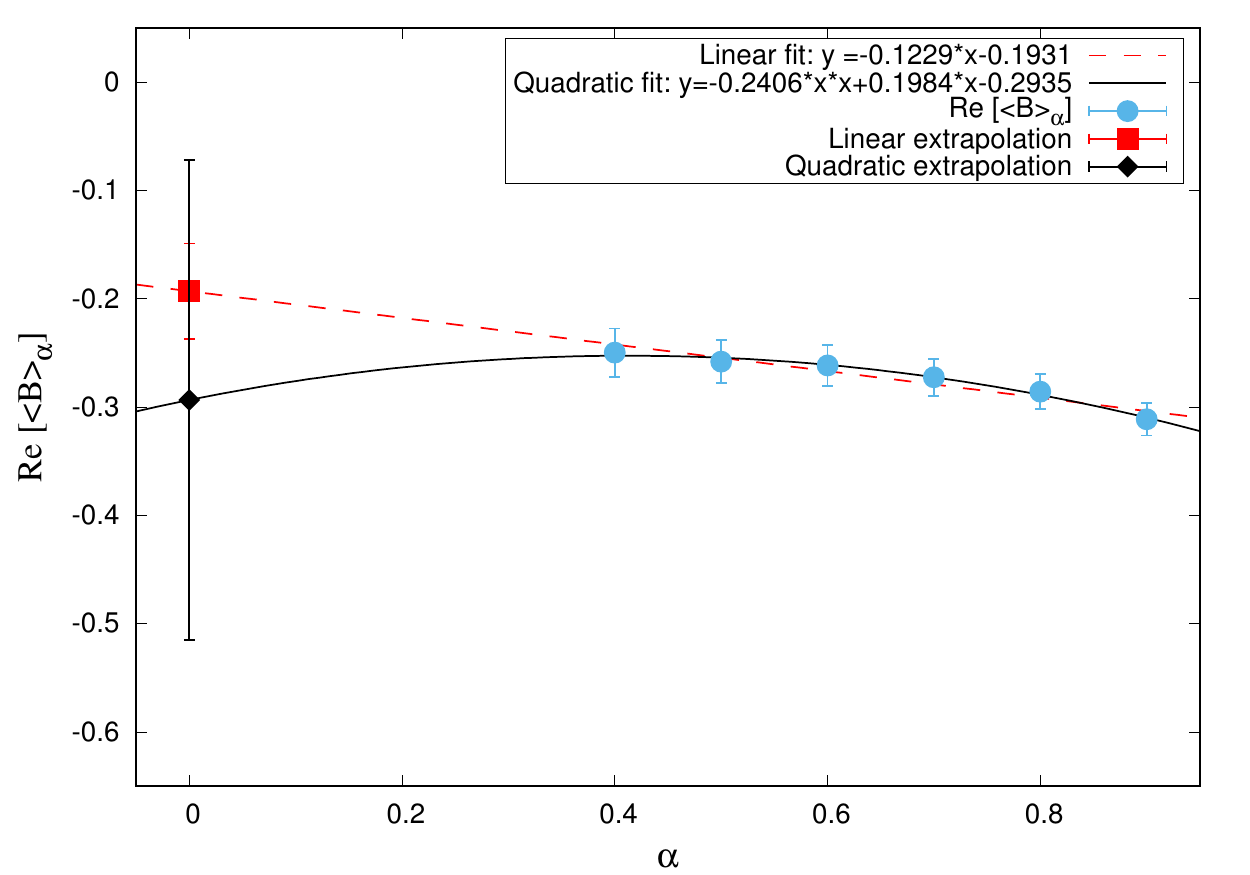}}
\subfloat[Imaginary part of $ \langle B\rangle_\alpha$ for $\delta = 1$]{\includegraphics[width=3.0in]{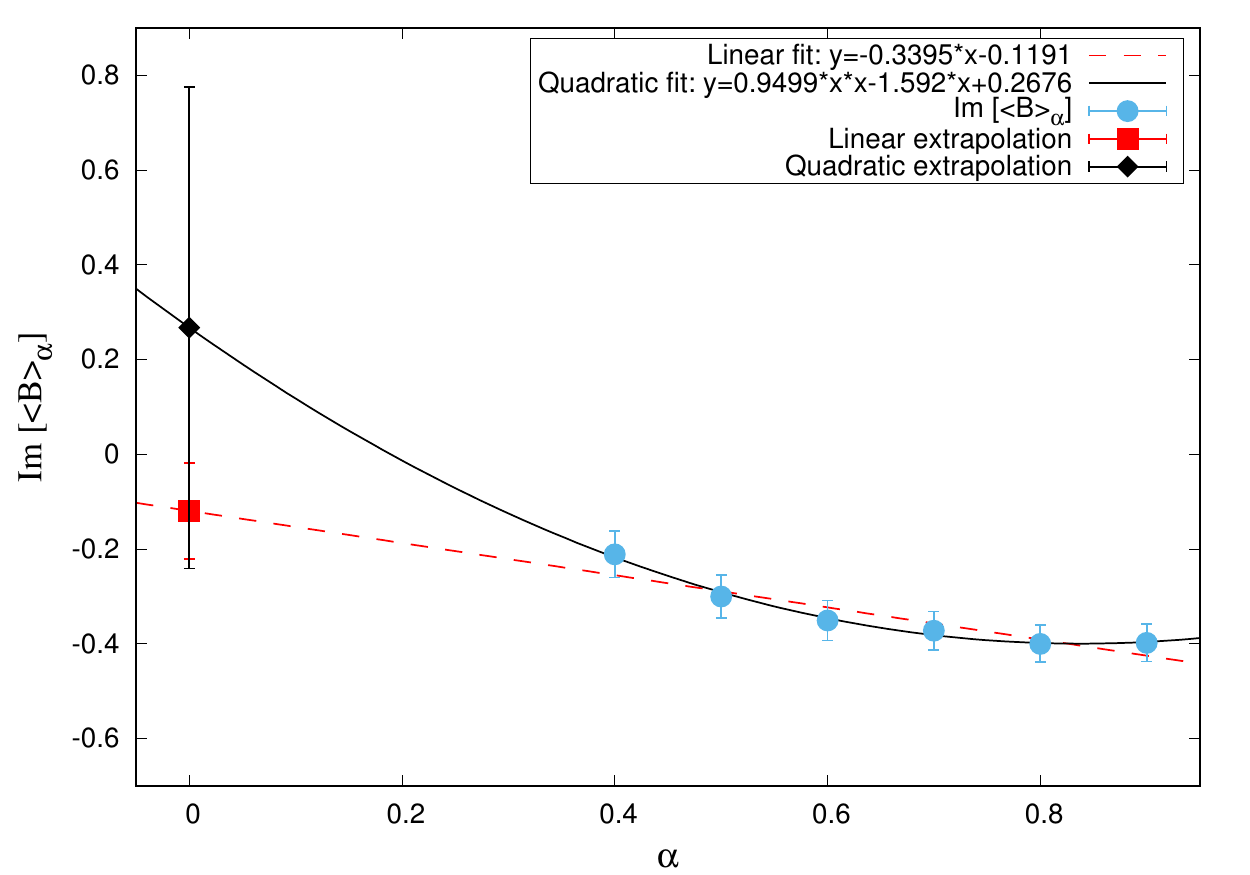}}

\subfloat[Real part of $ \langle B\rangle_\alpha$ for $\delta = 3$]{\includegraphics[width=3.0in]{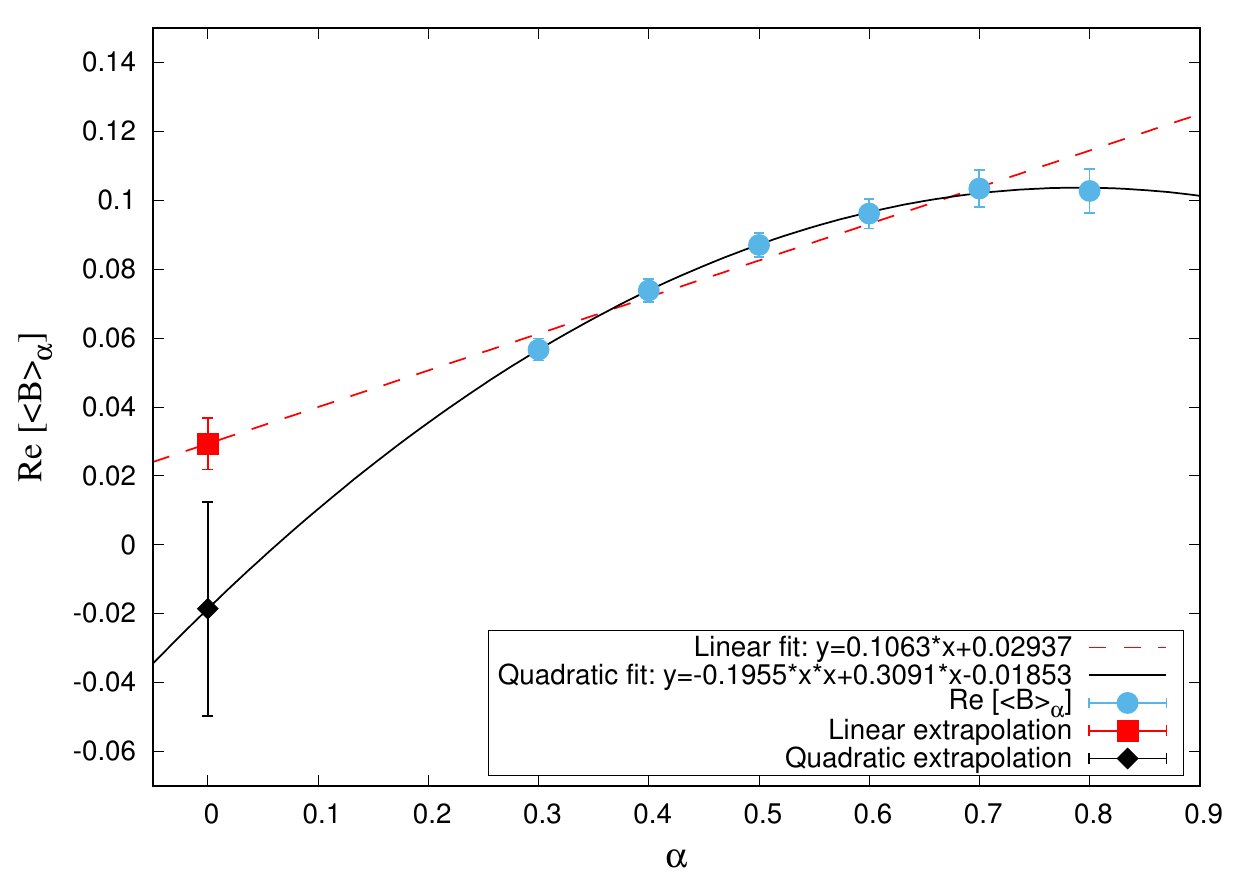}}
\subfloat[Imaginary part of $ \langle B\rangle_\alpha$ for $\delta = 3$]{\includegraphics[width=3.0in]{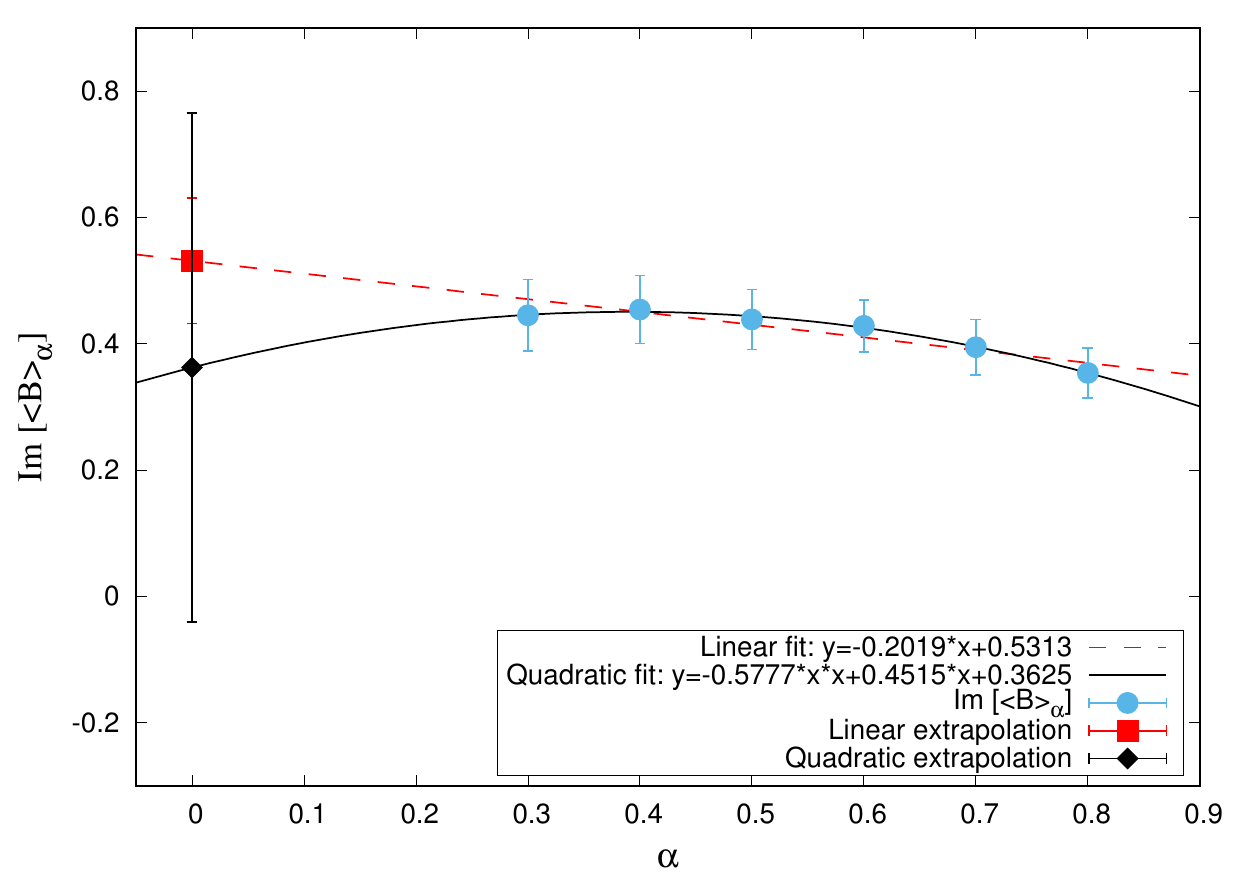}}

\caption{The expectation values of $B$ against the regularization parameter, $\alpha$ for superpotential $W'(\phi) = -ig (i \phi)^{(1+\delta)}$ with $g=0.5$. (Top-Left) Real part and (Top-Right) imaginary part of $\langle B \rangle_\alpha$ for $\delta = 1$. (Bottom-Left) Real part and (Bottom-Right) imaginary part of $\langle B \rangle_\alpha$ for $\delta = 3$. The simulations were performed with adaptive Langevin step size $\Delta \tau \leq 5 \times 10^{-5}$, thermalization steps $N_{\rm therm} =  5 \times  10^4$, generation steps $N_{\rm gen} = 10^7$ and measurements taken every $500$ steps. The dashed red lines are the linear fits to $\langle B \rangle_\alpha$ in $\alpha$, and red dots are the linear extrapolation value at $\alpha = 0$.  The solid black lines represent the quadratic fits to $\langle B \rangle_\alpha$ in $\alpha$, and black dots are the quadratic extrapolation value at $\alpha = 0$ . The $\alpha \to 0$ limit values obtained from these plots are given in Table  \ref{tab:delta_1p0_3p0}.}
\label{fig:delta_fit_1p0_3p0}

\end{figure*}

\begin{figure*}[htp]

\subfloat[Real part of $ \langle B\rangle_\alpha$ for $\delta = 2$]{\includegraphics[width=3.0in]{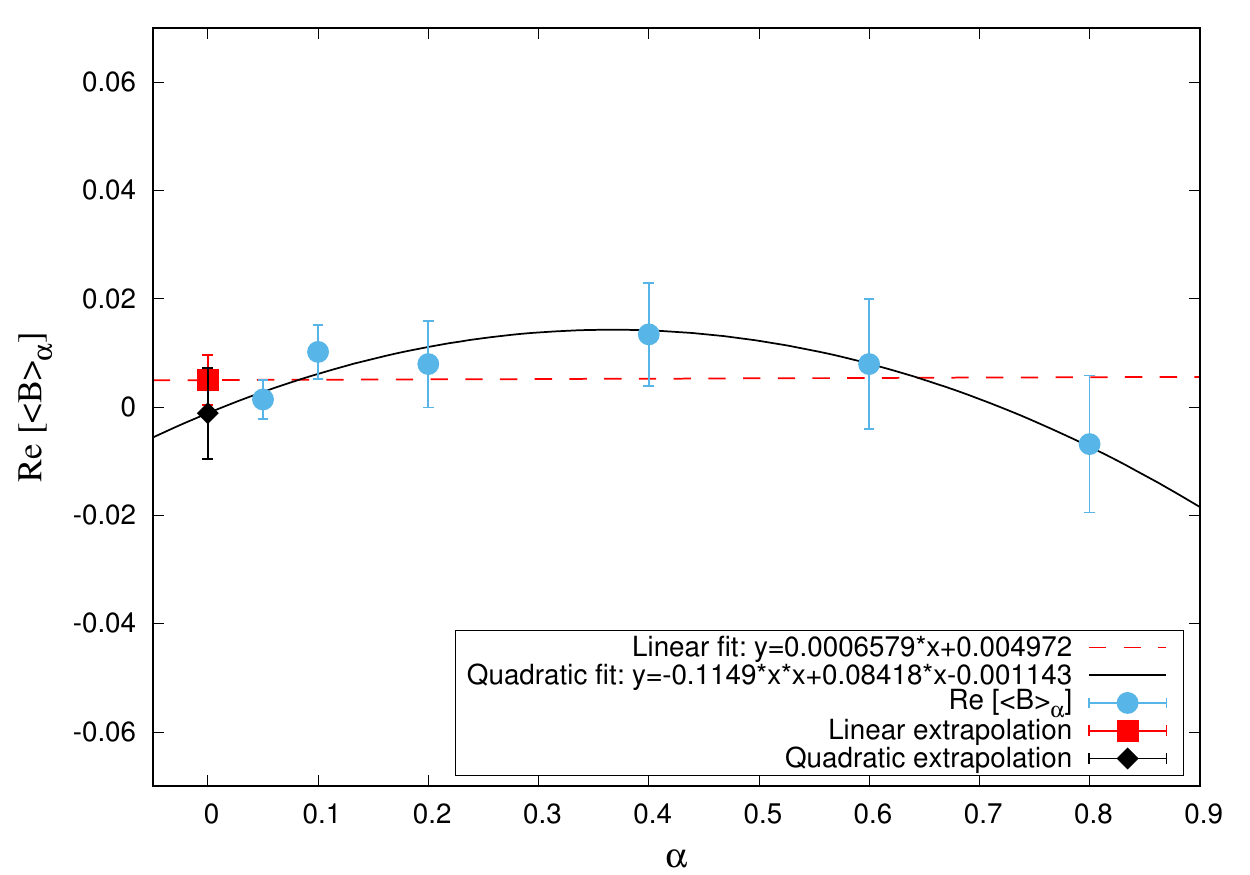}}
\subfloat[Imaginary part of $ \langle B\rangle_\alpha$ for $\delta = 2$]{\includegraphics[width=3.0in]{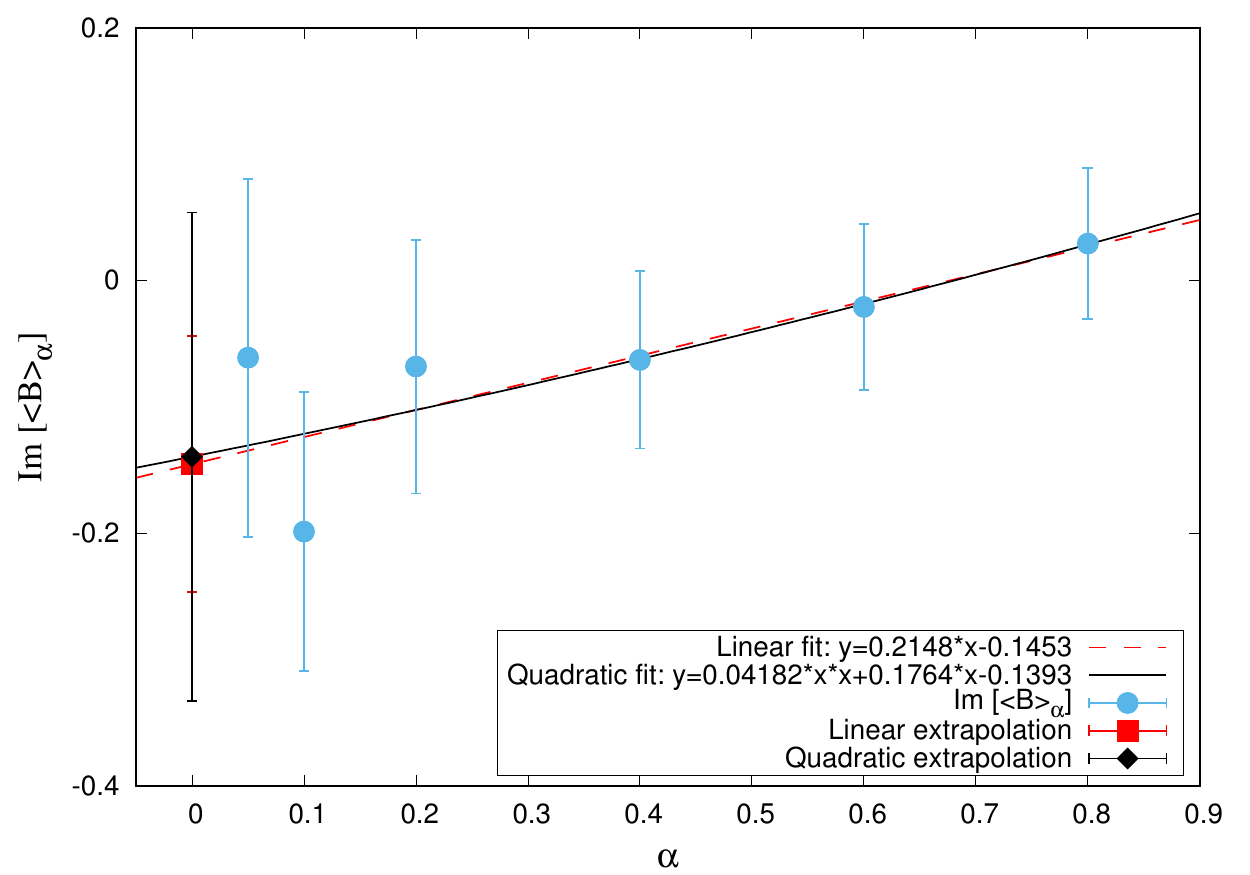}}

\subfloat[Real part of $ \langle B\rangle_\alpha$ for $\delta = 4$]{\includegraphics[width=3.0in]{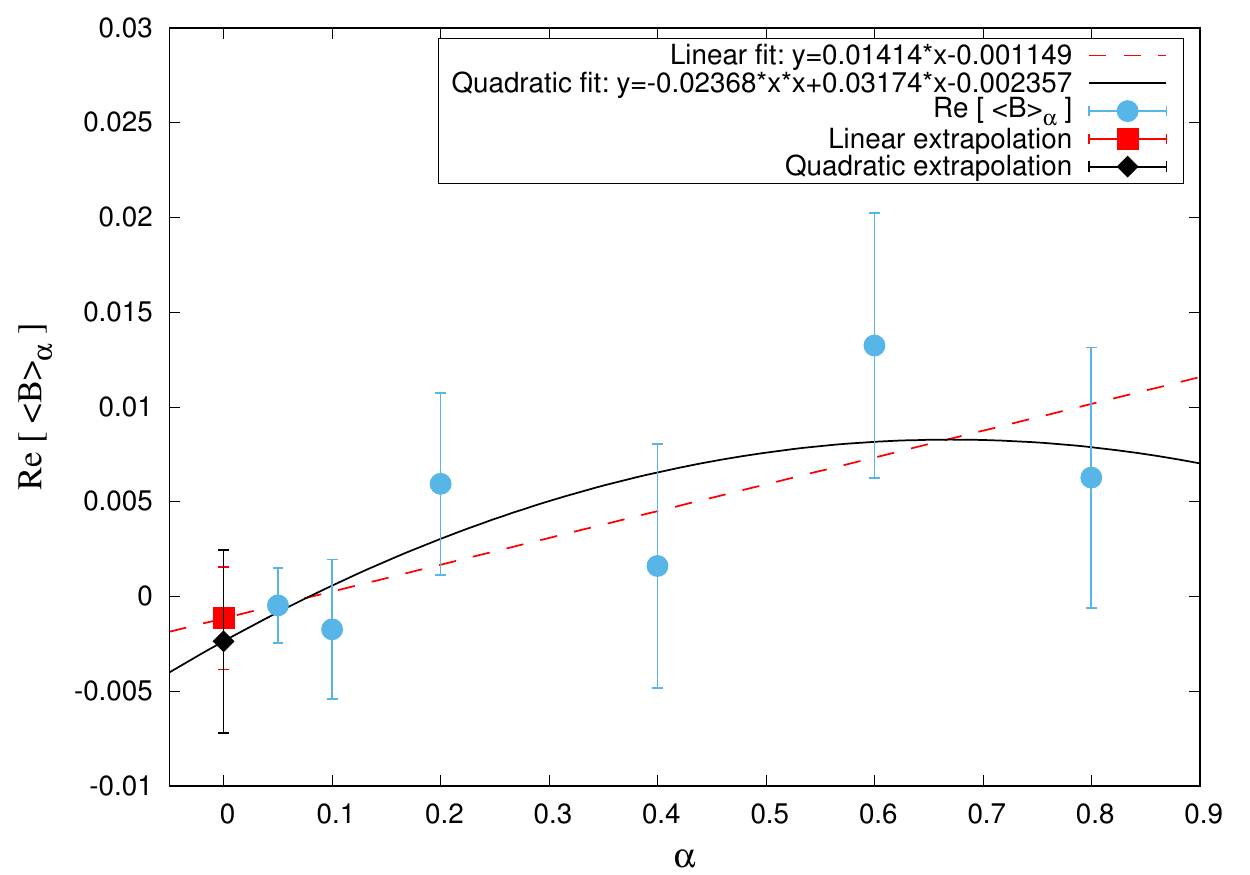}}
\subfloat[Imaginary part of $ \langle B\rangle_\alpha$ for $\delta = 4$]{\includegraphics[width=3.0in]{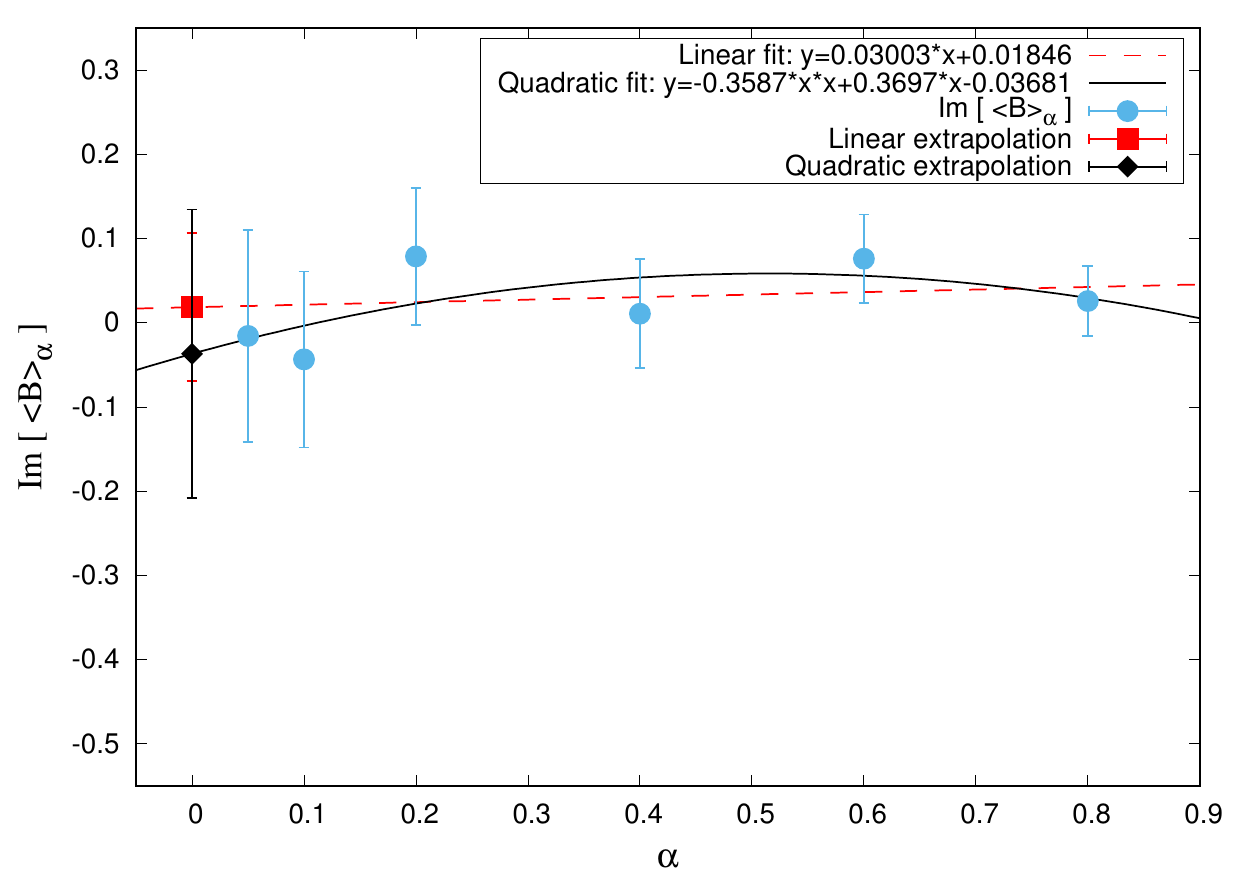}}

\caption{The expectation values of $B$ against the regularization parameter, $\alpha$ for superpotential $W'(\phi) = -ig (i \phi)^{(1+\delta)}$ with $g=0.5$.(Top-Left) Real part and (Top-Right) imaginary part of $\langle B \rangle_\alpha$ for $\delta = 2$. (Bottom-Left) Real part and (Bottom-Right) imaginary part of $\langle B \rangle_\alpha$ for $\delta = 4$. The simulations were performed with adaptive Langevin step size $\Delta \tau \leq 5 \times 10^{-5}$, thermalization steps $N_{\rm therm} = 5 \times 10^4$, generation steps $N_{\rm gen} = 10^7$ and measurements were taken every $500$ steps. The dashed red lines are the linear fits to $\langle B \rangle_\alpha$ in $\alpha$, and red dots are the linear extrapolation value at $\alpha = 0$.  The solid black lines represent the quadratic fits to $\langle B \rangle_\alpha$ in $\alpha$, and black dots are the quadratic extrapolation value at $\alpha = 0$. The $\alpha \to 0$ limit values obtained from these plots are given in Table \ref{tab:delta_2p0_4p0}.}
\label{fig:delta_fit_2p0_4p0}

\end{figure*}

In Fig. \ref{fig:B-delta-history} we show the Langevin time history of the auxiliary $B$ field for $\delta = 1, 2, 3$ and $4$. We show linear and quadratic extrapolations to $\alpha \to 0$ limit in Fig. \ref{fig:delta_fit_1p0_3p0} for $\delta = 1, 3$ and Fig. \ref{fig:delta_fit_2p0_4p0} for $\delta = 2, 4$, respectively. The results are tabulated in Table \ref{tab:delta_1p0_3p0} and \ref{tab:delta_2p0_4p0}. It is clear from our simulation results that the expectation value of auxiliary field, $\langle B \rangle_\alpha$, vanishes in the limit $\alpha \to 0$. Hence we conclude that SUSY is not broken in the model with $\delta$-potential for values of $\delta = 1, 2, 3, 4$.

\section{Conclusions and future directions}
\label{sec:concl-f-dirs}

We have successfully used complex Langevin dynamics with stochastic quantization to investigate supersymmetry breaking in a class of zero-dimensional $\cN = 2$ supersymmetric models with real and complex actions. We looked at double-well superpotential, general polynomial superpotential and also $\mathcal{PT}$-symmetric models inspired $\delta$-potentials. In some cases we were able to cross check the presence or absence of supersymmetry breaking wherever analytical results were available. Our simulations strongly suggest that SUSY is preserved for $\mathcal{PT}$-symmetric models inspired $\delta$-potentials. We have also investigated the reliability of complex Langevin simulations by monitoring Fokker-Planck equation as correctness criterion (in Appendix. \ref{app:FP-correctness}) and also by looking at the probability distributions of the magnitude of the drift terms (in Appendix. \ref{app:drift-decay}).

It would be interesting to study complex Langevin dynamics in the above models, generalized to non-Abelian cases, for example with $SU(N)$ symmetry. Supersymmetry may be restored in the large-$N$ limit of these models. It would also be interesting to explore spontaneous SUSY breaking when $\delta$ in the superpotential is a continuous parameter. Other possibilities include extending our investigations to 1- and 2-dimensional models with same superpotentials. These results will appear in an upcoming work \cite{ToAppear:2019}.      

\acknowledgments 

Work of AJ was partially supported by the Seed Grant from IISER Mohali. AK was partially supported by IISER Mohali and a CSIR Research Fellowship (Fellowship No. 517019).  

\appendix

\section{Reliability of complex Langevin simulations}
\label{app:reliability}

In this section we would like to justify the simulations used in this work. We look at two of the methods proposed in the recent literature. One is based on the Fokker-Planck equation as a correctness criterion and the other is based on the probability distribution of the magnitude of the drift term. 

\subsection{Fokker-Planck equation as correctness criterion}
\label{app:FP-correctness}

The holomorphic observables of the theory $\cO[\phi, \tau]$ evolve according to \cite{Aarts:2009uq, Aarts:2011ax, Aarts:2013uza}
\beq
\frac{\partial \cO[\phi, \tau]}{\partial \tau} = \widetilde{L} \cO[\phi, \tau], 
\eeq
where
$\widetilde{L}$ is the Langevin operator
\beq
\widetilde{L} = \left[ \frac{\partial}{\partial \phi}  - \frac{\partial}{\partial \phi} S[\phi] \right] \ \frac{\partial}{\partial \phi}.
\eeq

Once the equilibrium distribution is reached, assuming that it exists, we can remove the $\tau$ dependence from the observables. Then we have
\beq
C_\cO \equiv \langle \widetilde{L} \cO[\phi] \rangle = 0,
\eeq
and this can be used as a {\it criterion for correctness} of the complex Langevin method. This criterion has been investigated in various models in Refs. \cite{Aarts:2009uq, Aarts:2011ax, Aarts:2013uza}. The criterion for correctness, in principle, needs to be satisfied for a complete set of observables $\cO[\phi]$, in a suitably chosen basis \cite{Aarts:2011ax}. It leads to an infinite tower of identities, which as a collection, resembles to the Schwinger-Dyson equations. \\

For the observable $\cO$, as the auxiliary $B$ field, we have
\bea
\widetilde{L} \cO &=& \widetilde{L} B \nn \\
&=&  -i W''' + i W' {W''}^2 - \frac{i W'' W'''}{\Big ( e^{i \alpha} - 1 + W''\Big)}.
\eea

\begin{figure*}[htp]

\subfloat[$g = 1$]{\includegraphics[width=3.0in]{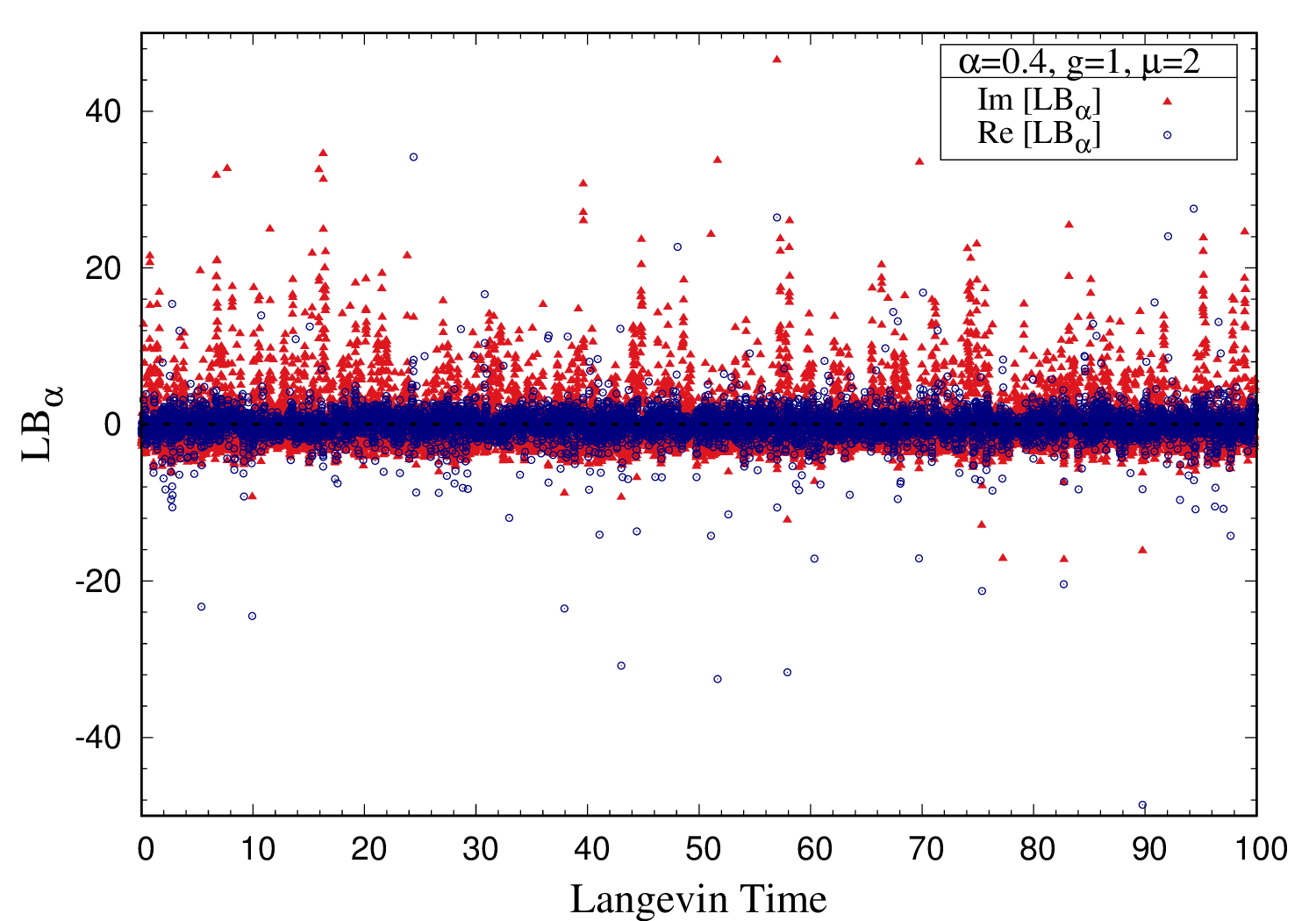}}
\subfloat[$g = 3$]{\includegraphics[width=3.0in]{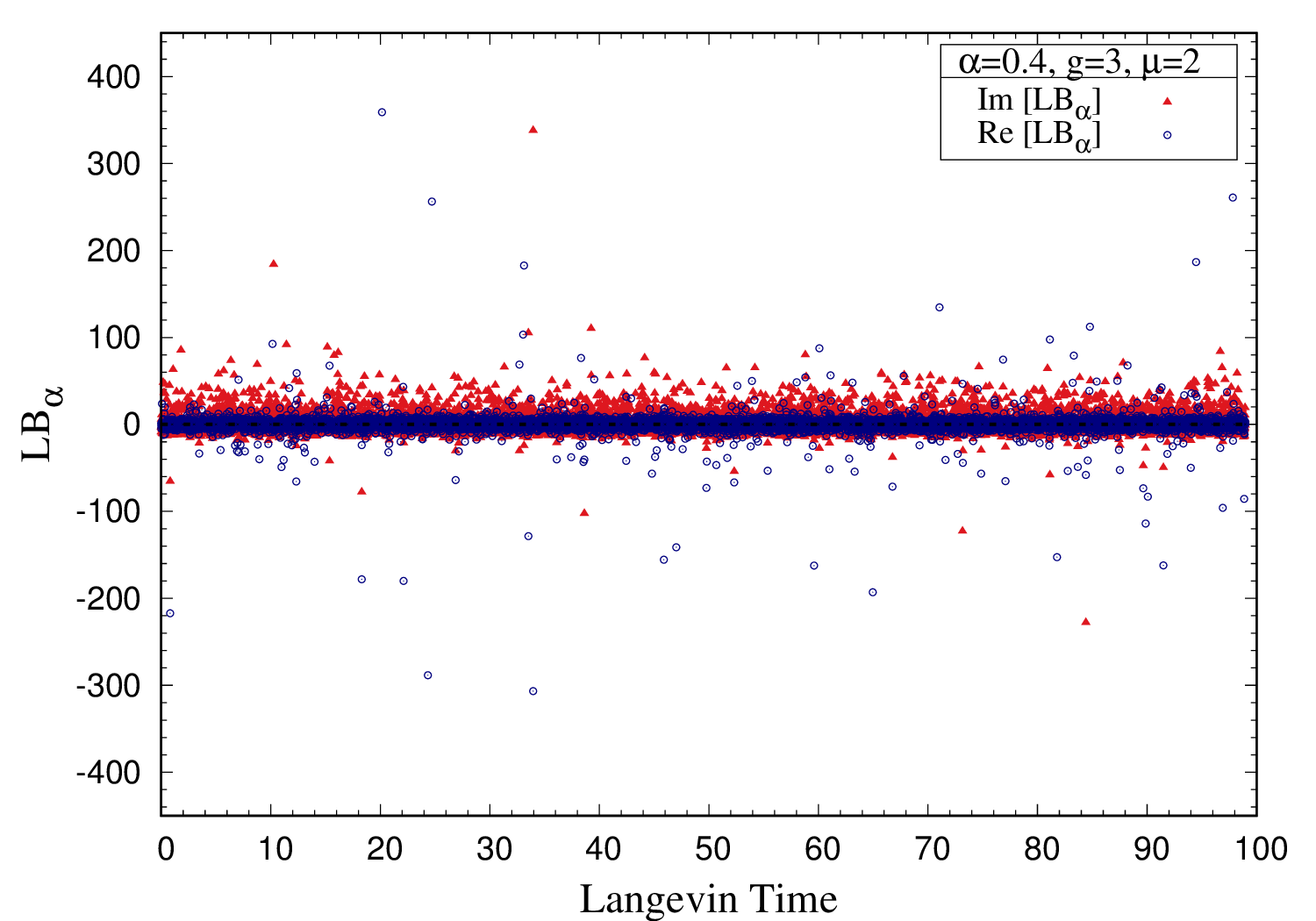}}

\caption{The Langevin time history of $\widetilde{L} B$ for regularization parameter, $\alpha=0.4$ . Simulations were performed for superpotential $W' = g \ (\phi^2 + \mu^2)$ with $\mu = 2$, $g = 1$ (Left) and $g=3$ (Right). In these simulations, we have used adaptive Langevin step size $\Delta \tau \leq 10^{-4}$, generation steps $N_{\rm gen} = 10^6$ and measurements taken every $100$ steps. The exact value is $\widetilde{L} B = 0$ at equilibrium distribution.}
\label{fig:LO_dw}

\end{figure*}

\begin{figure*}[htp]

\subfloat[$\delta = 1$]{\includegraphics[width=3.0in]{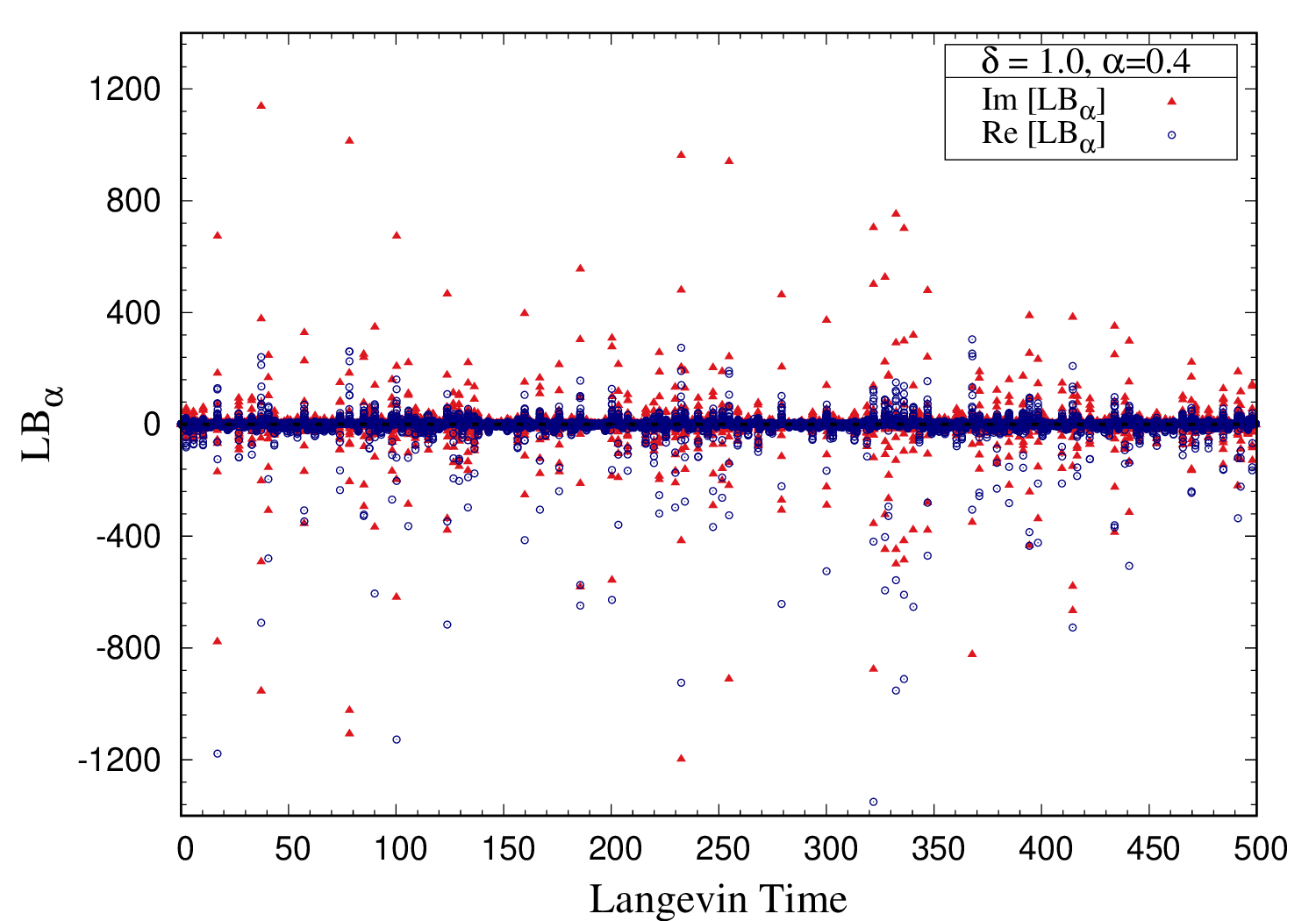}}
\subfloat[$\delta = 2$]{\includegraphics[width=3.0in]{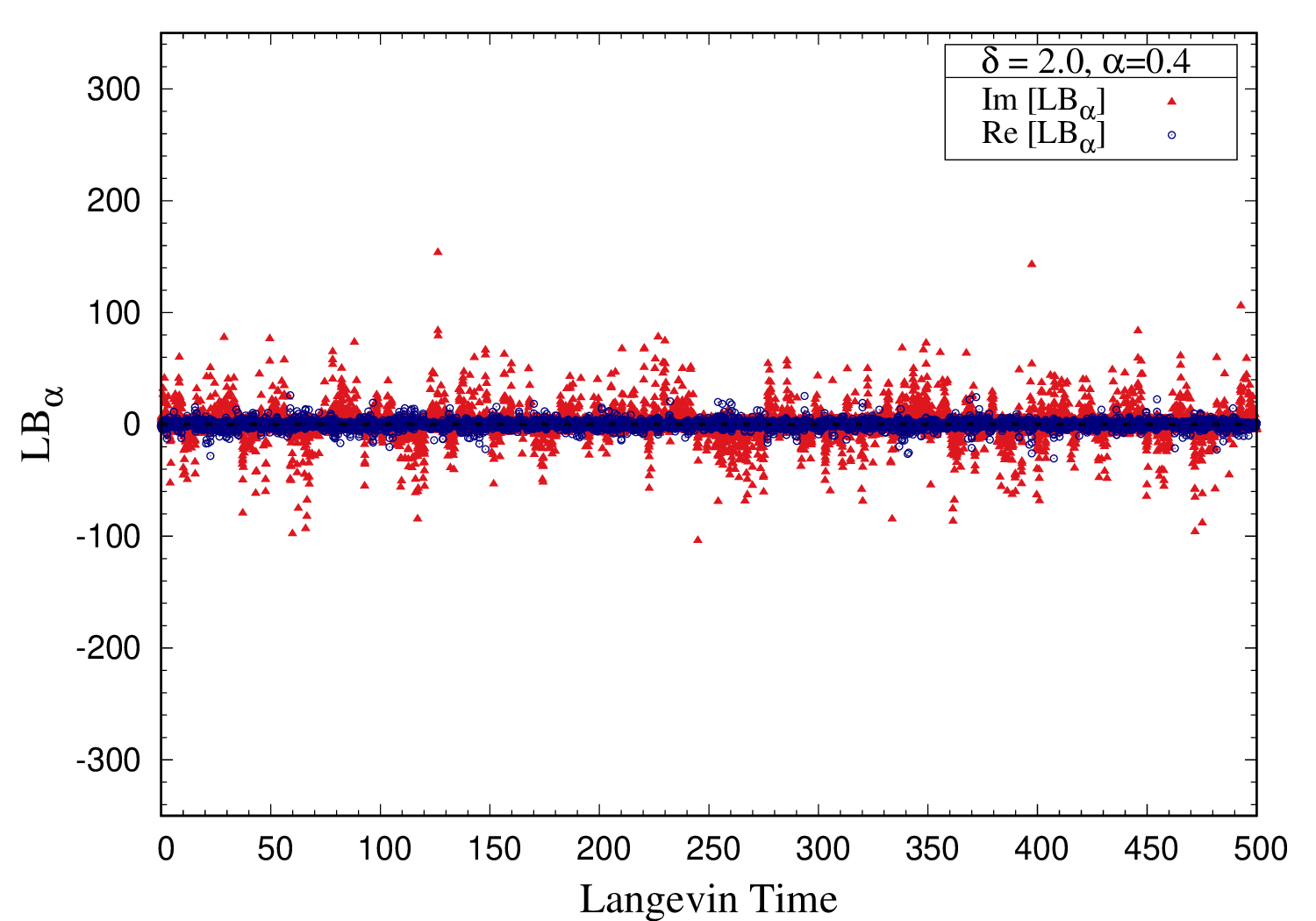}}

\subfloat[$\delta = 3$]{\includegraphics[width=3.0in]{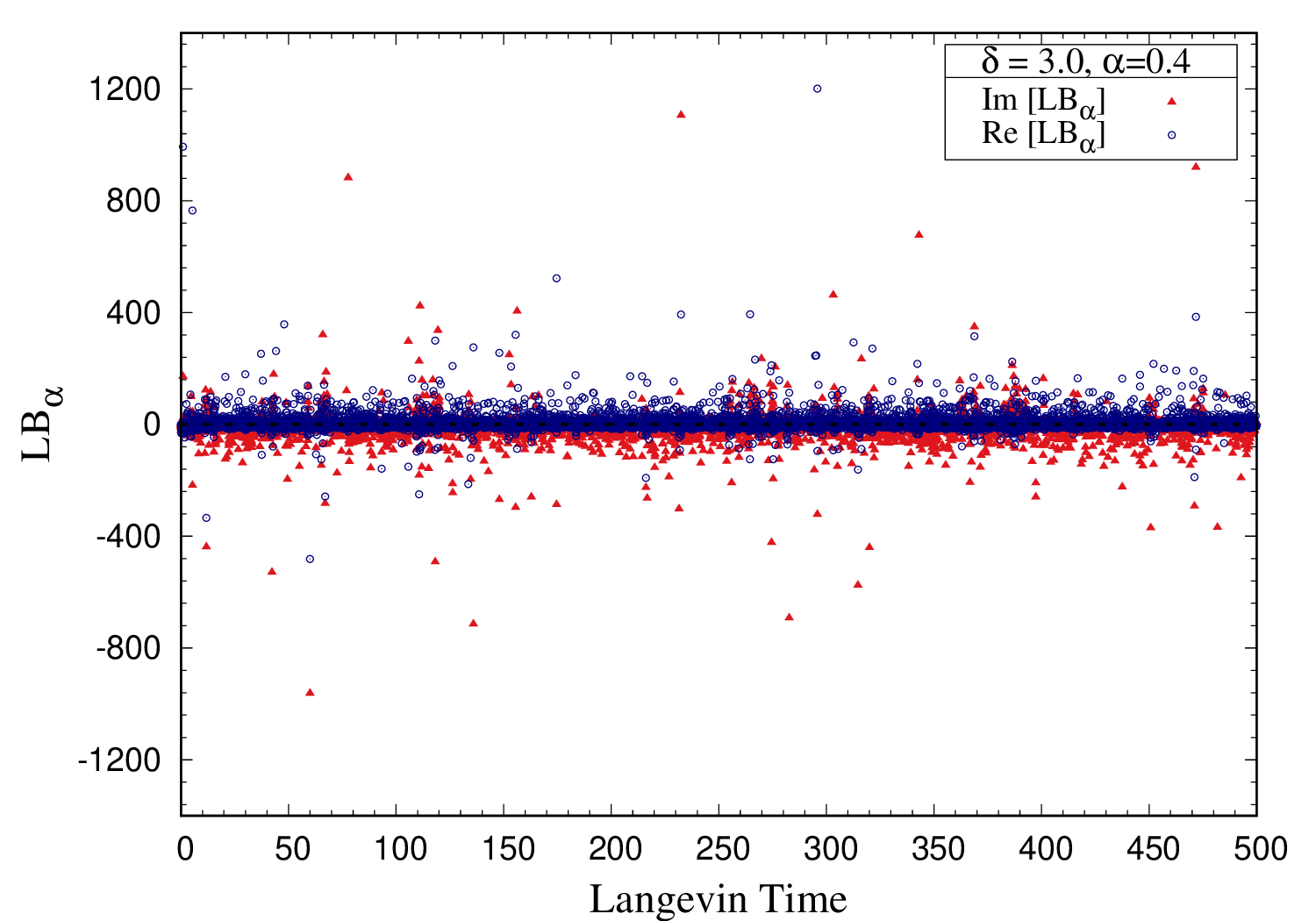}}
\subfloat[$\delta = 4$]{\includegraphics[width=3.0in]{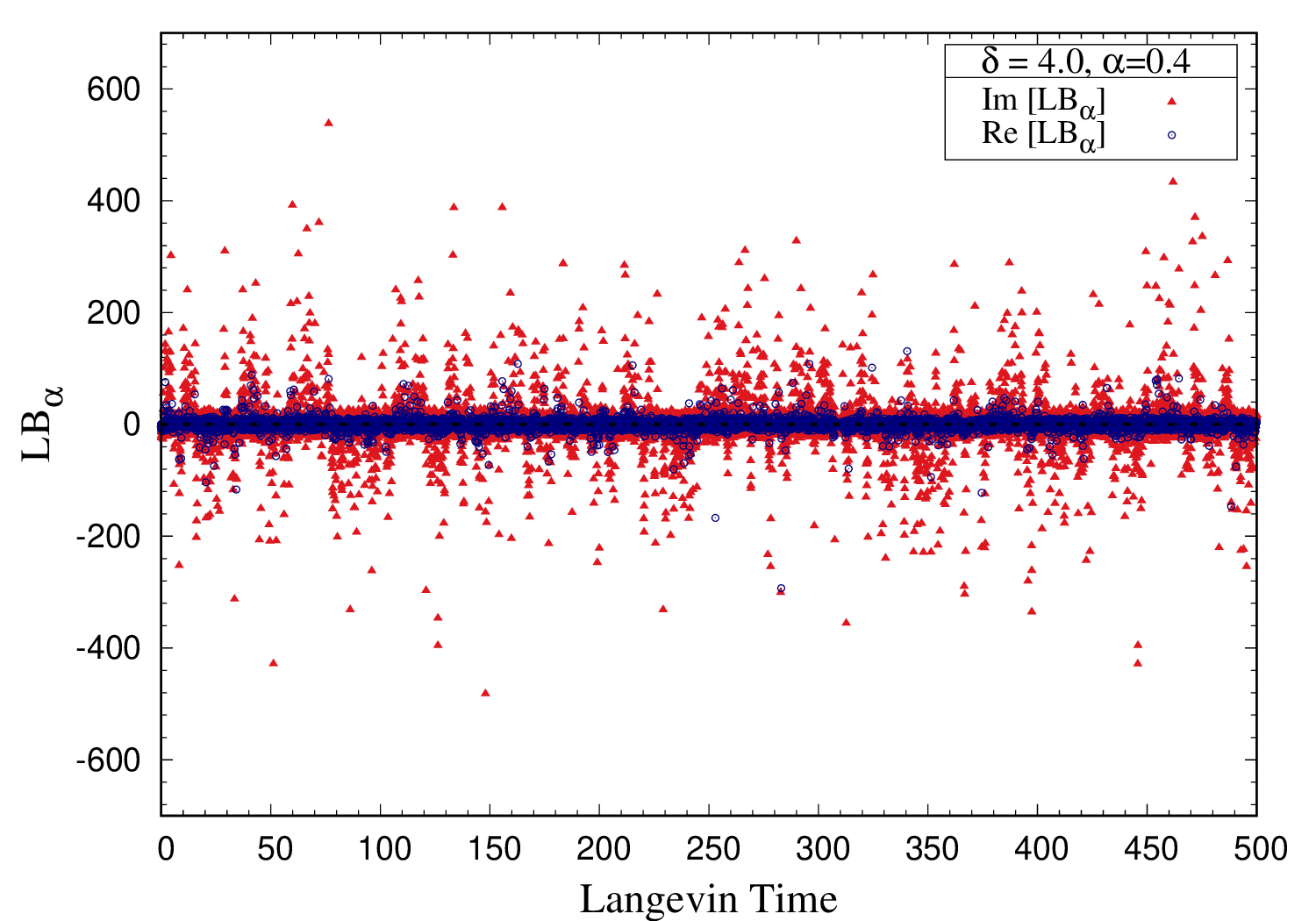}}

\caption{The Langevin time history of $\widetilde{L} B$ for regularization parameter, $\alpha=0.4$ . Simulations were performed for superpotential $W'(\phi) = -ig (i \phi)^{(1+\delta)}$ with $g=0.5$ for various values of delta: $\delta = 1$ (Top-Left), $\delta = 2$ (Top-Right), $\delta = 3$ (Bottom-Left) and $\delta = 4$ (Bottom-Right). In these simulations, we have used adaptive Langevin step size $\Delta \tau \leq 5\times10^{-5}$, generation steps $N_{\rm gen} = 10^7$ and measurements taken every $500$ steps. The exact value at equilibrium distribution is $\widetilde{L}B = 0$.}
\label{fig:LO_delta}

\end{figure*}

We show the Langevin history of the above mentioned correctness criterion, $\widetilde{L} B$, with regularization parameter $\alpha=0.4$ for the superpotentials $W' = g (\phi^2 + \mu^2)$ and $W' = -ig (i\phi)^{1 + \delta}$ in Figs. \ref{fig:LO_dw} and \ref{fig:LO_delta}, respectively. In Table \ref{tab:sqw_LO} we provide the simulated values of $\langle \widetilde{L} B \rangle_\alpha$ for superpotential $W' = g (\phi^2 + \mu^2)$ with coupling parameter $g = 1,3$ and various values of regularization parameter, $\alpha$. In Tables. \ref{tab:delta_LO_1p0_3p0} and \ref{tab:delta_LO_2p0_4p0}, we tabulate the simulated values of $\langle \widetilde{L} B \rangle_\alpha$ for superpotential $W'(\phi) = -ig (i \phi)^{(1+\delta)}$ with coupling parameter $g = 0.5$ and various values of regularization parameter, $\alpha$.

\subsection{Decay of the drift terms}
\label{app:drift-decay}

Another method to check the correctness of the complex Langevin dynamics, as proposed in Refs. \cite{Nagata:2016vkn, Nagata:2018net}, is to look at the probability distribution $P(u)$ of the magnitude of the drift term $u$ at large values of the drift. We have the magnitude of the drift term
\beq
u = \left| \frac{\partial S}{\partial \phi} \right|.
\eeq

\begin{figure*}[htp]

\subfloat[$g = 1.0$]{\includegraphics[width=3.0in]{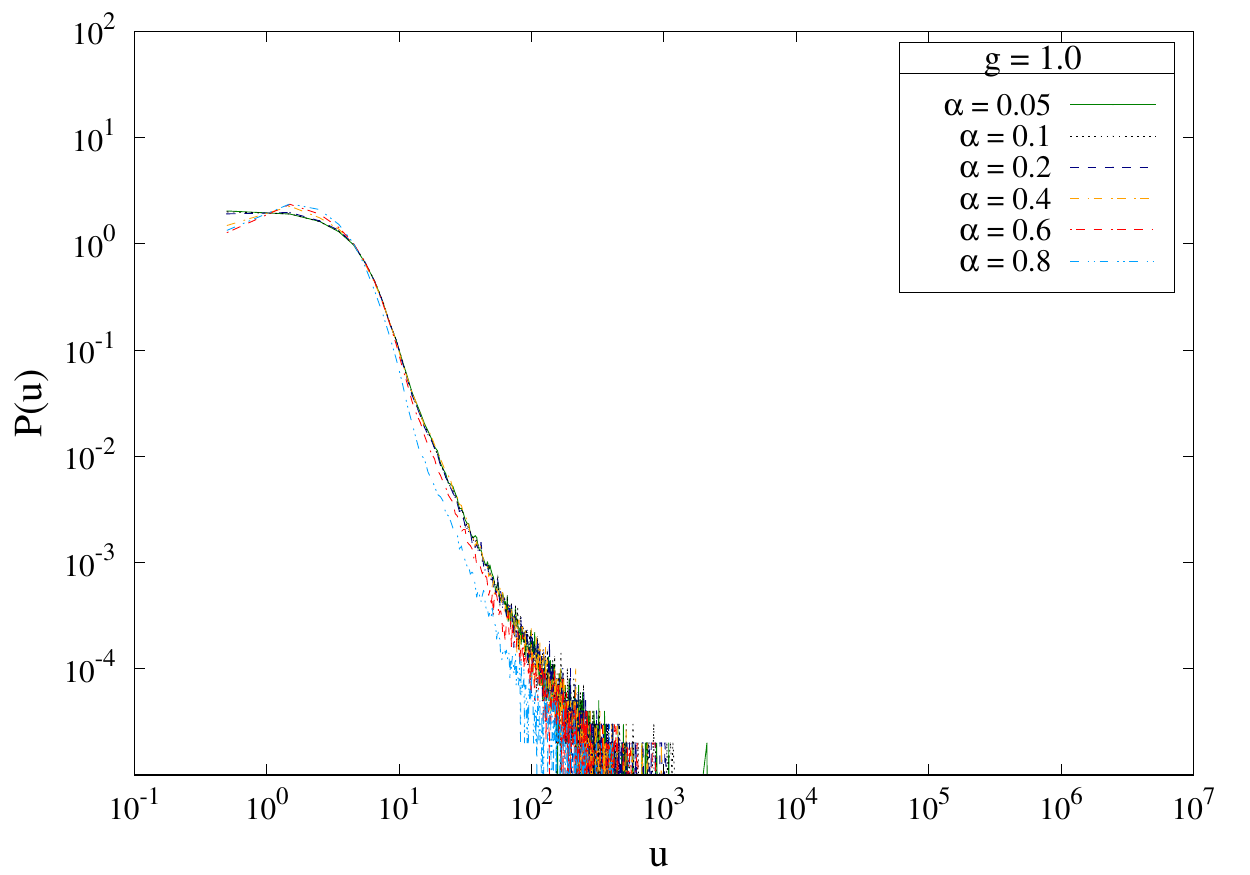}}
\subfloat[$g = 3.0$]{\includegraphics[width=3.0in]{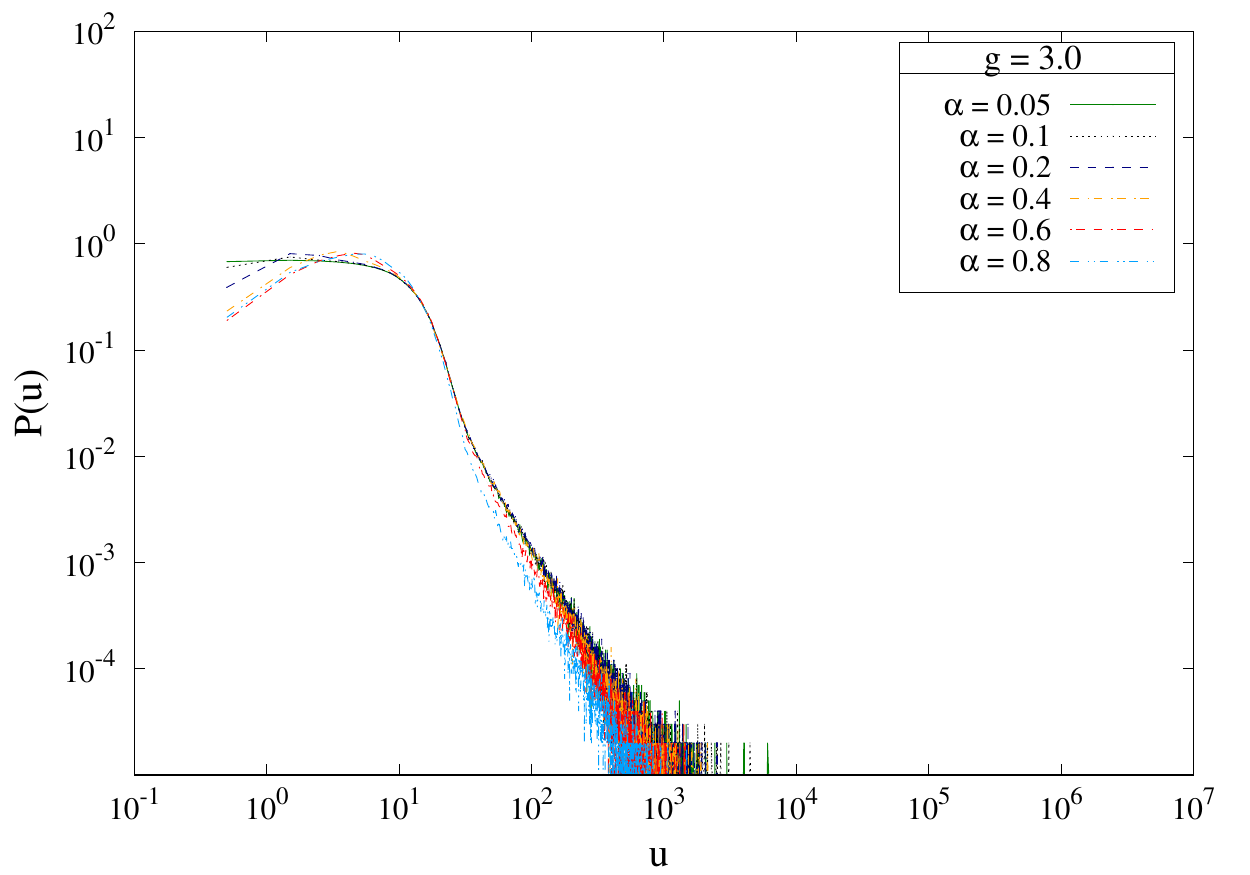}}

\caption{The probability distribution $P(u)$ of the magnitude of the drift term $u$ for the superpotential $W'(\phi) = g (\phi^2 +\mu^2)$ on a log-log plot. Simulations were performed for $g = 1.0$ (Left) and $g = 3.0$ (Right) with $\mu= 2.0$. We used adaptive Langevin step size $\Delta \tau \leq 10^{-4}$, and generation steps $N_{\rm gen} = 10^6$.}
\label{fig:drift_dw}

\end{figure*}

\begin{figure*}[htp]

\subfloat[$\delta = 1$]{\includegraphics[width=3.0in]{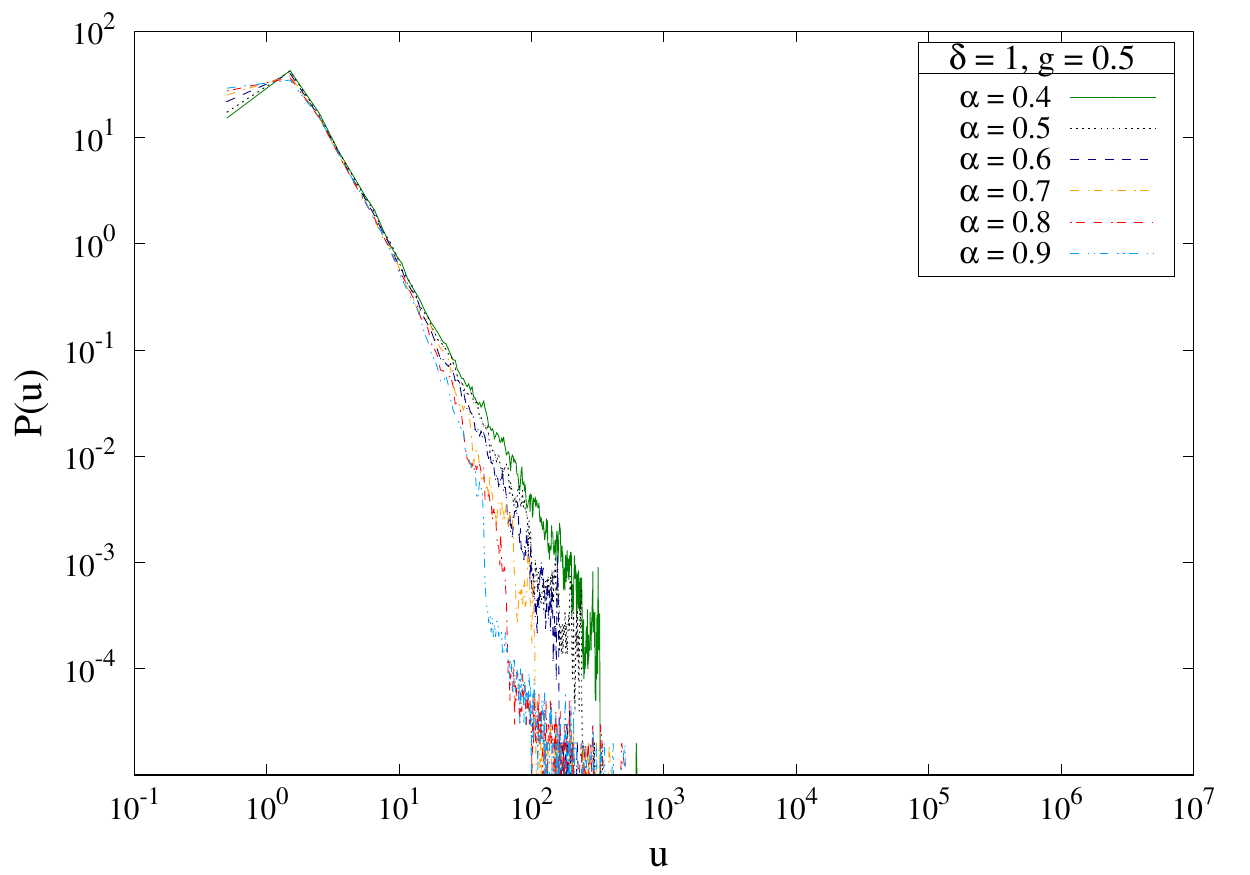}}
\subfloat[$\delta = 2$]{\includegraphics[width=3.0in]{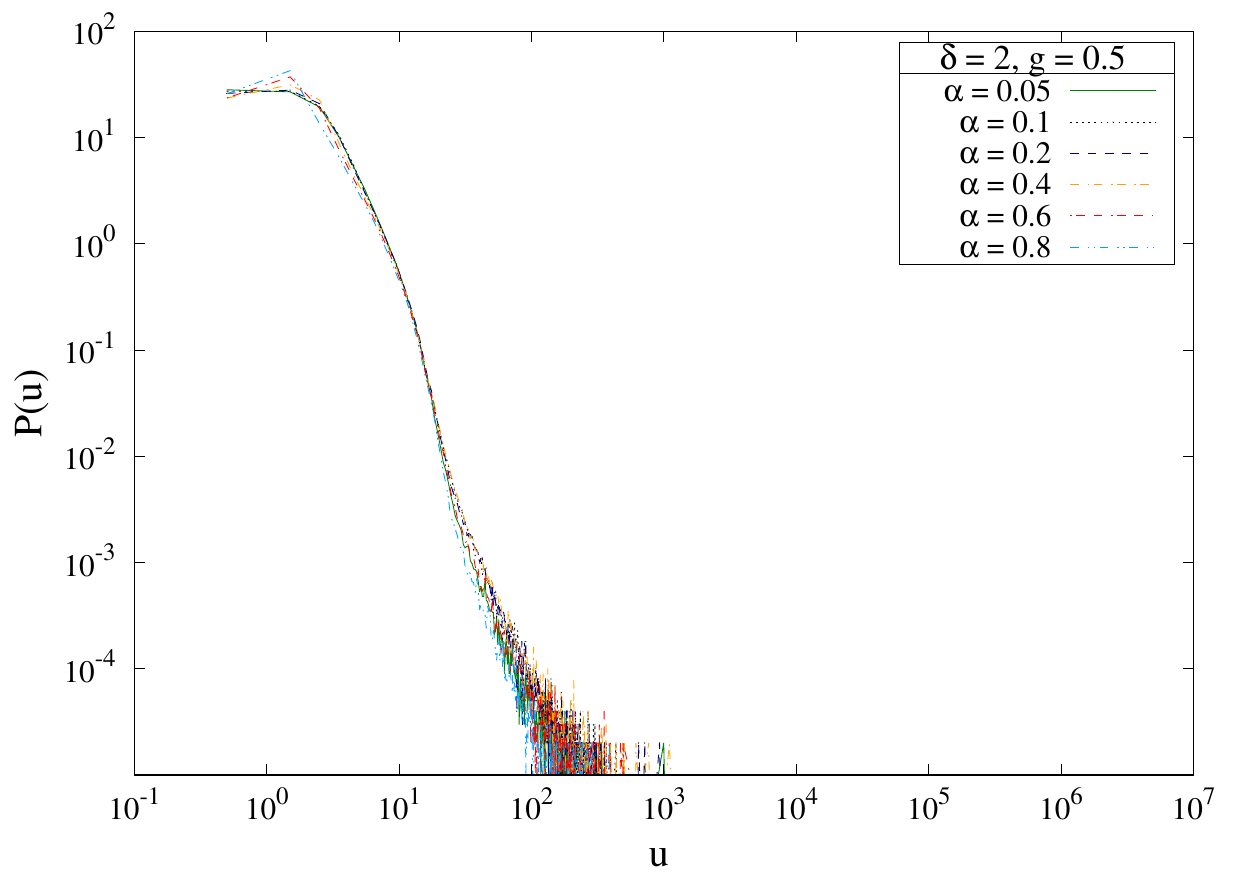}}

\subfloat[$\delta = 3$]{\includegraphics[width=3.0in]{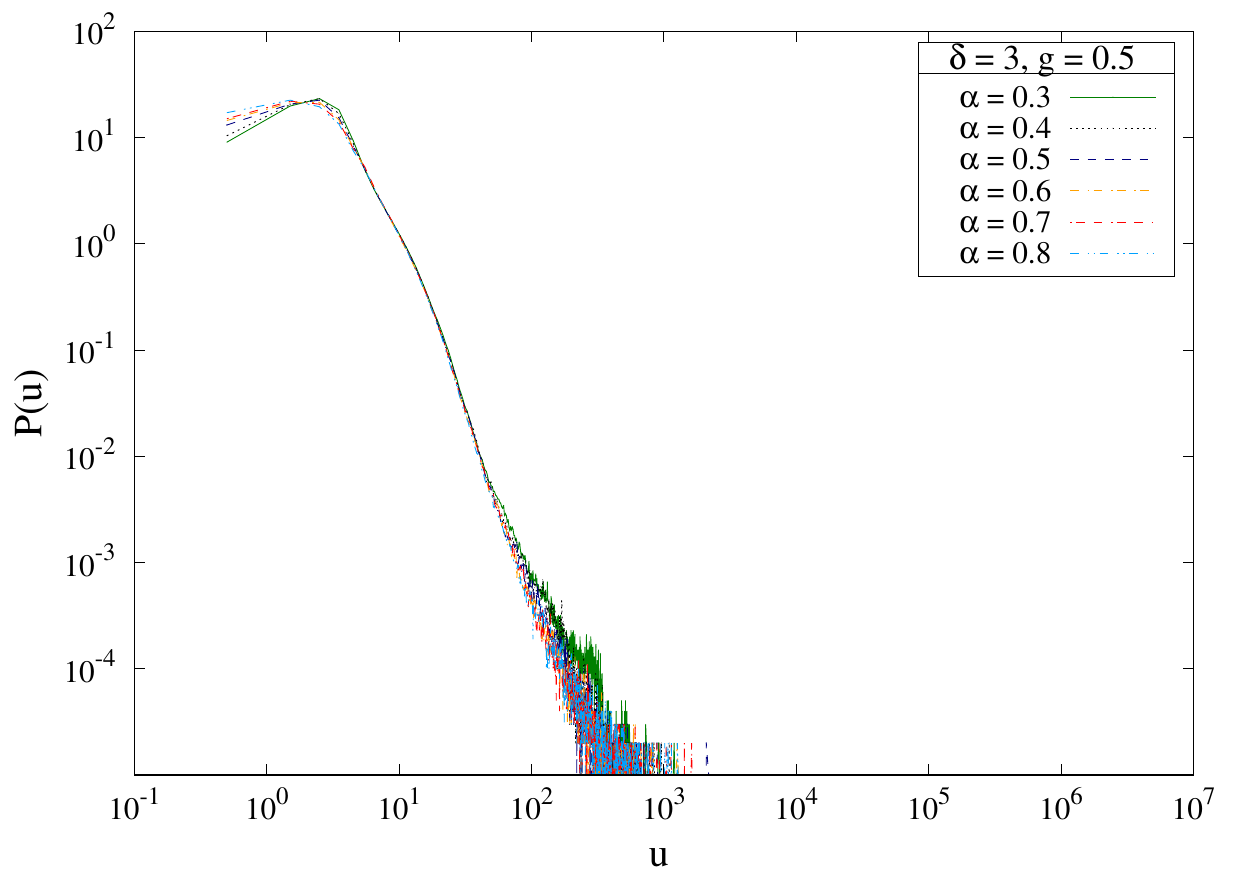}}
\subfloat[$\delta = 4$]{\includegraphics[width=3.0in]{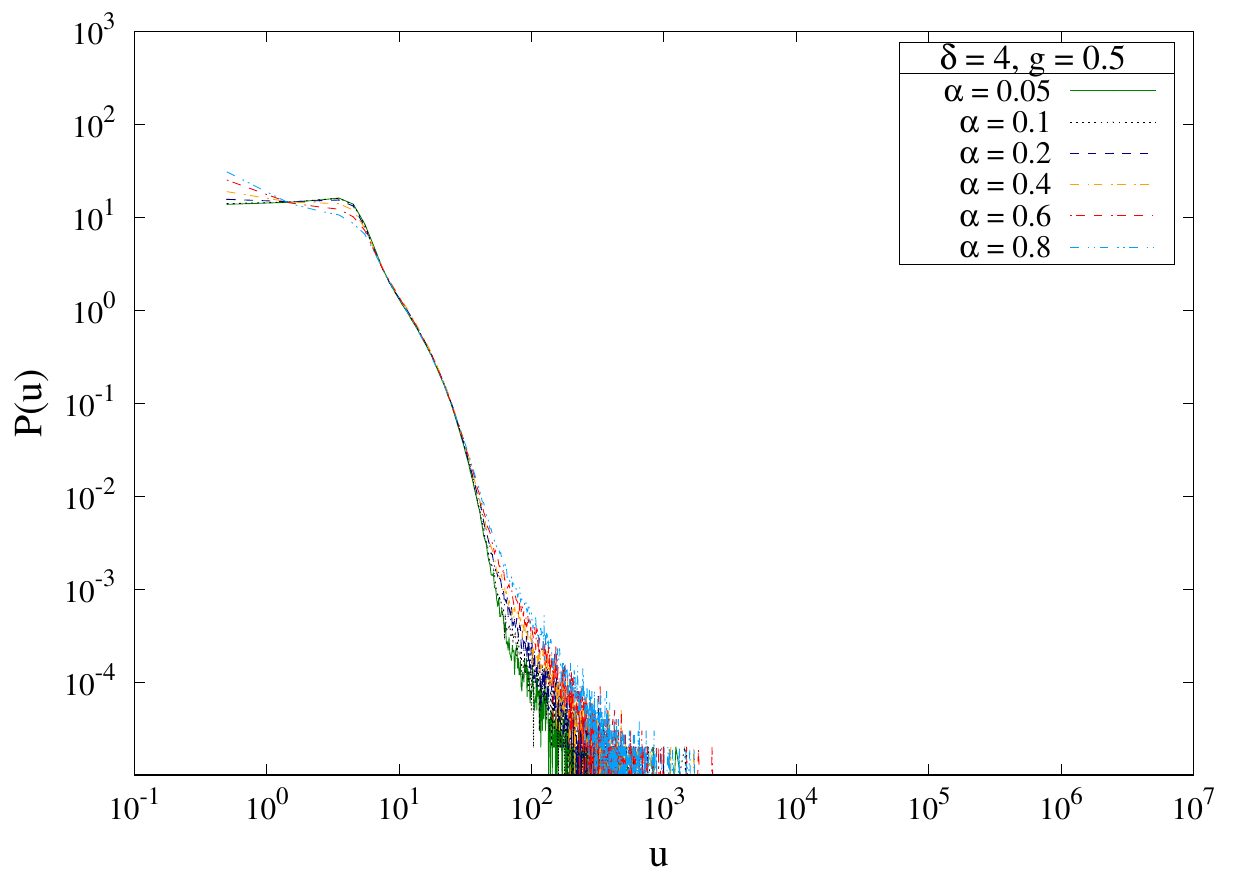}}

\caption{The probability distribution $P(u)$ of the magnitude of the drift term $u$ for the superpotential $W'(\phi) = -ig (i \phi)^{(1+\delta)}$ on a log-log plot. Simulations were performed for $\delta = 1$ (Top-Left), $\delta = 2$ (Top-Right), $\delta = 3$ (Bottom-Left) and $\delta = 4$ (Bottom-Right) with coupling constant $g = 0.5$. We used adaptive Langevin step size $\Delta \tau \leq 5 \times 10^{-5}$ and generation steps $N_{\rm gen} = 10^7$.}
\label{fig:drift_delta}

\end{figure*}

In Refs. \cite{Nagata:2016vkn, Nagata:2018net} the authors demonstrated, in a few simple models, that the probability of the drift term should be suppressed exponentially at larger magnitudes in order to guarantee the correctness of complex Langevin method. However, in the models we investigated in this work we see that the probability distribution falls off like a power law with $u$, even though we have excellent agreements with corresponding analytical results, wherever applicable. In Fig. \ref{fig:drift_dw} we show the probability distribution $P(u)$ against $u$ for the superpotential $W'(\phi) = g (\phi^2 +\mu^2)$ on a log-log plot. In Fig. \ref{fig:drift_delta} we show the probability distribution $P(u)$ of the magnitude of drift term $u$ for superpotential $W'(\phi) = -ig (i \phi)^{(1+\delta)}$ on a log-log plot. In both cases we see that the distribution falls off like a power law for large $u$ values. This needs further investigations, and we save it for future work.

\section{Simulation Data Tables}
\label{app:data-tables}

\begin{table*}[h]
\begin{tabular}{| c | c | c | c  | c |} 
	\hline \hline
	$W'$& $~~\mu~~$  & $~~g~~$ & $~~~~~\alpha~~~~~$ & $~~~~~~\langle B \rangle |_{\alpha}~~~~~~$ \\ [1.5ex] \hline\hline
	\multirow{10}{*}{$g\Big(\phi^2 + \mu^2\Big)$} &\multirow{10}{*}{$2.0$} & \multirow{5}{*}{$1.0$}
	
			&  	0.05	 &	 $ -0.0003(12) - i  4.2250 (72)$ \\ \cline{4-5}
	&&		&  	0.1	 	 &	 $ -0.0015(23) - i  4.2283 (72)$ \\ \cline{4-5}
	&&		&  	0.2		 &	 $ -0.0056(37) - i  4.2261 (72)$ \\ \cline{4-5}
	&&	 	&  	0.4		 &	 $ -0.0025(65) - i  4.2065 (72)$  \\ \cline{4-5}
	&&	 	&  	0.6	 	 &	 $ 0.0076(62)  - i  4.1820 (74)$  \\ \cline{4-5}
	&&	 	&  	0.8		 &	 $ 0.0077(61)  - i  4.1537 (74)$   \\ \cline{4-5} 
	&&	&  	$\alpha \rightarrow 0$ &	 $ -0.0003(35) - i 4.2340(123)$  \\ \cline{3-5} 
	&&	\multirow{5}{*}{$3.0$}	
			&  	0.05	 &	 $ 0.0001(1)  - i  12.0820 (11)$ \\ \cline{4-5}
	&&		&  	0.1		 &	 $ 0.0001(2)  - i  12.0813 (11)$ \\ \cline{4-5}
	&&		&  	0.2		 &	 $ 0.0000(4)  - i  12.0796 (11)$	\\ \cline{4-5}
	&&	 	&  	0.4		 &	 $ 0.0002(7)  - i  12.0735 (11) $   \\ \cline{4-5}
	&&	 	&  	0.6		 &	 $ 0.0006(9)  - i  12.0662 (11)$  \\ \cline{4-5}
	&&	 	&  	0.8		 &	 $ 0.0004(10) - i  12.0567 (11) $ \\ \cline{4-5}
	&&	&  	$\alpha \rightarrow 0$ &	 $ 0.0001 (4) - i 12.0840(18)$ \\ \hline
\end{tabular}
\caption{\label{tab:sqw_B_mu2p0}The expectation values $\langle B \rangle_\alpha$ obtained using complex Langevin simulations for the model with superpotential $W' = g \left( \phi^2 + \mu^2 \right)$. In the limit $\alpha \to 0, \langle B \rangle_\alpha \neq 0 $. Thus SUSY is broken in this model.}
\end{table*}

\begin{table*}[h]
	\begin{tabular}{| c | c | c | c  | c |} 
		\hline \hline
		$W'$& $~~\mu~~$  & $~~g~~$ & $~~~~~\alpha~~~~~$ & $~~~~~~\langle B \rangle |_{\alpha}~~~~~~$ \\ [1.5ex] \hline\hline
		\multirow{10}{*}{$ig\Big(\phi^2 + \mu^2\Big)$} &\multirow{10}{*}{$2.0$} & \multirow{5}{*}{$1.0$}
		
				&  	0.05	 &	 $ -0.0018 (41) - i  0.0006 (337)$ \\ \cline{4-5}
		&&		&  	0.1		 &	 $ -0.0020 (41) + i  0.0008 (337)$ \\ \cline{4-5}
		&&		&  	0.2		 &	 $ -0.0026 (41) + i  0.0035 (336)$ \\ \cline{4-5}
		&&	 	&  	0.4		 &	 $ -0.0049 (41) + i  0.0084 (336)$  \\ \cline{4-5}
		&&	 	&  	0.6	 	 &	 $ -0.0084 (40) + i  0.0123 (338)$  \\ \cline{4-5}
		&&	 	&  	0.8		 &	 $ -0.0125 (40) + i  0.0150 (337)$   \\ \cline{4-5} 
		&&	&  	$\alpha \rightarrow 0$ &	 $ -0.0009(70) - i 0.0017(576)$  \\ \cline{3-5} 
		&&	\multirow{5}{*}{$3.0$}	
				&  	0.05	 &	 $ 0.0002(5)   + i  0.0009 (133)$ \\ \cline{4-5}
		&&		&  	0.1		 &	 $ 0.0002(5)   + i  0.0011 (133)$ \\ \cline{4-5}
		&&		&  	0.2		 &	 $ 0.0001(5)   + i  0.0014 (133)$	\\ \cline{4-5}
		&&	 	&  	0.4		 &	 $ -0.0001 (5) + i  0.0021 (133) $   \\ \cline{4-5}
		&&	 	&  	0.6		 &	 $ -0.0005 (5) + i  0.0026 (133)$  \\ \cline{4-5}
		&&	 	&  	0.8		 &	 $ -0.0009 (5) + i  0.0031 (133) $ \\ \cline{4-5}
		&&	&  	$\alpha \rightarrow 0$ &	 $ 0.0003(9) + i 0.0008(227)$ \\ \hline
	\end{tabular}
	\caption{The expectation values $\langle B \rangle_\alpha$ obtained using complex Langevin simulations for the model with superpotential $W' = ig\Big(\phi^2 + \mu^2\Big)$. We see that, in the limit $\alpha \to 0, \langle B \rangle_\alpha = 0 $. Thus SUSY is preserved in this model.}
	\label{tab:isqw_mu2p0}
\end{table*}

\begin{table*}[h]
	\begin{tabular}{| c | c  | c | c |} 
		\hline \hline
		$~~~~k~~~~$& $~~~~~\alpha~~~~~$ & $~~~~~~~$ $~~~~~~\langle B \rangle |_{\alpha}~~~~~~$ $~~~~~~~~~$ & $~~~~$ SUSY $~~~~$ \\ [1.5ex] \hline\hline
		\multirow{6}{*}{$ 3 $}	
		&  	0.05	 &	 $ 0.0083 (15) - i  0.0018 (447)$  & \multirow{6}{*}{Preserved} \\ \cline{2-3}
		&  	0.1		 &	 $ 0.0162 (24) - i  0.0023 (443)$  &\\ \cline{2-3}
		&  	0.2		 &	 $ 0.0275 (37) - i  0.0030 (454)$  &\\ \cline{2-3}
		&  	0.4		 &	 $ 0.0531 (57) + i  0.0121 (440)$  &\\ \cline{2-3}
		&  	0.6	 	 &	 $ 0.0677 (71) - i  0.0078 (428)$  &\\ \cline{2-3}
		&  	0.8		 &	 $ 0.0789 (82) - i  0.0177 (437)$  & \\ \cline{2-3}
		&  	$\alpha \rightarrow 0$ &	 $ 0.0025(40) - i 0.0024(761) $  &\\ \cline{1-4} 
		\multirow{6}{*}{$4 $}	
		&  	0.05	 &	 $ -0.0010 (10) - i  1.2774 (70)$ & \multirow{6}{*}{Broken}\\ \cline{2-3}
		&  	0.1		 &	 $ -0.0032 (20) - i  1.2738 (71)$ &\\ \cline{2-3}
		&  	0.2		 &	 $ -0.0158 (36) - i  1.2649 (76)$ &\\ \cline{2-3}
		&  	0.4		 &	 $ -0.0425 (62) - i  1.2571 (80)$ & \\ \cline{2-3}
		&  	0.6	 	 &	 $ -0.0519 (81) - i  1.2373 (86)$ & \\ \cline{2-3}
		&  	0.8		 &	 $ -0.0719 (85) - i  1.2044 (98)$ &  \\ \cline{2-3}
		&  	$\alpha \rightarrow 0$ &	 $ 0.0044(31) - i 1.2800(126) $  &\\ \hline
	\end{tabular}
	\caption{\label{tab:real-gen-poly}The expectation values $\langle B \rangle_\alpha$ obtained using complex Langevin simulations for the models with superpotentials $W' = g_k \phi^k + g_{k-1} \phi^{k-1} + \cdots + g_0$ with $g_k = g_{k-1} = \cdots = g_0 = 1$ and $k = 3, 4$.}
\end{table*}

\begin{table*}[h]
	\begin{tabular}{| c | c  | c | c | c |} 
		\hline \hline
		$W'$& $~~\mu~~$  & $~~g~~$ & $~~~~~\alpha~~~~~$ & $~~~~~~\langle B \rangle |_{\alpha}~~~~~~$ \\ [1.5ex] \hline\hline
		\multirow{10}{*}{$ig\phi \Big(\phi^2 + \mu^2\Big)$} &\multirow{10}{*}{$2.0$} & \multirow{5}{*}{$1.0$}
				&  	0.05	 &	 $-0.0002(3)  + i 3.3561(23) $ \\ \cline{4-5}
		&&		&  	0.1		 &	 $-0.0003(4)  + i 3.3562(23) $ \\ \cline{4-5}
		&&		&  	0.2		 &	 $-0.0008(7)  + i 3.3553(23) $ \\ \cline{4-5}
		&&	 	&  	0.4		 &	 $-0.0015(12) + i 3.3482(24) $  \\ \cline{4-5}
		&&	 	&  	0.6	 	 &	 $-0.0026(15) + i 3.3428(24) $  \\ \cline{4-5}
		&&	 	&  	0.8		 &	 $-0.0037(17) + i 3.3322(24) $   \\ \cline{4-5} 
		&&	&  	$\alpha \rightarrow 0$ &	 $ 0.0000(8) + i 3.3585(40)$  \\ \cline{3-5} 
		&&	\multirow{5}{*}{$3.0$}	
				&  	0.05	 &	 $ 0.0000(0)  + i 9.3434(7) $ \\ \cline{4-5}
		&&		&  	0.1		 &	 $ 0.0000(0)  + i 9.3430(7) $ \\ \cline{4-5}
		&&		&  	0.2		 &	 $-0.0000(0)  + i 9.3425(7) $ \\ \cline{4-5}
		&&	 	&  	0.4		 &	 $-0.0002(2)  + i 9.3408(7) $  \\ \cline{4-5}
		&&	 	&  	0.6	 	 &	 $-0.0005(2)  + i 9.3380(7) $  \\ \cline{4-5}
		&&	 	&  	0.8		 &	 $-0.0007(3)  + i 9.3352(8) $   \\ \cline{4-5} 
		&&	&  	$\alpha \rightarrow 0$ &	 $ 0.0000(1) + i 9.3440(13)$  \\ \hline
		
	\end{tabular}
	\caption{The expectation values $\langle B \rangle_\alpha$ obtained using complex Langevin simulations for the model with superpotential $W' = ig\phi \Big(\phi^2 + \mu^2\Big)$ with $g = 1, 3$ and $\mu = 2$. We see that SUSY is broken in this model.}
	\label{tab:sqw_iphi_mu2p0}
\end{table*}

\begin{table*}
	\begin{tabular}{| c | c | c | c |} 
		\hline\hline
		$ ~~~~~\delta~~~~~$  & $~~~~ \alpha ~~~~$ &  $ \langle B \rangle |_{\alpha} $  & $~~~~$ SUSY $~~~~$ \\[1.6ex]
		\hline
		\hline
		\multirow{7}{*}{$1.0$}
		&0.4	& $-0.2498 (224)   - i 0.2109 (487) $ & \multirow{7}{*}{Preserved}\\ \cline{2-3}
		&0.5	& $-0.2580 (202)   - i 0.2998 (450) $ & \\ \cline{2-3}
		&0.6	& $-0.2617 (186)   - i 0.3504 (420)$ & \\ \cline{2-3}
		&0.7	& $-0.2726 (172)   - i 0.3719 (403) $ & \\ \cline{2-3}
		&0.8	& $-0.2858 (160)   - i 0.3998 (391) $& \\  \cline{2-3}
		&0.9	& $-0.3113 (149)   - i 0.3978 (391) $ & \\  \cline{2-3}	
		& $\alpha \rightarrow 0$ &	 $ - 0.2433(2213) + i 0.0742(5080)  $  &\\ [1.5ex]
		\hline

		\multirow{7}{*}{$3.0$}
		&0.3	& $ 0.0567 (32) + i 0.4452 (566) $ &  \multirow{7}{*}{Preserved} \\ \cline{2-3}
		&0.4	& $ 0.0738 (32) + i 0.4544 (538) $ & \\ \cline{2-3}
		&0.5	& $ 0.0870 (34) + i 0.4387 (475) $ &   \\  \cline{2-3}
		&0.6	& $ 0.0961 (43) + i 0.4284 (416) $ & \\  \cline{2-3}
		&0.7	& $ 0.1034 (53) + i 0.3946 (441) $ & \\  \cline{2-3}
		&0.8	& $ 0.1027 (64) + i 0.3539 (398) $ & \\  \cline{2-3}	
		& $\alpha \rightarrow 0$ &	 $ 0.0054(311) + i 0.3625(4025)  $ & \\ [1.5ex]
		\hline
	\end{tabular}
	\caption{The expectation values $\langle B \rangle_\alpha$ obtained using complex Langevin dynamics for the models with superpotential $W'(\phi) = -ig (i \phi)^{(1+\delta)}$ with $g = 0.5$ ad $\delta = 1, 3$, respectively.}
	\label{tab:delta_1p0_3p0}
\end{table*}

\begin{table*}
\begin{tabular}{| c | c | c | c |} 
	\hline\hline
	$ ~~~~~~\delta ~~~~~~$  & $~~~~ \alpha ~~~~$ &  $ ~~~~~~~~~~~~~~~~~~~\langle B \rangle |_{\alpha}~~~~~~~~~~~~~~~~ $  & $~~~~$ SUSY $~~~~$ \\[1.6ex]
	\hline
	\hline
	\multirow{7}{*}{$2.0$}
	&0.05	& $ 0.0014 (36)  - i 0.0609 (1416) $ & \multirow{7}{*}{Preserved}\\ \cline{2-3}
	&0.1	& $ 0.0102 (50)  - i 0.1986 (1101) $ & \\ \cline{2-3}
	&0.2	& $ 0.0079 (80)  - i 0.0679 (1004) $ & \\ \cline{2-3}
	&0.4	& $ 0.0134 (96)  - i 0.0627 (701) $ & \\ \cline{2-3}
	&0.6	& $ 0.0079 (120) - i 0.0208 (655)$ & \\  \cline{2-3}
	&0.8	& $-0.0068 (126) + i 0.0294 (595) $ & \\  \cline{2-3}	
	& $\alpha \rightarrow 0$ &	 $ 0.0019(84) - i 0.1423(1932)$  &\\ [1.5ex]
	\hline
	
	\multirow{7}{*}{$4.0$}
	&0.05	& $-0.0005 (20) - i 0.0155 (1257) $ & \multirow{7}{*}{Preserved}\\ \cline{2-3}
	&0.1	& $-0.0017 (37) - i 0.0435 (1043) $ & \\ \cline{2-3}
	&0.2	& $ 0.0059 (48) + i 0.0787 (817) $ & \\ \cline{2-3}
	&0.4	& $ 0.0016 (64) + i 0.0108 (648) $ & \\ \cline{2-3}
	&0.6	& $ 0.0132 (70) + i 0.0761 (526)$ & \\  \cline{2-3}
	&0.8	& $ 0.0063 (68) + i 0.0258 (418) $ & \\  \cline{2-3}	
	& $\alpha \rightarrow 0$ &	 $ -0.0018(48) - i 0.0092(1712)$  &\\ [1.5ex]
	\hline

\end{tabular}
\caption{The expectation values $\langle B \rangle_\alpha$ obtained using complex Langevin dynamics for the models with superpotential $W'(\phi) = -ig (i \phi)^{(1+\delta)}$ with $g = 0.5$ and $\delta = 2, 4$, respectively.}
\label{tab:delta_2p0_4p0}
\end{table*}

\begin{table*}[h]
	\begin{tabular}{| c | c | c | c  | c |} 
		\hline \hline
		$W'$& $~~\mu~~$  & $~~g~~$ & $~~~~~\alpha~~~~~$ & $~~~~~~\langle \widetilde{L}B \rangle |_{\alpha}~~~~~~$ \\ [1.5ex] \hline\hline
		\multirow{10}{*}{$g\Big(\phi^2 + \mu^2\Big)$} &\multirow{10}{*}{$2.0$} & \multirow{5}{*}{$1.0$}
		
				&  	0.05	 &	 $ -0.0019 (78)   +  i  0.0020 (1379)$ \\ \cline{4-5}
		&&		&  	0.1		 &	 $ -0.0133 (130)  +  i  0.0792 (1388)$ \\ \cline{4-5}
		&&		&  	0.2		 &	 $ -0.0322 (264)  +  i  0.0996 (1368)$ \\ \cline{4-5}
		&&	 	&  	0.4		 &	 $ -0.0090 (420)  +  i  0.0486 (1329)$  \\ \cline{4-5}
		&&	 	&  	0.6	 	 &	 $ -0.0852 (685)  -  i  0.0191 (1444)$  \\ \cline{4-5}
		&&	 	&  	0.8		 &	 $ -0.0252 (539)  +  i  0.0264 (1258)$   \\ \cline{4-5} 
		&&	&  	$\alpha \rightarrow 0$ &	 $ 0.0023 (230) + i 0.0555 (2357) $  \\[1.5ex] \cline{3-5} 
		&&	\multirow{5}{*}{$3.0$}	
				&  	0.05	 &	 $ 0.0257 (250)    - i  0.0304 (1561)$ \\ \cline{4-5}
		&&		&  	0.1		 &	 $-0.0682 (724)    + i  0.0222 (1660)$ \\ \cline{4-5}
		&&		&  	0.2		 &	 $ 0.0678 (966)    - i  0.0088 (1712)$	\\ \cline{4-5}
		&&	 	&  	0.4		 &	 $ 0.1330 (1656)   + i  0.2933 (2790) $   \\ \cline{4-5}
		&&	 	&  	0.6		 &	 $ 0.0816 (2031)   + i  0.4755 (2733)$  \\ \cline{4-5}
		&&	 	&  	0.8		 &	 $-0.2429 (1627)   + i  0.1306 (1682) $ \\ \cline{4-5}
		&&	&  	$\alpha \rightarrow 0$ & $ 0.0098 (778) - i 0.0840 (3020) $ \\[1.5ex] \hline
	\end{tabular}
	\caption{\label{tab:sqw_LO}The expectation values $\langle \widetilde{L}B \rangle_\alpha$ obtained using complex Langevin simulations for the models with superpotential $W' = g\Big(\phi^2 + \mu^2\Big)$.}
\end{table*}

\begin{table*}[h]
	\begin{tabular}{| c | c | c |}
		\hline 
		\hline
		$~~~~~~\delta~~~~~~$  & $~~~~~~~\alpha~~~~~~~$  & $~~~~~~~~~~~~~~~~~~~~~~~\langle \widetilde{L} B \rangle |_\alpha~~~~~~~~~~~~~~~~~~~~~$  \tabularnewline
		\hline 
		\hline 
		\multirow{4}{*}{1.0}
		& 0.4  & $-0.6263 (3592)    + i 0.0042 (3062) $  \tabularnewline
		\cline{2-3} 
		& 0.5  & $-0.1442 (2127)    + i 0.0202 (1752)$  \tabularnewline
		\cline{2-3}
		& 0.6  & $-0.0239 (1517)    + i 0.0400 (1375)$  \tabularnewline
		\cline{2-3} 
		& 0.7  & $~~0.0198 (1192)    + i 0.0387 (1171)$  \tabularnewline
		\cline{2-3} 
		& 0.8  & $-0.0107 (1169)    + i 0.0494 (988)$  \tabularnewline
		\cline{2-3} 
		& 0.9  & $-0.0401 (990)     + i 0.0104 (915)$  \tabularnewline
		\cline{2-3} 
		& $\alpha \rightarrow 0$ 
		& $ -1.2716 (2.421) - i 0.1173 (2.122) $  \tabularnewline [1.5ex]
		\hline 
		\multirow{4}{*}{3.0}
		& 0.3  & $ 0.1846 (5176)    + i 0.1366 (3738) $  \tabularnewline
		\cline{2-3} 
		& 0.4  & $-0.3282 (1845)    + i 0.0443 (3164)$  \tabularnewline
		\cline{2-3}
		& 0.5  & $-0.2215 (1856)    + i 0.1869 (2377)$  \tabularnewline
		\cline{2-3} 
		& 0.6  & $-0.2046 (1456)    + i 0.2870 (1969)$  \tabularnewline
		\cline{2-3} 
		& 0.7  & $ 0.0022 (1476)    + i 0.2841 (2076)$  \tabularnewline
		\cline{2-3} 
		& 0.8  & $-0.0483 (1412)    + i 0.1976 (1960)$  \tabularnewline
		\cline{2-3} 
		& $\alpha \rightarrow 0$
		& $-0.3031 (2.181) - i 0.2210 (2.335)  $   \tabularnewline  [1.5ex]
		\hline 
	\end{tabular}
	\caption{\label{tab:delta_LO_1p0_3p0}The simulated values of $\widetilde{L} B_\alpha$ for the models with superpotential $W'(\phi) = -ig (i \phi)^{(1+\delta)}$, with coupling parameter $g = 0.5$ and $\delta = 1, 3$, respectively.}
\end{table*}

\clearpage

\begin{table*}
\begin{tabular}{| c | c | c |}
	\hline 
	\hline
		$~~~~~~\delta~~~~~~$  & $~~~~~~~\alpha~~~~~~~$  & $~~~~~~~~~~~~~~~~~~~~~~~\langle \widetilde{L} B \rangle |_\alpha~~~~~~~~~~~~~~~~~~~~~$  \tabularnewline
	\hline 
	\hline 
	\multirow{4}{*}{2.0} 
	& 0.05 & $ 0.0036 (49)  - i 0.1572 (1315)$  \tabularnewline
	\cline{2-3}  
	& 0.1 	& $ 0.0082 (94)  - i 0.2145 (1273)$  \tabularnewline
	\cline{2-3}  
	& 0.2  & $ 0.0113 (156) - i 0.1480 (1359)$  \tabularnewline
	\cline{2-3} 
	& 0.4  & $ 0.0066 (246) - i 0.1409 (1300)$  \tabularnewline
	\cline{2-3} 
	& 0.6  & $-0.0014 (312) - i 0.1029 (1280)$  \tabularnewline
	\cline{2-3} 
	&  0.8 & $-0.0023 (348) - i 0.1132 (1245)$  \tabularnewline
	\cline{2-3} 
	& $\alpha \rightarrow 0$ 
	& $  0.0034 (142) - i 0.1906 (2223) $  \tabularnewline
	\hline 
	\multirow{4}{*}{4.0}
	& 0.05 & $-0.0086 (127)  + i 0.3919 (2944)$  \tabularnewline
	\cline{2-3}
	& 0.1	& $-0.0292 (202)  + i 0.3050 (2945)$  \tabularnewline
	\cline{2-3}  
	& 0.2  & $-0.0127 (310)  + i 0.5222 (2910)$  \tabularnewline
	\cline{2-3} 
	& 0.4  & $ 0.0295 (503)  + i 0.4377 (2889)$  \tabularnewline
	\cline{2-3} 
	& 0.6  & $ 0.0497 (595)  + i 0.3674 (2690)$  \tabularnewline
	\cline{2-3} 
	& 0.8  & $-0.0781 (1796) + i 0.1504 (3194)$  \tabularnewline
	\cline{2-3} 
	& $\alpha \rightarrow 0$
	& $  -0.0171 (361) + i 0.3794 (5019) $   \tabularnewline
	\hline 
\end{tabular}
\caption{\label{tab:delta_LO_2p0_4p0}The simulated values of $\widetilde{L} B_\alpha$ for the models with superpotential $W'(\phi) = -i g (i \phi)^{(1+\delta)}$, with coupling parameter $g = 0.5$ and $\delta = 2, 4$, respectively.}
\end{table*}

\bibliographystyle{utphys.bst}
\bibliography{susy_lang}
\end{document}